\documentclass{aa}
\pdfoutput=1
\usepackage[varg]{txfonts}
\bibpunct{(}{)}{;}{a}{}{,}
\usepackage{amssymb}
\usepackage{graphicx}
\usepackage{url}
\usepackage{mathrsfs}
\usepackage[usenames,dvipsnames]{color}
\usepackage{lscape}

\defcitealias{2012MNRAS.425.1007B}{BN12}

\begin{document}

\title{Type Ia supernova Hubble diagram with near-infrared and optical observations\thanks{Partly based on observations made
          with ESO telescopes at the Paranal Observatory under program IDs 079.A-0192 and 081.A-0734.}\fnmsep\thanks{The tables with the near-infrared $J$ and $H$ $K$-corrections are only available in electronic form at the CDS via anonymous ftp to 
{\tt cdsarc.u-strasbg.fr (130.79.128.5)}
or via {\tt http://cdsweb.u-strasbg.fr/cgi-bin/qcat?J/A+A/}}}

\author{V.~Stanishev\inst{1,2}, A.~Goobar\inst{3,4}, R.~Amanullah\inst{3,4}, B.~Bassett\inst{5,6,7}, Y.T.~Fantaye\inst{8}, P.~Garnavich\inst{9}, R.~Hlozek\inst{10}, J.~Nordin\inst{11}, P.M.~Okouma\inst{6,12}, L.~\"Ostman\inst{3}, M.~Sako\inst{13}, R.~Scalzo\inst{14}, and M.~Smith\inst{15}
   }

\institute{
Department of Physics, Chemistry and Biology, IFM, Link\"oping University, SE-581 83 Link\"oping,
Sweden 
\and
CENTRA - Centro Multidisciplinar de Astrof\'isica, Instituto Superior T\'ecnico, Av.
Rovisco Pais 1, 1049-001 Lisbon, Portugal  
\and
Department of Physics, Stockholm University, Albanova University Center, S--106 91 Stockholm, Sweden 
\and
The Oskar Klein Centre, Stockholm University, S--106 91 Stockholm, Sweden  
\and 
African Institute for Mathematical Sciences, 6-8 Melrose Road, Muizenberg, Cape Town, Republic of South Africa
\and
South African Astronomical Observatory, Observatory, Cape Town, Republic of South Africa
\and
Department of Maths and Applied Maths, University of Cape Town, Rondebosch 7701, Republic of South Africa
\and 
Department of Mathematics, University of Rome Tor Vergata, Rome, Italy
\and 
Department of Physics, University of Notre Dame, Notre Dame, IN 46556, USA
\and 
Department of Astrophysical Sciences, Princeton University, Peyton Hall, 4 Ivy Lane, Princeton, NJ 08544, USA
\and
Institut f\"ur Physik, Humboldt-Universit\"at zu Berlin, Newtonstra{\ss}e 15, 12489 Berlin, Germany
\and
Department of Physics, University of the Western Cape, Belleville, Cape Town,  Republic of South Africa
\and
Research School of Astronomy and Astrophysics, The Australian National University,
Mount Stromlo Observatory, Cotter Road, Weston ACT 2611, Australia
\and
Department of Physics and Astronomy, University of Pennsylvania,
209 South 33rd Street, Philadelphia, PA 19104, USA
\and
School of Physics \& Astronomy, University of Southampton, Highfield, Southampton SO17 1BJ, U.K.
}

\offprints{vallery.stanishev@gmail.com}

\date{Received date; Accepted date}

\abstract{ Type Ia Supernovae (SNe~Ia) have been used as
  standardizable candles in the optical wavelengths to measure
  distances with an accuracy of $\sim7$\% out to redshift
  $z\sim1.5$. There is evidence that in the near-infrared (NIR)
  wavelengths SNe~Ia are even better standard candles, however, NIR observations are much more time-consuming.
}
{We aim to test whether the NIR peak magnitudes could be accurately estimated with only a single observation obtained close to maximum light, provided that the time of $B$ band maximum, the $B-V$ color at maximum and the optical stretch parameter are known.
}
{
We present multi-epoch  $UBVRI$ and single-epoch $J$ and $H$ photometric observations of 16 SNe~Ia in the redshift range $z=0.037-0.183$, doubling the leverage of the current SN~Ia NIR Hubble diagram and the number of SNe beyond redshift 0.04. This sample was analyzed together with 102 NIR and 458 optical light curves (LCs) of normal SNe~Ia from the literature.}   
{
The analysis of 45 NIR LCs with well-sampled first maximum shows that
a single template accurately describes the LCs if its time axis is stretched with the optical stretch parameter. This allows us to estimate the peak NIR magnitudes of SNe with only few observations obtained within ten days from $B$-band maximum.
The NIR Hubble residuals show weak correlation with $\Delta M_{15}$ and the color excess $E(B-V)$,  and for the first time we report a potential dependence  on the  $J_\mathrm{max}-H_\mathrm{max}$ color. With these corrections, 
the intrinsic NIR luminosity scatter of SNe~Ia is estimated to be $\sim0.10$ mag, which 
is smaller than what can be derived for a similarly heterogeneous sample at optical wavelengths.
Analysis of both NIR and optical data shows that the dust extinction in the host galaxies corresponds to a low $R_V\simeq1.8-1.9$.
}
{
We conclude that SNe~Ia are at least as good standard candles in the NIR as in the optical and are potentially less affected by systematic uncertainties. We extended the NIR SN~Ia Hubble diagram to its nonlinear part at $z\sim0.2$ and  confirmed that it is feasible to accomplish this result with very modest sampling of the NIR LCs, if complemented by well-sampled optical LCs. 
With future facilities it will be possible to extend the NIR Hubble diagram beyond redshift $z\simeq1,$ and our results suggest that the most efficient way to achieve this would be to obtain a single observation close to the NIR maximum. 
}


\titlerunning{Type Ia Supernova Cosmology in the Near-Infrared}
\authorrunning{Stanishev et al.}

\maketitle

\section{Introduction}

Since the discovery of the accelerated expansion of the universe
\citep{riess98,p99} the world sample of
Type Ia supernovae (SNe~Ia) at cosmological distances has grown substantially. Mainly
thanks to the combined effort from the Sloan Digital Sky Survey II (SDSS-II) and SNLS surveys, the current optical SN~Ia Hubble diagram is built from 740 well measured objects
\citep[][and references therein]{2014A&A...568A..22B} and 1049 SNe with the addition of PanSTARSS1 \citep[][and references therein]{2017arXiv171000845S}.
  These measurements provide the most accurate study of the expansion history of the universe
to date, especially when combined with baryon acoustic oscillation (BAO) observations from 6dFGS, SDSS, BOSS and WiggleZ \citep{2011MNRAS.416.3017B,2014MNRAS.441...24A,2014MNRAS.441.3524K,2015MNRAS.449..835R,2017MNRAS.470.2617A}
and CMB constraints \citep{2016A&A...594A..13P}.  However, it
remains unclear at this point whether the precision of the measurements of cosmological
parameters from, for example, LSST \citep{lsst}, WFIRST \citep{2013arXiv1305.5422S}, or the proposed DESIRE survey exploiting the \emph{Euclid} satellite \citep{2014A&A...572A..80A}, most notably the equation of state parameter of the dark energy, will
improve significantly with further increase in the sample of SN~Ia
at rest-frame optical wavelengths. Systematic uncertainties -- instrumental, astrophysical and cosmological -- may pose severe limitations
\citep[see, e.g.,][]{2011ARNPS..61..251G,2011ApJS..192....1C,2014A&A...568A..22B,2017arXiv171000845S}.

  Recent SN~Ia cosmological analyses assume the same $R_V$ for all SNe and recover significantly lower value compared to the typical Milky Way $R_V\simeq3.1$. We lack solid understanding of what causes this. Furthermore, the assumption of universal $R_V$ is almost certainly not correct. First, in the Milky Way $R_V$ varies between 2 and 5.5 and there is no reason why this should not also  be the case for other galaxies. Second,  there is growing evidence that $R_V$ varies between different supernovae (SNe) in the nearby universe \citep[][and references therein]{2010AJ....139..120F,2014ApJ...789...32B,2015MNRAS.453.3300A}. There are also indications that the most heavily reddened SNe tend to have low $R_V\sim1-2$, while in the more moderately reddened events a range of values up to $R_V\sim3$ is observed \citep{2014ApJ...789...32B}. This raises  concerns for possible systematic uncertainties related to the extinction corrections. For example, that the current assumption of a universal reddening law independent of redshift may not hold. Even if the reddening law were to remain unchanged, it is already problematic that different populations would be observed as the redshift increases. At high redshift, the  most reddened SNe will be missed and the sample will be dominated by low-reddening events. Applying color cuts to the nearby sample will not completely prevent the distribution of colors to differ between the low- and high-redshift sample, which can introduce systematic differences. Similarly, the  "luminosity -- light curve (LC) shape" relation used to calibrate SNe~Ia lacks strong theoretical explanation and its evolution with redshift cannot be excluded at this stage.

One alternative that has been proposed is to explore the use of near-infrared (NIR; 1-2.5~$\mu$m)
observations of SNe~Ia. This has recently been used to place constraints on the Hubble constant 
by \citet{2018A&A...609A..72D}.
Observing SN~Ia in the NIR has several important advantages. Firstly,
the extinction corrections are by a factor of four to six smaller than in the optical $B$ band. 
Secondly, there is evidence, both from observations \citep{meikle00,kri_04a,kri_04,2008ApJ...689..377W} and theoretical calculations \citep{2006ApJ...649..939K}, that at maximum SNe~Ia are natural standard candles in the NIR
and no correction for the LC shape is needed for normal SNe.
Although more recent works by  \cite{2010AJ....139..120F} and \cite{2012PASP..124..114K} 
do suggest that such a correction may also be needed in the NIR, it is much smaller than in the optical. Thirdly,
\cite{kri_04,kri_ir_temp} and later \cite{2008ApJ...689..377W} demonstrated that smooth LC 
templates, the time-axis of which is stretched according to the SN optical stretch parameter 
$s_\mathrm{opt}$ \citep{1997ApJ...483..565P},   accurately describe rest-frame $J$ and $H$-band LCs within ten days from $B$-band
maximum. This raises the possibility of accurately measuring the peak NIR magnitudes of SNe~Ia \emph{even with a single observation} within a week from $B$-band maximum, provided the time of $B$-band maximum and $s_\mathrm{opt}$ are known,
and \citet{2012MNRAS.425.1007B} and \citet{2014ApJ...784..105W} have shown indications that 
this may indeed be a viable approach. If further verified, this
could be very important because, due to the large atmospheric background, ground-based observations in the NIR rapidly become
prohibitively time-consuming when observing faint objects and  building well-sampled LCs of
high-redshift SNe~Ia becomes nearly impossible. 
The main goal of this work is to further test the feasibility to measure the 
NIR peak magnitudes with  single observation and extend the NIR Hubble diagram to intermediate 
redshifts. We report on the optical $UBVRI$ and NIR  $J$ and $H$ measurements of 16 SNe~Ia in the
redshift range $z=0.037-0.183$. We present the results of the combined analysis of this 
new data-set together with 102 NIR and 458 optical LCs of normal SNe Ia from the literature.

The paper is organized as follows. In Sect.~\ref{sec:obs} we present the target selection, observations, data reduction, photometry and calibration. Section~\ref{sec:lcfit} provides details on the LC fitting, including the derivation of new NIR LC templates and NIR $K$-corrections. In  Sect.~\ref{sec:res} we present our results and in Sect.~\ref{sec:disc_concl} a discussion followed by our conclusions. Throughout the paper we have assumed the concordance cosmological model with $\Omega_M=0.27$, $\Omega_\Lambda=0.73$ and $h=0.708$.

\section{Observations and data reduction}

\begin{table*}[!t]

\caption{Details of the SNe observed in this work.} 
\label{t:sne}
\begin{tabular}{@{}lcccccc@{}}
\hline
\hline\noalign{\smallskip}
SN ID   &    RA (J2000)         & Decl. (J2000)   & $z_\mathrm{helio}$ & $z_\mathrm{CMB}$ & $E(B-V)_\mathrm{MW}$ & Discovery  \\
\hline\noalign{\smallskip}
SN2007hz           &      21:03:08.95  & $-$01:01:45.1  &    0.1393 &   0.1381  &  0.070  &  SDSS-II \\ 
SN2007ih           &      21:33:10.77  & $-$00:57:36.5  &    0.1710 &   0.1697  &  0.042  & $\cdots$ \\
SN2007ik           &      22:38:53.72  & $-$01:10:01.5  &    0.1830 &   0.1815  &  0.043  & $\cdots$ \\
SN2008bz           &      12:38:57.74  & +11:07:46.2    &    0.0603 &   0.0614  &  0.025  &  ROTSE   \\
SNF20080510-005    &      13:38:24.29  & +09:40:11.3    &    0.0840 &   0.0850  &  0.024  & SNFactory\\ 
SNF20080512-008    &      17:03:35.14  & +31:21:11.7    &    0.0840 &   0.0840  &  0.030  &  $\cdots$\\
SNF20080512-010    &      16:11:04.35  & +52:27:09.9    &    0.0630 &   0.0631  &  0.018  &  $\cdots$\\
SNF20080516-000    &      11:22:29.71  & +56:24:26.5    &    0.0720 &   0.0726  &  0.009  &  $\cdots$\\
SNF20080516-022    &      14:59:27.97  & +58:00:09.8    &    0.0740 &   0.0742  &  0.013  &  $\cdots$\\ 
SNF20080517-010    &      12:36:23.88  & +39:30:43.8    &    0.1200 &   0.1209  &  0.014  &  $\cdots$\\
SNF20080522-000    &      13:36:47.59  & +05:08:30.4    &    0.0450 &   0.0460  &  0.023  &  $\cdots$\\
SNF20080522-001    &      12:51:39.88  & $-$11:59:23.7  &    0.0490 &   0.0502  &  0.039  &  $\cdots$\\
SNF20080522-004    &      13:11:32.70  & $-$18:39:45.3  &    0.1050 &   0.1062  &  0.069  &  $\cdots$\\
SNF20080522-011    &      15:19:58.91  & +04:54:16.9    &    0.0370 &   0.0376  &  0.037  &  $\cdots$\\
SNF20080606-012    &      17:43:57.23  & +51:22:18.8    &    0.0790 &   0.0788  &  0.031  &  $\cdots$\\
SNF20080610-003    &      16:17:37.73  & +03:28:28.7    &    0.0930 &   0.0933  &  0.054  &  $\cdots$\\
\hline 
\end{tabular}
\end{table*}

\label{sec:obs}

\subsection{Target selection and observing strategy}

Between 2007 and 2008 we observed sixteen SNe Ia at redshifts between $z=0.037$ and $z=0.183$. Details of the targets are given in Table~\ref{t:sne}. The targets were selected from the SNe discovered by SDSS-II -- 3 SNe, SNfactory \citep{2002SPIE.4836...61A} -- 12 SNe, and ROTSE -- 1 SN, to be spectroscopically classified before maximum light.  The early classification was essential in order to obtain the NIR observations within a week from maximum and to obtain good enough optical observation to derive the LC parameters. For each SN, a single NIR observation and multiple $UBVRI$ observations  with cadence of  between five and seven days were obtained, except for SNe 2007hz, 2007ih and 2007ik for which the $gri$ photometry obtained by SDSS-II was used \citep{2014arXiv1401.3317S}.  For most SNe the observations started earlier than ten days after the $B$-band maximum light.

Most of the SNe in our sample are close to their host galaxy nuclei and  the underlying nonuniform galaxy background may significantly degrade the photometry. To remove the galaxy background contamination we used images of the host galaxies of the SNe that we obtained about a year after the main survey, when the SN had faded. In the NIR, template images were obtained for all three instruments used. In the optical, template images were only obtained with ALFOSC at NOT. The galaxy template images were aligned with the SN image\footnote{Fourth-order B-spline interpolation was used for all image interpolations in this paper because it has one of the best interpolation properties of all known interpolation schemes, e.g. \cite{bspline}.}, convolved  with a suitable kernel so that the point-spread functions (PSF) of the two images are the same, then scaled to match the flux level of the SN image and subtracted. The SN flux can then be correctly measured on the background-subtracted image.  The image subtraction was carried out with Alard's optimal image subtraction software \citep{optsub1,optsub2}, slightly modified and kindly made available to us by B. Schmidt. SNe 2007hz, SNF20080512-008 and SNF20080516-022 were found to be away of their host galaxies and the galaxy background in the NIR was not a problem. For these SNe no galaxy reference images were obtained and the photometry was performed without image subtraction.

\subsection{Near infrared imaging}

The NIR photometry of the SNe was obtained at three facilities. $J$ and $H$ imaging of the 
three SDSS-II SNe was obtained at VLT/ISAAC \citep{moorwood98} and nine of the others were observed with
NOTCAM \citep{2000SPIE.4008..714A} at the Nordic Optical Telescope. Six SNe were observed at the Japanese InfraRed Survey Facility (IRSF)
1.4m telescope located at SAAO  and equipped with  the imaging camera SIRIUS \citep{2003SPIE.4841..459N}, which can simultaneously observe in $JHK_\mathrm{s}$ bands. Two of the SNe were observed both at  IRSF and NOT. All three instruments are equipped with 1024$\times$1024 pixel Hawaii detectors and filters from the 
 Mauna Kea Observatories Near-Infrared Filter Set (MKO)\footnote{At ISAAC the $J_\mathrm{s}$ filter was used, which corresponds to MKO $J$.} \citep{irfilt}.

The observations consisted of series of dithered images in order to facilitate the night sky background subtraction. 
Twilight sky flat-field images were also obtained. The image reduction followed with small modifications the scheme outlined in 
\cite{2009A&A...507...61S}. All ISAAC images were first corrected for the effect of "electrical ghost"
with the recipe \emph{ghost} within the ESO software library \emph{Eclipse}\footnote{available at \url{http://www.eso.org/projects/aot/eclipse/}}. Then dark current and flat-field corrections were applied to all images. The sky background subtraction was performed with the XDIMSUM package in IRAF.\footnote{All data reduction and calibration was done in IRAF and with our own programs written in IDL. IRAF is distributed by the National Optical Astronomy Observatories, which are operated by the Association of Universities for Research in Astronomy, Inc., under cooperative agreement with the National Science Foundation.}
To estimate the sky background for given image, the eight exposures closest in time were scaled to the same mode and combined with median. The combined sky image was scaled to the mode of the image and subtracted from it. The pedestal on NIR arrays is highly variable in time and may be very difficult to remove accurately. In some of the images, significant residuals were found along the bottom and the middle rows of the array, where the readout of the two halves of  the array starts. To correct for this effect the median value of each row was computed with $2\sigma$ clipping and subtracted. The sky-subtracted images were combined and the combined image was used to create an object mask with the IRAF OBJMASK task. The mask was then applied to each individual image and the sky subtraction  was repeated, but now all pixels belonging to objects were excluded during the background estimation and the residual pedestal subtraction. The only difference was that instead of the median, the mode along the image rows was used to remove the residual pedestal.
  
Before the final image combination residual cosmic
ray hits were identified and cleaned with the Laplacian detection
algorithm of \citet{crs}. ISAAC and NOTCam have significant optical 
image distortions. We derived corrections for this effect 
using observations of dense stellar fields with astrometry from SDSS and combined the 
images with the DRIZZLE algorithm \citep{2002PASP..114..144F}.
For the final combination, the individual images were also assigned weights 
in order to optimize the signal-to-noise ratio (S/N) for faint point sources. The weights
are inversely proportional to the product of the square of the seeing,
the variance of the sky background and the multiplicative factor that
brings the images to the same flux scale.

\subsubsection{Photometry and calibration}

 \begin{table*}[!t]
\caption{NIR photometry of the SNe observed in this work. The 1$\sigma$ uncertainties are given in parentheses.} 
\label{t:nirphot}
\begin{tabular}{@{}lcccrc@{}}
\hline
\hline\noalign{\smallskip}
    SN ID       &       JD$-$2400000 	&     $J$ 	&     $H$ 		&	     Phase\tablefootmark{a} & Instrument \\
\hline\noalign{\smallskip}

SN 2007hz         &   54359.59  &  20.904    (0.056)  &  20.974    (0.096)  &   7.2  & ISAAC        \\
SN 2007ih         &   54361.63  &  21.641    (0.086)  &   $-$               &   6.5  &   $\cdots$   \\ 
SN 2007ih         &   54359.67  &   $-$               &  21.587    (0.184)  &   5.1  &   $\cdots$   \\
SN 2007ik         &   54362.55  &   $-$               &  21.444    (0.226)  &   4.5  &   $\cdots$   \\

SNF20080510-005   &   54609.48  &  20.215    (0.195)  &  19.919    (0.171)  &   4.0  &  NOTCAM      \\  
SNF20080512-008   &   54609.55  &  20.579    (0.095)  &  20.037    (0.102)  &  13.7  &   $\cdots$   \\
SNF20080512-010   &   54608.55  &  19.459    (0.044)  &  19.475    (0.064)  &   7.2  &   $\cdots$   \\
SNF20080516-000   &   54608.50  &  19.077    (0.051)  &  19.222    (0.067)  &   1.9  &   $\cdots$   \\
SNF20080516-022   &   54609.50  &  19.200    (0.045)  &  19.213    (0.084)  &$-$0.3  &   $\cdots$   \\
SNF20080517-010   &   54610.46  &  20.697    (0.102)  &  20.888    (0.270)  &   4.0  &   $\cdots$   \\
SNF20080522-000   &   54627.46  &  18.319    (0.033)  &  18.190    (0.031)  &   5.5  &   $\cdots$   \\ 
SNF20080606-012   &   54633.47  &  19.183    (0.027)  &  19.461    (0.086)  &   1.5  &   $\cdots$   \\
SNF20080610-003   &   54633.54  &  19.890    (0.051)  &  19.770    (0.080)  &   0.1  &   $\cdots$   \\
                                             	                     	     
SN 2008bz         &   54592.31  &  20.023    (0.265)  &  18.780    (0.210)  &  13.0  &  SIRIUS      \\
SNF20080510-005   &   54609.30  &  20.063    (0.141)  &  19.564    (0.150)  &   3.8  & $\cdots$     \\
SNF20080522-000   &   54614.28  &  17.703    (0.018)  &  17.930    (0.036)  &$-$6.0  &   $\cdots$   \\
SNF20080522-001   &   54613.31  &  18.965    (0.062)  &  18.719    (0.091)  &   8.8  &   $\cdots$   \\
SNF20080522-004   &   54614.46  &  19.861    (0.130)  &  19.988    (0.276)  &   5.8  &   $\cdots$   \\
SNF20080522-011   &   54614.37  &  17.284    (0.017)  &  17.547    (0.036)  &$-$1.9  &   $\cdots$   \\
\hline
\end{tabular}\\
\tablefoot{
\tablefoottext{a}{stretch-corrected rest-frame phase (Eq~\ref{eq:ph}).}
}
\end{table*}

The instrumental magnitudes were measured by PSF photometry with the DAOPHOT  
package \citep{1987PASP...99..191S} in IRAF. The photometric calibration of the SNe observed at ISAAC was obtained with standard
stars from \cite{persson98}, which were observed as part of ISAAC standard calibration plan. The zero-points (ZPs) estimated from the standards 
were transferred to the SN images assuming atmospheric extinction coefficients of $k_\mathrm{J}=0.05$ and $k_\mathrm{H}=0.04$ mag\,airmass$^{-1}$ \citep{2011A&A...528A..43L}.  The image ZPs for the observations  obtained at NOT and SAAO were computed with 2MASS stars in the field of view. 
For each SN image, a ZP was calculated for each 2MASS star. The image ZPs and their uncertainty are, respectively, the average of the individual ZPs (with $3\sigma$ outliers removed if present) and the standard deviation of the mean. No color terms were applied for the following reasons. Firstly, it is quite difficult to derive accurate NIR color-terms observationally because of the lack of suitable standard star fields containing stars with different colors, for example, like those used in the optical \citep{2009AJ....137.4186L}. Secondly, for the photometric system used we expect small color terms. Synthetic photometry of the stars in \cite{pickles} with the passbands that we adopted for the MKO and 2MASS systems (see the Appendix) shows only small color-terms  ($\leq0.03$) between these two systems. \cite{2011A&A...528A..43L} also found zero color-term for ISAAC $H$ band and only a small one $\sim0.02\,(J-H)$ for $J$ band.  The final calibrated magnitudes are given in Table~\ref{t:nirphot}.

\begin{table*}[!t]
\caption{Instrument color-terms and their 1$\sigma$ uncertainties.}
\label{t:cts}
\begin{tabular}{@{}lccccccc@{}}
\hline
\hline\noalign{\smallskip}
Instrument & $ct_U$ & $ct_B$ & $ct_V$ & $ct_R$ & $ct_I$ \\
\hline\noalign{\smallskip}
ALFOSC   & 0.095 (0.002) &    0.004 (0.002) & $-$0.053 (0.002) & $-$0.102 (0.003) & $-$0.050 (0.003) \\
ANDICAM  &   $-$         &    0.077 (0.004) &    0.001 (0.003) &    0.042 (0.004) & $-$0.028 (0.003) \\
MOSCA    & 0.070 (0.020) & $-$0.020 (0.012) & $-$0.081 (0.012) & $-$0.070 (0.011) &    0.021 (0.020) \\
STANCam  & 0.077 (0.009) & $-$0.011 (0.006) & $-$0.052 (0.006) & $-$0.110 (0.008) &  $-$0.049 (0.003)\\ 
\hline
\end{tabular} 
\end{table*}

\subsection{Optical imaging}

For SNe 2007hz, 2007ih and 2007ik the photometry obtained by SDSS-II was used \citep{2014arXiv1401.3317S}. For the remaining 13 SNe, 
the optical observations  were obtained with four different instruments, ALFOSC, MOSCA and STANCam at the Nordic Optical Telescope (NOT)
 equipped with broadband $UBVRI$ filters and ANDICAM at CTIO 1.3m telescope operated by the SMARTS Consortium equipped with $BVRI$
filters. Standard fields from \citet{2009AJ....137.4186L} were also observed with each instrument.
The observations were collected between May and July 2008. On May 23rd, 2009 the SN fields were re-observed with ALFOSC 
to obtain reference images for the galaxy subtraction. All CCD images were bias and flat-field corrected, and trimmed.  Cosmic
ray hits were identified and cleaned with the Laplacian detection algorithm of \citet{crs}. When multiple exposures per filter were obtained, they were combined into a single image.

\subsubsection{Photometry and calibration}

For the optical observations, aperture photometry was chosen over PSF photometry because the main instrument used for the survey - ALFOSC - has significant PSF variation over the image and in most SN fields there were not enough bright stars to model this effect. The instrumental magnitudes were measured through digital apertures with diameter of double the mean full width at half maximum (FWHM)\ of the stars in the field. It is known that aperture diameter of approximately FWHM maximizes the S/N, but double the FWHM aperture was chosen in order to minimize the effect of PSF variation.

The standard star fields observed in eight photometric nights with ALFOSC at NOT were used to calibrate local sequences of stars around the SNe.  Aperture corrections from two- to five- times FWHM were computed  with the IRAF.PHOTCAL package and applied to the measured magnitudes. All standard star measurements were fitted simultaneously (with 3$\sigma$ clipping) with linear equations of the form \citep{harris81}:
\begin{eqnarray}
U&=&u+ct_U\,(U-B)-k'_U\,X-k''_U\,X\,(U-B)+zp_U \nonumber \\
B&=&b+ct_B\,(B-V)-k'_B\,X-k''_B\,X\,(B-V)+zp_B \nonumber \\
V&=&v+ct_V\,(B-V)-k'_V\,X+zp_V \label{eq:tran} \\
R&=&r+ct_R\,(V-R)-k'_R\,X+zp_R \nonumber \\
I&=&i+ct_I\,(V-I)-k'_I\,X+zp_I, \nonumber  
\end{eqnarray}
where upper-case and lower-case letters denote the standard and instrumental magnitudes, respectively, $ct$ the color-terms, $X$ the airmass, $k'$ and $k''$ the first- and the second-order extinction coefficients and the $zp$ the instrument ZPs. The derived instrument color-terms are given in Table~\ref{t:cts}. The second-order extinction coefficients $k''$ were fixed to $-0.03$ based on our synthetic photometry calculations and past photometric works \citep[see also][]{2005AJ....129.2914B}. Because the observations were obtained during a short period of time the data was fitted with common color-terms and instrument ZPs. The standard star observations with the other three instruments we only used to derive the color-terms, again assuming $k''=-0.03$ for $U$ and $B$ bands.

With the derived transformation equations, the standard magnitudes of the stars in the SN fields were computed. The weighted mean and standard deviation of the calibrations from different nights (weighted by the estimated photometric uncertainties) were taken as standard magnitude and its uncertainty. The stars that showed scatter larger than that expected from the photometric errors were discarded. The final calibrated local sequences are indicated on the finding charts in Fig.~\ref{f:fc1} and the magnitudes are listed in Table~\ref{t:stds}.

\longtab{
\begin{landscape}
\begin{longtable}{@{}cccccccccccc@{}}
\caption{$S$-corrected SN optical photometry and the corresponding $S$ corrections $S_{UBVRI}$ (to be applied as defined by Eq.~\ref{eq:corr}). 
The 1$\sigma$ uncertainties are given in parentheses.}
\label{t:optphot} \\
\hline\hline\noalign{\smallskip}
JD$-$2400000 &  $U$ & $S_U$  &  $B$ & $S_B$ &   $V$ & $S_V$     &     $R$ & $S_R$   & $I$ & $S_I$ &  Instrument  \\
\hline\noalign{\smallskip}
\endfirsthead
\caption{continued.}\\
\hline\hline\noalign{\smallskip}
JD$-$2400000 &  $U$ & $S_U$  &  $B$ & $S_B$ &   $V$ & $S_V$     &     $R$ & $S_R$   & $I$ & $S_I$ &  Instrument  \\
\hline\noalign{\smallskip}
\endhead
\hline
\endfoot

\multicolumn{12}{c}{SN 2008bz}\\
\hline\noalign{\smallskip}

54586.409  & 18.499 (0.032)& 0.189     & 18.284 (0.033)& 0.024     & 18.182 (0.029)& $-$0.013  & 18.169 (0.019)& 0.030     & 18.765 (0.030)& 0.038     & ALFOSC     \\
54592.441  & 19.425 (0.055)& 0.168     & 18.935 (0.015)& $-$0.003  & 18.499 (0.014)& $-$0.012  & 18.596 (0.029)& 0.022     & 19.188 (0.077)& $-$0.010  & MOSCA     \\
54598.436  & 20.284 (0.069)& 0.260     & 19.631 (0.019)& 0.005     & 18.930 (0.012)& $-$0.026  & 18.766 (0.013)& 0.008     & 19.308 (0.024)& $-$0.009  & ALFOSC     \\
54610.476  & $-$ & $-$                 & 20.803 (0.042)& 0.000     & 19.739 (0.036)& $-$0.069  & 19.177 (0.029)& $-$0.017  & 19.139 (0.032)& $-$0.021  & STANCam     \\

\hline\noalign{\smallskip}
\multicolumn{12}{c}{SNF20080510-005}\\
\hline\noalign{\smallskip}

54599.395  & 18.583 (0.033)& 0.019     & 19.055 (0.022)& 0.039     & 18.996 (0.020)& 0.011     & 18.927 (0.015)& $-$0.002  & 19.166 (0.022)& 0.028      & ALFOSC     \\
54610.463  & 19.206 (0.097)& 0.143     & 19.139 (0.036)& 0.016     & 18.875 (0.047)& 0.001     & 18.781 (0.019)& 0.017     & 19.246 (0.040)& 0.068      & STANCam     \\
54618.569  & 20.201 (0.058)& 0.201     & 19.644 (0.025)& 0.025     & 19.219 (0.021)& 0.003     & 19.186 (0.020)& 0.027     & 19.940 (0.044)& 0.060      & ALFOSC     \\
54624.408  & 20.837 (0.121)& 0.238     & 20.155 (0.029)& 0.009     & 19.577 (0.022)& $-$0.002  & 19.421 (0.020)& 0.025     & 20.154 (0.052)& 0.042      & ALFOSC     \\
54636.493  & $-$ & $-$                 & 21.393 (0.075)& 0.011     & 20.307 (0.043)& $-$0.061  & 19.782 (0.058)& $-$0.000  & 19.745 (0.034)& 0.014      & ALFOSC     \\

\hline\noalign{\smallskip}
\multicolumn{12}{c}{SNF20080512-008}\\
\hline\noalign{\smallskip}

54609.616  & 19.734 (0.057)& 0.156     & 19.315 (0.032)& $-$0.002  & 18.851 (0.018)& 0.004     & 18.747 (0.019)& 0.029     & 19.306 (0.029)& 0.048      & STANCam     \\
54613.610  & $-$ & $-$                 & $-$ & $-$                 & 19.061 (0.037)& $-$0.004  & 18.931 (0.013)& 0.024     & $-$ & $-$                  & ALFOSC     \\
54618.612  & 20.953 (0.070)& 0.248     & 20.175 (0.018)& 0.003     & 19.330 (0.013)& $-$0.026  & 19.070 (0.013)& 0.013     & 19.519 (0.023)& 0.030      & ALFOSC     \\
54624.443  & 21.133 (0.146)& 0.236     & 20.699 (0.037)& 0.009     & 19.617 (0.016)& $-$0.059  & 19.145 (0.013)& 0.002     & 19.369 (0.020)& 0.015      & ALFOSC     \\
54635.675  & $-$ & $-$                 & 21.033 (0.146)& 0.008     & 20.219 (0.067)& $-$0.066  & 19.629 (0.027)& $-$0.020  & 19.559 (0.046)& $-$0.014   & STANCam     \\
54641.549  & $-$ & $-$                 & 21.521 (0.127)& 0.008     & 20.515 (0.055)& $-$0.070  & 19.918 (0.037)& 0.006     & 19.905 (0.032)& $-$0.018   & ALFOSC     \\

\hline\noalign{\smallskip}
\multicolumn{12}{c}{SNF20080512-010}\\
\hline\noalign{\smallskip}

54612.562  & 19.413 (0.057)& 0.212     & 18.966 (0.025)& 0.013     & 18.568 (0.013)& $-$0.009  & 18.545 (0.013)& 0.028     & 19.197 (0.029)& 0.026      & ALFOSC     \\
54618.597  & 20.396 (0.050)& 0.264     & 19.843 (0.018)& 0.005     & 19.033 (0.012)& $-$0.026  & 18.784 (0.013)& 0.007     & 19.116 (0.020)& $-$0.003   & ALFOSC     \\
54626.470  & 21.239 (0.111)& 0.159     & 20.813 (0.028)& 0.005     & 19.751 (0.017)& $-$0.063  & 19.111 (0.015)& $-$0.009  & 18.978 (0.016)& $-$0.012   & ALFOSC     \\
54635.658  & $-$ & $-$                 & 21.047 (0.168)& 0.005     & 20.279 (0.070)& $-$0.072  & 19.772 (0.040)& 0.007     & 19.602 (0.048)& $-$0.032   & STANCam     \\
54641.523  & $-$ & $-$                 & 21.414 (0.115)& 0.003     & 20.560 (0.053)& $-$0.050  & 20.000 (0.036)& 0.014     & 19.947 (0.050)& $-$0.034   & ALFOSC     \\

\hline\noalign{\smallskip}
\multicolumn{12}{c}{SNF20080516-000}\\
\hline\noalign{\smallskip}

54611.490  & $-$ & $-$                 & 18.422 (0.047)& 0.023     & 18.261 (0.020)& $-$0.006  & 18.179 (0.013)& 0.021     & 18.595 (0.022)& 0.051      & STANCam     \\
54618.507  & 19.114 (0.035)& 0.196     & 18.813 (0.012)& 0.024     & 18.537 (0.012)& 0.002     & 18.473 (0.011)& 0.026     & 19.086 (0.023)& 0.057      & ALFOSC     \\
54626.425  & 20.322 (0.073)& 0.254     & 19.628 (0.014)& 0.010     & 19.019 (0.011)& $-$0.005  & 18.827 (0.012)& 0.018     & 19.321 (0.022)& 0.022      & ALFOSC     \\
54635.401  & $-$ & $-$                 & 20.529 (0.104)& 0.003     & 19.535 (0.041)& $-$0.058  & 19.032 (0.023)& $-$0.002  & 19.216 (0.099)& 0.003      & STANCam     \\
54641.401  & $-$ & $-$                 & 20.897 (0.059)& 0.009     & 19.873 (0.024)& $-$0.061  & 19.173 (0.014)& $-$0.009  & 19.141 (0.026)& $-$0.005   & ALFOSC     \\
54656.432  & $-$ & $-$                 & 21.256 (0.100)& 0.014     & 20.523 (0.048)& $-$0.062  & 19.936 (0.023)& 0.014     & 20.021 (0.056)& $-$0.025   & ALFOSC     \\

\hline\noalign{\smallskip}
\multicolumn{12}{c}{SNF20080516-022}\\
\hline\noalign{\smallskip}

54611.509  & 18.229 (0.033)& 0.090     & 18.469 (0.019)& 0.024     & 18.393 (0.024)& $-$0.004  & 18.321 (0.027)& 0.018     & 18.751 (0.027)& 0.041      & STANCam     \\
54618.583  & 19.278 (0.032)& 0.195     & 18.895 (0.011)& 0.023     & 18.620 (0.012)& 0.001     & 18.573 (0.014)& 0.025     & 19.127 (0.019)& 0.062      & ALFOSC     \\
54624.459  & 20.013 (0.062)& 0.237     & 19.498 (0.018)& 0.010     & 18.980 (0.013)& 0.003     & 18.889 (0.014)& 0.023     & 19.502 (0.025)& 0.032      & ALFOSC     \\
54636.515  & $-$ & $-$                 & 20.820 (0.072)& 0.007     & 19.780 (0.045)& $-$0.059  & 19.224 (0.019)& $-$0.002  & 19.257 (0.030)& 0.003      & ALFOSC     \\
54641.497  & $-$ & $-$                 & 21.166 (0.050)& 0.007     & 20.235 (0.032)& $-$0.061  & 19.358 (0.023)& $-$0.010  & 19.361 (0.029)& $-$0.008   & ALFOSC     \\
54648.564  & $-$ & $-$                 & 21.453 (0.147)& 0.014     & $-$ & $-$                 & $-$ & $-$                 & $-$ & $-$                  & ALFOSC     \\
54656.455  & $-$ & $-$                 & 21.691 (0.078)& 0.005     & 20.857 (0.039)& $-$0.054  & 20.170 (0.028)& 0.014     & 20.298 (0.068)& $-$0.030   & ALFOSC     \\

\hline\noalign{\smallskip}
\multicolumn{12}{c}{SNF20080517-010}\\
\hline\noalign{\smallskip}

54610.471  & $-$ & $-$                 & 19.907 (0.046)& 0.015     & 19.603 (0.017)& $-$0.003  & 19.479 (0.029)& 0.010     & 19.615 (0.023)& 0.085      & STANCam     \\
54618.532  & 21.150 (0.101)& 0.100     & 20.601 (0.023)& 0.042     & 19.965 (0.020)& $-$0.032  & 19.810 (0.019)& 0.032     & 20.239 (0.086)& 0.099      & ALFOSC     \\
54626.443  & 22.413 (0.258)& 0.179     & 21.496 (0.042)& 0.023     & 20.534 (0.022)& $-$0.058  & 20.132 (0.019)& 0.016     & 20.314 (0.074)& 0.084      & ALFOSC     \\
54636.451  & $-$ & $-$                 & $-$ & $-$                 & $-$ & $-$                 & $-$ & $-$                 & 20.273 (0.068)& 0.037      & ALFOSC     \\
54641.416  & $-$ & $-$                 & 22.729 (0.152)& 0.025     & 21.460 (0.061)& $-$0.102  & 20.868 (0.034)& $-$0.028  & 20.460 (0.061)& 0.020      & ALFOSC     \\
54648.460  & $-$ & $-$                 & 22.946 (0.356)& 0.043     & 21.864 (0.130)& $-$0.108  & 21.261 (0.135)& $-$0.010  & 20.810 (0.155)& 0.015      & ALFOSC     \\

\hline\noalign{\smallskip}
\multicolumn{12}{c}{SNF20080522-000}\\
\hline\noalign{\smallskip}

54618.556  & 16.486 (0.021)& $-$0.027) & 17.147 (0.013)& 0.016     & 17.126 (0.012)& $-$0.014  & 17.134 (0.011)& 0.019     & 17.330 (0.012)& $-$0.008   & ALFOSC     \\
54625.591  & $-$ & $-$                 & 17.269 (0.011)& 0.033     & 17.130 (0.021)& 0.012     & 17.009 (0.033)& 0.031     & 17.409 (0.075)& $-$0.012   & ANDICAM\\
54626.457  & 17.073 (0.024)& 0.131     & 17.267 (0.017)& 0.022     & 17.123 (0.009)& $-$0.020  & 17.029 (0.012)& 0.034     & 17.491 (0.014)& 0.008      & ALFOSC     \\
54628.555  & $-$ & $-$                 & 17.389 (0.022)& 0.037     & 17.161 (0.017)& 0.009     & 17.075 (0.018)& 0.027     & 17.495 (0.066)& $-$0.017   & ANDICAM\\
54631.544  & $-$ & $-$                 & 17.520 (0.025)& 0.046     & 17.260 (0.016)& 0.006     & 17.241 (0.046)& 0.017     & 17.706 (0.138)& $-$0.022   & ANDICAM\\
54635.560  & $-$ & $-$                 & 17.846 (0.033)& 0.050     & 17.462 (0.029)& $-$0.003  & $-$ & $-$                 & $-$ & $-$                  & ANDICAM\\
54636.478  & 18.231 (0.050)& 0.227     & 17.904 (0.023)& 0.012     & 17.586 (0.028)& $-$0.024  & 17.602 (0.016)& 0.027     & 18.141 (0.022)& $-$0.027   & ALFOSC     \\
54645.584  & $-$ & $-$                 & 19.059 (0.037)& 0.095     & 18.106 (0.019)& 0.002     & 17.801 (0.022)& 0.013     & 17.903 (0.059)& $-$0.042   & ANDICAM\\
54650.536  & $-$ & $-$                 & 19.538 (0.026)& 0.103     & 18.381 (0.016)& 0.017     & 17.916 (0.016)& 0.002     & 17.923 (0.035)& $-$0.031   & ANDICAM\\
54656.444  & $-$ & $-$                 & 19.818 (0.018)& 0.020     & 18.662 (0.014)& $-$0.062  & 18.079 (0.011)& $-$0.018  & 18.076 (0.012)& $-$0.045   & ALFOSC     \\
54658.547  & $-$ & $-$                 & 19.929 (0.103)& 0.112     & 18.770 (0.049)& 0.028     & 18.172 (0.051)& $-$0.017  & 18.138 (0.108)& $-$0.028   & ANDICAM\\
54663.537  & $-$ & $-$                 & 20.190 (0.205)& 0.097     & 19.018 (0.065)& 0.033     & 18.438 (0.044)& $-$0.036  & 18.378 (0.067)& $-$0.033   & ANDICAM\\
54664.409  & $-$ & $-$                 & $-$ & $-$                 & 18.980 (0.047)& $-$0.068  & 18.487 (0.021)& $-$0.003  & 18.447 (0.051)& $-$0.059   & STANCam     \\
54671.400  & $-$ & $-$                 & 20.303 (0.019)& 0.012     & 19.305 (0.013)& $-$0.065  & 18.805 (0.014)& 0.007     & 18.854 (0.015)& $-$0.046   & ALFOSC     \\

\hline\noalign{\smallskip}
\multicolumn{12}{c}{SNF20080522-001}\\
\hline\noalign{\smallskip}

54618.492  & 18.926 (0.046)& 0.240     & 18.460 (0.015)& 0.016     & 18.052 (0.011)& $-$0.022  & 17.984 (0.012)& 0.025     & 18.608 (0.028)& $-$0.016   & ALFOSC     \\
54626.412  & 19.892 (0.051)& 0.263     & 19.308 (0.017)& 0.013     & 18.419 (0.011)& $-$0.039  & 18.211 (0.009)& 0.004     & 18.567 (0.018)& $-$0.040   & ALFOSC     \\
54639.589  & $-$ & $-$                 & 20.195 (0.161)& 0.115     & 19.183 (0.129)& 0.033     & 18.427 (0.066)& $-$0.006  & 18.429 (0.106)& $-$0.013   & ANDICAM\\
54640.606  & $-$ & $-$                 & 20.230 (0.124)& 0.118     & 19.212 (0.062)& 0.032     & 18.643 (0.051)& $-$0.008  & 18.520 (0.070)& $-$0.016   & ANDICAM\\
54651.522  & $-$ & $-$                 & 20.643 (0.044)& 0.083     & 19.781 (0.045)& 0.036     & 19.104 (0.045)& $-$0.026  & 19.204 (0.106)& $-$0.017   & ANDICAM\\
54656.397  & $-$ & $-$                 & 20.702 (0.126)& 0.015     & 19.785 (0.067)& $-$0.065  & 19.270 (0.034)& 0.008     & 19.407 (0.063)& $-$0.042   & ALFOSC     \\
54661.531  & $-$ & $-$                 & 20.870 (0.224)& 0.098     & 20.049 (0.141)& 0.021     & 19.385 (0.105)& $-$0.033  & 19.715 (0.200)& $-$0.004   & ANDICAM\\
54671.388  & $-$ & $-$                 & 21.076 (0.099)& 0.023     & 20.200 (0.037)& $-$0.050  & 19.820 (0.034)& 0.012     & 19.976 (0.071)& $-$0.024   & ALFOSC     \\

\hline\noalign{\smallskip}
\multicolumn{12}{c}{SNF20080522-004}\\
\hline\noalign{\smallskip}

54618.475  & 20.326 (0.131)& 0.162     & 19.740 (0.031)& 0.036     & 19.529 (0.023)& $-$0.004  & 19.383 (0.026)& 0.018     & 19.932 (0.096)& 0.079      & ALFOSC     \\
54626.398  & $-$ & $-$                 & 20.364 (0.030)& 0.029     & 19.921 (0.023)& $-$0.021  & 19.762 (0.026)& 0.031     & 20.294 (0.078)& 0.073      & ALFOSC     \\
54636.464  & $-$ & $-$                 & $-$ & $-$                 & 20.635 (0.076)& $-$0.061  & 20.098 (0.045)& 0.008     & 20.174 (0.058)& 0.053      & ALFOSC     \\
54641.388  & $-$ & $-$                 & 21.563 (0.141)& 0.023     & 20.820 (0.067)& $-$0.085  & 20.263 (0.045)& $-$0.006  & 20.292 (0.073)& 0.038      & ALFOSC     \\
54648.396  & $-$ & $-$                 & 22.009 (0.558)& 0.027     & 21.151 (0.195)& $-$0.093  & 20.317 (0.115)& $-$0.021  & $-$ & $-$                  & ALFOSC     \\
54656.386  & $-$ & $-$                 & $-$ & $-$                 & $-$ & $-$                 & 20.913 (0.161)& $-$0.008  & 20.637 (0.125)& 0.007      & ALFOSC     \\

\hline\noalign{\smallskip}
\multicolumn{12}{c}{SNF20080522-011}\\
\hline\noalign{\smallskip}

54613.600  & $-$ & $-$                 & 16.845 (0.012)& 0.006     & 16.888 (0.012)& $-$0.018  & 16.835 (0.012)& 0.019     & 17.061 (0.024)& $-$0.016   & ALFOSC     \\
54624.393  & 17.182 (0.023)& 0.174     & 17.085 (0.015)& 0.014     & 16.954 (0.012)& $-$0.025  & 16.911 (0.013)& 0.037     & 17.431 (0.016)& $-$0.019   & ALFOSC     \\
54638.723  & $-$ & $-$                 & 18.458 (0.068)& 0.095     & 17.800 (0.074)& $-$0.000  & 17.501 (0.032)& 0.024     & 17.763 (0.044)& $-$0.073   & ANDICAM\\
54650.585  & $-$ & $-$                 & 19.556 (0.043)& 0.098     & 18.386 (0.024)& 0.028     & 17.819 (0.027)& $-$0.018  & 17.702 (0.034)& $-$0.046   & ANDICAM\\
54656.468  & $-$ & $-$                 & 19.720 (0.022)& 0.015     & 18.717 (0.014)& $-$0.058  & 18.194 (0.011)& $-$0.015  & 18.111 (0.024)& $-$0.076   & ALFOSC     \\
54661.632  & $-$ & $-$                 & 20.178 (0.209)& 0.075     & 18.971 (0.062)& 0.033     & 18.458 (0.075)& $-$0.044  & 18.460 (0.073)& $-$0.042   & ANDICAM\\
54664.419  & $-$ & $-$                 & 19.942 (0.093)& 0.010     & 19.036 (0.057)& $-$0.062  & 18.571 (0.025)& 0.002     & 18.648 (0.105)& $-$0.054   & STANCam     \\
54671.413  & $-$ & $-$                 & 20.108 (0.019)& 0.014     & 19.208 (0.014)& $-$0.054  & 18.798 (0.012)& 0.000     & 18.850 (0.023)& $-$0.039   & ALFOSC     \\

\hline\noalign{\smallskip}
\multicolumn{12}{c}{SNF20080606-012}\\
\hline\noalign{\smallskip}

54636.530  & 19.298 (0.064)& 0.172     & 18.944 (0.023)& 0.031     & 18.824 (0.019)& 0.002     & 18.724 (0.019)& 0.017     & 18.954 (0.034)& 0.057      & ALFOSC     \\
54641.511  & 19.861 (0.067)& 0.186     & 19.337 (0.017)& 0.022     & 19.017 (0.015)& 0.004     & 18.951 (0.014)& 0.023     & 19.283 (0.024)& 0.065      & ALFOSC     \\
54648.501  & $-$ & $-$                 & 20.154 (0.041)& 0.010     & 19.435 (0.025)& $-$0.001  & $-$ & $-$                 & 19.396 (0.075)& 0.034      & ALFOSC     \\
54649.677  & $-$ & $-$                 & 20.264 (0.064)& 0.078     & 19.519 (0.052)& 0.047     & 19.331 (0.060)& $-$0.000  & 19.669 (0.149)& 0.035      & ANDICAM\\
54654.664  & $-$ & $-$                 & 20.965 (0.170)& 0.082     & 19.862 (0.065)& 0.050     & 19.415 (0.062)& 0.040     & 19.387 (0.141)& 0.034      & ANDICAM\\
54663.406  & $-$ & $-$                 & 21.650 (0.110)& 0.004     & 20.532 (0.192)& $-$0.064  & 19.778 (0.089)& $-$0.011  & 19.558 (0.095)& $-$0.004   & ALFOSC     \\
54671.427  & $-$ & $-$                 & 21.901 (0.060)& 0.008     & 20.958 (0.036)& $-$0.069  & 20.337 (0.025)& 0.009     & 20.127 (0.033)& $-$0.022   & ALFOSC     \\

\hline\noalign{\smallskip}
\multicolumn{12}{c}{SNF20080610-003}\\
\hline\noalign{\smallskip}

54641.536  & 20.012 (0.079)& 0.179     & 19.456 (0.021)& 0.026     & 19.129 (0.016)& 0.003     & 18.992 (0.015)& 0.016     & 19.548 (0.027)& 0.073      & ALFOSC     \\
54648.441  & 20.706 (0.167)& 0.192     & 20.013 (0.036)& 0.020     & 19.553 (0.026)& $-$0.005  & 19.321 (0.028)& 0.029     & 20.093 (0.138)& 0.061      & ALFOSC     \\
54655.527  & $-$ & $-$                 & 20.840 (0.022)& $-$0.007  & 19.998 (0.021)& $-$0.023  & 19.596 (0.016)& 0.012     & 19.903 (0.060)& 0.048      & STANCam     \\
54663.391  & $-$ & $-$                 & $-$ & $-$                 & 20.412 (0.106)& $-$0.073  & 19.779 (0.071)& $-$0.005  & 19.883 (0.152)& 0.022      & ALFOSC     \\
54671.443  & $-$ & $-$                 & 21.962 (0.052)& 0.008     & 20.985 (0.031)& $-$0.075  & 20.217 (0.019)& $-$0.014  & 20.114 (0.029)& $-$0.002   & ALFOSC     \\

\end{longtable}
\end{landscape}
}

The calibrated local sequence stars, measured with two-times FWHM diameter apertures, were used to compute the image ZPs.
For each SN image, a ZP $zp^X\equiv zp-k'\,X$ was calculated for each calibrated star  by applying Eqs.~\ref{eq:tran}.  The final image ZPs and their uncertainty are, respectively, the average of the individual ZPs (with $3\sigma$ outliers removed if present) and the standard deviation. The image ZPs were added to the SN magnitudes measured in the galaxy-subtracted images to obtain the SN magnitudes in the natural systems of the instruments used, $m^\mathrm{nat}$.

It is known that Eqs.~\ref{eq:tran} cannot accurately transform SN photometry into the standard Johnson-Cousins $UBVRI$ photometric system \citep{sun_scorr,phot99ee,kri_01el}. For this reason, the optical photometry was calibrated with the method described in \citet{stan_03du}, which is based on the $S$-correction introduced by \citet{phot99ee}. Because we took into account the second-order extinction, the calibration equations were slightly modified.  The calibration correction from natural magnitudes at airmass $X$, $m^\mathrm{nat}_X$, to standard magnitudes $m^\mathrm{std}$ is computed by means of synthetic photometry:
\begin{eqnarray}
m^\mathrm{std}=m^\mathrm{nat} & -2.5\log\left(\int\,f_\lambda^\mathrm{phot}(\lambda)R^\mathrm{std}(\lambda)d\lambda\right) & \nonumber\\
 & +2.5\log\left(\int\,f_\lambda^\mathrm{phot}(\lambda)R^\mathrm{nat}(\lambda)p(\lambda)^X\,d\lambda\right) & + const.
\label{eq:scorr}
\end{eqnarray}
Here  $f_\lambda^\mathrm{phot}(\lambda)$ is the photon spectral energy distributions (SED) of the SN,
$R^\mathrm{nat}(\lambda)$ and  $R^\mathrm{std}(\lambda)$ are the responses of the natural and the standard systems, 
respectively. We note that $R^\mathrm{nat}(\lambda)$ does not include the atmospheric transmission and thus 
the transmission of the atmosphere at airmass $X=1$ above the site of observation, $p(\lambda)$, is also included in the calculation.
Because the SNe in our sample do not have extensive spectrophotometry we used \cite{2007ApJ...663.1187H} spectral template as 
 $f_\lambda^\mathrm{phot}(\lambda)$, corrected for the Milky Way reddening and red-shifted to the redshift of each SN.
 The constant in Eq.~\ref{eq:scorr} is such that the correction is
zero for A0\,V stars with all color indexes zero. This ensures that
for normal stars the synthetic $S$-correction gives the same results as
the linear color-term corrections (Eq.~\ref{eq:tran}). The constant
can be determined from synthetic photometry of stars for which both
photometry and spectrophotometry is available. In the Appendix we provide full account of the calibration procedure, including a discussion of the 
uncertainties introduced when the second-order extinction is not taken into account.
The final photometry of the SNe is given in Table~\ref{t:optphot}.

\section{SNe~Ia as standardizable candles}
\label{sec:hubblefit}

The rest-frame peak magnitudes SN~Ia in band $X$, $m_X$, can by standardized with the following two-parameter relation:
\begin{equation}
m_X^\ast=m_X-a_X \times (\Delta M_{15}-1.1)-R_X^{YZ} \times E(Y-Z),
\label{eq:b_std}
\end{equation}
where $E(Y-Z)\geq 0$ is the color excess in bands $Y$ and $Z$ caused by dust and which can be converted into absorption in band $X$ via $R_X^{YZ} \times E(Y-Z)$. The second correction term in the equation accounts for the well-known "luminosity -- LC shape" relation and  as a LC shape parameter we use $\Delta M_{15}$, which measures the rest-frame $B$-band magnitude decline during the first 15 days after the time of $B$-band maximum light  $t_{B_\mathrm{max}}$ \citep{dm15}. Throughout this work we use $E(B-V)$, which is calculated from the observed color excess at the time of $B$-band maximum light $E(B-V)_\mathrm{obs}=(B-V)_{B_\mathrm{max}}-(B-V)_{B_\mathrm{max},0}$ by setting the negative values to zero. The intrinsic unreddened SN~Ia color $(B-V)_{B_\mathrm{max},0}$ is a nonlinear function of $\Delta M_{15}$ that we determined (see Appendix~\ref{sec:colbv0}).

\subsection{Measuring the light curve parameters}

\label{sec:lcfit}

SN~Ia observations rarely allow direct measurement of the peak magnitudes and $\Delta M_{15}$. Thus, to obtain these parameters a multitude of indirect methods that fit optical and/or NIR template LCs to the observations have been developed \citep{1997ApJ...483..565P,2001ApJ...558..359G,2005A&A...443..781G,2007ApJ...659..122J,2007A&A...466...11G,2008ApJ...681..482C,2006ApJ...647..501P,kri_ir_temp,2008ApJ...689..377W,2009ApJ...704..629M,2011ApJ...731..120M,2011AJ....141...19B}. For the needs of our analysis we developed our own \emph{stretch} parametrization-based fitter \citep{1997ApJ...483..565P,2001ApJ...558..359G} for the following reasons. First, our primary goal was to test the accuracy of the estimate of SN~Ia NIR peak magnitudes from a single observation obtained close to maximum.  \cite{kri_ir_temp} have presented evidence that the stretch approach could be applied to the first maximum of $J$ band LCs and that the NIR and the optical stretch are tightly correlated. With the stretch fitting approach we were able to verify this result with a much larger sample of SNe with well-sampled LCs. Second, we used our own NIR templates and prescription to compute NIR $K$-corrections (Appendixes~\ref{sec:kcor} and \ref{sec:nirtemplates}). Third, from the optical LCs we need $t_{B_\mathrm{max}}$, $B$ and $V$ peak magnitudes, and a LC shape parameter. Considering the above, stretch-based fitter appears to be best suited for our purposes. 

\begin{table*}[!t]
\caption{Average difference between the $B$ and $V$ peak magnitudes, $(B-V)_{B_\mathrm{max}}$, and $\Delta M_{15}$ for the SNe in common between the large data sets. Only sets with more than ten SNe in common are shown. Numbers in parentheses are the 1$\sigma$ scatter in units of 0.001 mag.} 
\label{t:diff}
\setlength{\tabcolsep}{3pt}
\begin{tabular}{@{}lccccc@{}}
\hline
\hline\noalign{\smallskip}
Difference & $N_\mathrm{SN}$ & $B$ & $V$ & $(B-V)_{B_\mathrm{max}}$ & $\Delta M_{15}$ \\
\hline\noalign{\smallskip}
Berkeley$-$CSP  & 16 &    0.031 (38) &    0.045 (25) & $-$0.010 (39) & $-$0.016 (44) \\
Berkeley$-$CfA2 & 10 & $-$0.027 (38) & $-$0.023 (14) & $-$0.004 (38) &    0.025 (39) \\
Berkeley$-$CfA3 & 47 & $-$0.008 (43) &    0.008 (45) & $-$0.019 (49) & $-$0.017 (52) \\
CSP$-$CfA3      & 30 & $-$0.050 (35) & $-$0.050 (24) & $-$0.001 (39) & $-$0.007 (60) \\
\hline 
\end{tabular}
\end{table*}

Our fitting program is based on the IDL fitting package MPFIT\footnote{available at
\url{http://cow.physics.wisc.edu/~craigm/idl/idl.html}}. It can fit multicolor LCs simultaneously
 using two approaches. The first is to fit the data with rest-frame template LCs coupled with a spectral 
template to compute $K$-corrections \citep{kim96} in order to transform from rest-frame to observed magnitudes. In the second approach  
a spectral template is red-shifted and directly integrated through the observed filters. This requires that the spectral
template is calibrated to match the rest-frame LCs of SNe Ia, which is the case for \cite{2007ApJ...663.1187H} template.
This template is generated in such a way that it matches the $UBVRI$ LC templates of \cite{knop03} and
its $B-V$ color at maximum is $-0.058$ mag. In both approaches the fitted parameters are the common stretch $s$ that multiplies the time axis of the template, the time of $B$-band maximum light $t_{B_\mathrm{max}}$,
 and the magnitude offsets needed to fit each SN LC. The full covariance matrix of the fit parameters is also computed.  The LC templates are normalized to zero magnitude at $t_{B_\mathrm{max}}$ and because each template magnitude offset is fitted independently, the procedure returns the rest-frame SN magnitudes at $t_{B_\mathrm{max}}$. The program is flexible and one or several fit parameters can be fixed. It also allows the spectral template to be reddened by arbitrary reddening laws.

The stretch parametrization can only be used for $U$, $B$ or $V$ bands. In the redder bands SNe Ia show a second maximum (or just a shoulder in $R$) and the stretch approach can only be applied to the first maximum \citep[e.g.,][]{kri_ir_temp}. 
Therefore, the fits were limited to the observations obtained up to phase $+10$ \emph{stretch-corrected} days in $J$ and $H$ bands, and up to phase $+35$ days in the optical. The \emph{stretch-corrected} SN phase, $\phi$, is defined as
\begin{equation}
\phi=C\,\frac{t-t_{B_\mathrm{max}}}{s\,(1+z)},
\label{eq:ph}
\end{equation}
where $z$ is the SN host galaxy redshift, $s$ is the stretch and the constant $C$ is set to 1, unless specified otherwise. The stretch parameter $s$ can be either the optical stretch $s_\mathrm{opt}$ or the NIR stretch $s_\mathrm{NIR}$, which will be defined later. 

\subsubsection{Optical light curve parameters}

\begin{figure*}[!t]
\includegraphics*[width=18cm]{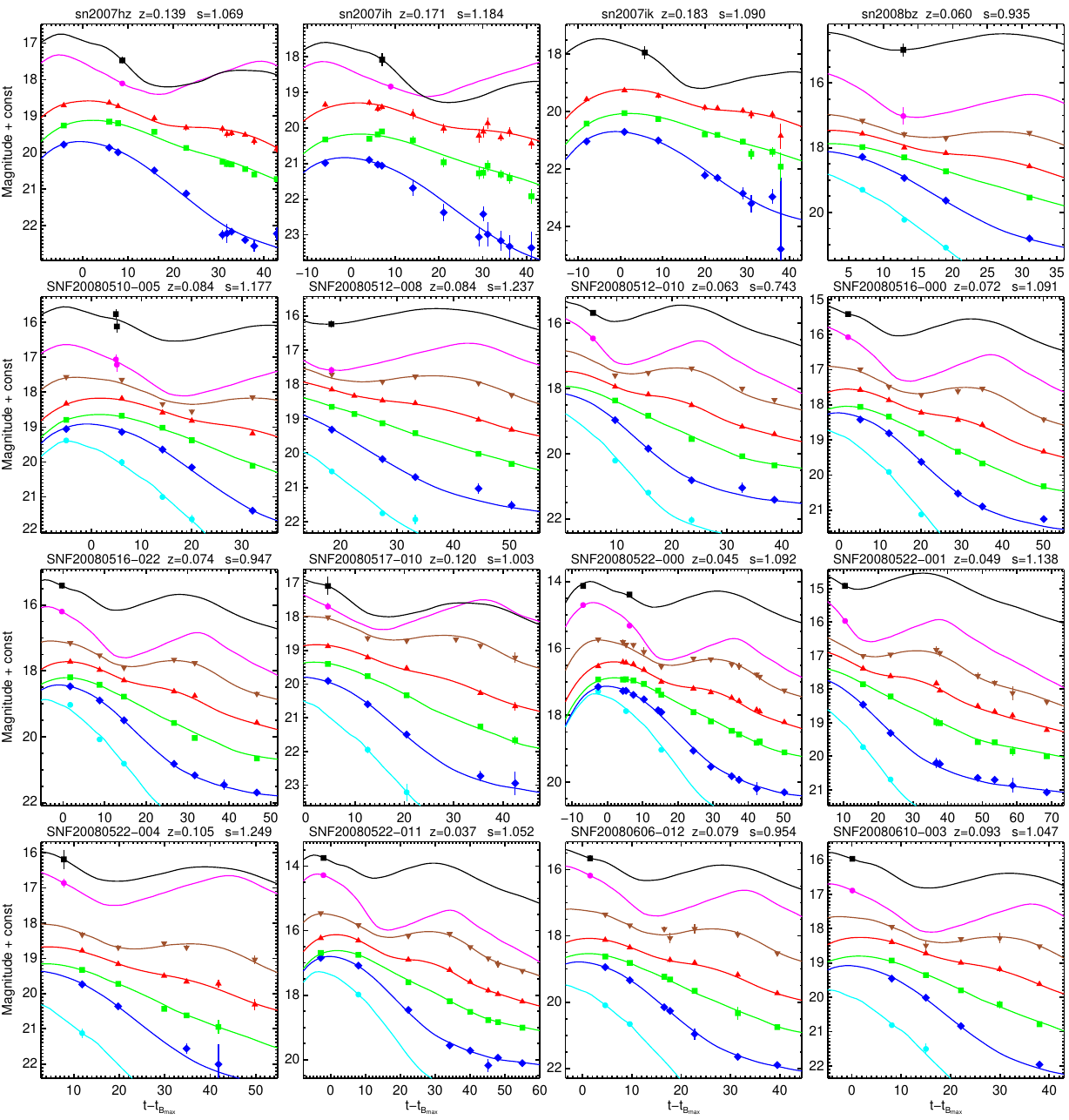}
\caption{$griJH$ and $UBVRIJH$ LCs and the best fits. For the three SDSS-II SNe (2007hz, 2007ih, 2007ik) from bottom to top are plotted $g$, $r-0.4$, $i-1$, $J-2.8$ and $H-3.8$ (no $J$ for SN~2007ik). For the other 13 SNe are plotted $U+0.8$, $B$, $V-0.2$, $R-0.6$, $I-1.6$, $J-3$ and $H-3.5$. We note that the simultaneous $gri$ or $UBVRI$ fits were only used for these plots. In the analysis we use the results from the simultaneous fits of the  $g$ and $r$, and $B$ and $V$  bands only.}
\label{f:fits}
\end{figure*} 

For the analysis of the NIR LCs we need the optical stretch $s_\mathrm{opt}$, $t_{B_\mathrm{max}}$, and $B$ and $V$ peak magnitudes. 
To determine them, the Milky Way extinction-corrected $g$ and $r$ LCs of SNe 2007hz, 2007ih and 2007ik by SDSS-II, and the 
$B$ and $V$-band  LCs of the remaining objects 
were fitted simultaneously\footnote{We would like to stress here that in our definition, $s_\mathrm{opt}$ is the common stretch that makes the template fit both $B$ and $V$-band LCs simultaneously} in a two-iteration procedure. In the first iteration the \cite{2007ApJ...663.1187H} spectral template 
was integrated through the relevant filters and fitted to the data to obtain the SN $B-V$ color at $t_{B_\mathrm{max}}$, $(B-V)_{B_\mathrm{max}}$. The spectral template was then modified to match the observed SN $(B-V)_{B_\mathrm{max}}$ color\footnote{The template was modified using  \cite{fitzpatrick99} extinction law, allowing for negative reddening when needed.} and  the fitting was repeated to obtain the final parameters.  This two-step procedure has the advantage that it accounts for the LC shape modification by the redshift of the filters' effective wavelength caused by reddening. In particular, the derived stretch parameters are largely free from this effect. The optical LC parameters of the SNe from our new sample are shown in Table~\ref{t:sne_o}. They show that there are no underluminous or highly reddened SNe in our sample. Further, the optical observations of all but two SNe started before phase +15 days and the NIR observations of only two SNe were obtained after phase +10 days.
 For illustration purposes, the simultaneous $gri$ or $UBVRI$ fits are shown in Fig.~\ref{f:fits}. In the analysis we use the results from the simultaneous fits of the  $g$ and $r$, and $B$ and $V$  bands only. 
 
 The optical LCs for the literature NIR sample (102 objects) were taken from the following sources:
 Cal\`{a}n-Tololo \citep{1996AJ....112.2408H}, CfA1,2,3 and 4 
\citep{1999AJ....117..707R,2006AJ....131..527J,2009ApJ...700..331H,2012ApJS..200...12H}, Berkeley \citep{2010ApJS..190..418G}, Carnegie Supernova Project (CSP) \citep{2010AJ....139..519C,2011AJ....142..156S}, Supernova Cosmology Project (SCP) \citep{2008ApJ...686..749K}, K. Krisciunas and collaborators \citep{kri_04,kri_ir_temp,kri_01el,kri_01,kri_00,2007AJ....133...58K}, SN~1998bu \citep{her98bu,1999ApJS..125...73J,1999AJ....117.1175S}, SN~2000E  \citep{2003ApJ...595..779V}, SN~2002bo \citep{02bo}, SN~2002dj \citep{2008MNRAS.388..971P}, SN~2003cg \citep{nancy06}, SN~2003du \citep{stan_03du}, SN~2005cf \citep{2009ApJ...697..380W,2007MNRAS.376.1301P}, SN~2008Q (Stanishev et al., in preparation), and SN~2011fe \citep{2013NewA...20...30M}. For the SNe observed in the  NIR by \cite{2012MNRAS.425.1007B} (hereafter referred as \citetalias{2012MNRAS.425.1007B} sample) the optical $g$ and $r$ LCs from \cite{2012MNRAS.426.2359M} were used. The CSP photometry is given in the natural photometric systems of the CSP telescopes and the observed $B$ and $V$ bands used were those from \cite{2011AJ....142..156S}. For the other SNe the standard Bessell  $B$ and $V$ bands \citep{2012PASP..124..140B} were used. If a given SN had optical LCs available from different sources, the optical parameters associated to it were derived from the LCs that had better phase coverage and sampling. In a few cases it was necessary to combine the optical photometry in order to obtain more reliable LC fits. 

To put the optical properties of the SN in our new sample in the context of other SNe~Ia, all objects with LCs that allow stretch fitting were also analyzed. This analysis was also performed because it may be relevant and of practical interest for SN cosmology. For many current and future high-redshift SN samples only the rest-frame $B$ and $V$ band are in the optical window.  The redder bands are red-shifted in the NIR  and are or will  be unavailable. The UV part of the spectrum is red-shifted to the optical, but the UV properties of SNe~Ia are much less understood and the peak magnitudes appear to have larger intrinsic scatter \citep{2010ApJ...721.1608B,2010ApJ...721.1627M,2013ApJ...779...23M}. Given this, it is interesting to study how well the SN~Ia standard candle can be calibrated when only rest-frame $B$ and $V$ observations are available, now with much larger sample of nearby SNe. In addition, with this large sample of high-quality LCs we can study the SN intrinsic $B-V$ color at $t_{B_\mathrm{max}}$, $(B-V)_{B_{\mathrm{max},0}}$, which is important for estimating the dust extinction in the SN host galaxies. 

In our analysis we found that the "luminosity -- LC shape" relation for normal SNe~Ia is linear (within the scatter) when $\Delta M_{15}$ is used as a LC shape parameter and not the stretch. Using a subsample of $\sim200$ SNe with well-sampled LCs we derived a new $s_\mathrm{opt}-\Delta M_{15}$ relation for normal SNe~Ia (see Appendix~\ref{ap:sdm15} for details) and used it to convert stretch to $\Delta M_{15}$.

In Table~\ref{t:diff} we report the weighted mean difference and 1$\sigma$ scatter between the $B$ and $V$ peak magnitudes, $(B-V)_{B_\mathrm{max}}$, and $\Delta M_{15}$ for the data sets that have more than ten SNe in common. One can see that there are small systematic differences in the peak magnitudes, which are significant to about 1$\sigma$ level. Similar small differences between the diferent datasets have also been noticed previously, for example, \cite{2013MNRAS.433.2240G}. The reasons for these systematic differences are likely related to the photometric calibration and different LC sampling.  $(B-V)_{B_\mathrm{max}}$ and $\Delta M_{15}$ shows relatively large scatter, but no significant bias. 
We also compare our results those obtained by \cite{2014A&A...568A..22B} with the SALT2 LC fitter.
For the 104 SNe in common the SALT2 $x1$ and our stretch parameters are tightly related as $s_\mathrm{opt}=1.0+0.108\times x1$, and $B_\mathrm{max}$ show only a small offset with a median value of 0.022 mag and a robust $1\sigma$ scatter of 0.042 mag. The median difference between our $(B-V)_{B_\mathrm{max}}$ color and SALT2 $c$ parameter is $0.047\pm0.042$ mag, which is consistent with the definition of the $c$ parameter \citep{2005A&A...443..781G}. Given the significant differences between the two LC fitting approaches, we conclude that the results compare rather well.

\subsubsection{NIR light curve parameters}

\begin{figure}[!t]
\includegraphics*[width=8.8cm]{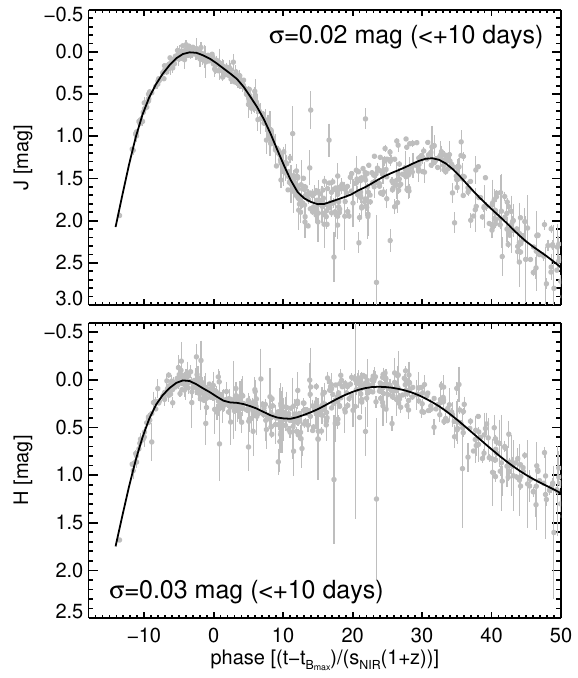}
\caption{Data used to create the $J$ and $H$ LC templates. The templates are overplotted with
red lines. $\sigma_\mathrm{w}$ is the weighted RMS of the data around the template during the phases earlier than 
$+$10 days. }
\label{f:templ}
\end{figure} 
 
The literature NIR LCs used in this work come from the following sources:  \citet{2012MNRAS.425.1007B} (BN12 set), 
 \citet{2008ApJ...689..377W} (CfA set), \citet{2010AJ....139..519C,2011AJ....142..156S} (CSP set), \citet{kri_04,kri_ir_temp,kri_01el,kri_01,kri_00,2007AJ....133...58K} (Kri set) and the following SNe from various sources (Var set)  SN~1998bu \citep{her98bu,1998IAUC.6907....2M,1999ApJS..125...73J}, SN~1999ee \citep{mario99ee}, SN~2000E  \citep{2003ApJ...595..779V},  SN~2002dj \citep{2008MNRAS.388..971P}, SN~2003cg \citep{nancy06}, SN~2003du \citep{stan_03du}, SN~2003hv \citep{2009A&A...505..265L}, SN~2004eo \citep{2007MNRAS.377.1531P}, SN~2005cf \citep{2009ApJ...697..380W,2007MNRAS.376.1301P}, SN~2008Q (Stanishev et al., in preparation), and SN~2011fe \citep{2012ApJ...754...19M}.

\paragraph{NIR light curve templates:}
\label{sec:nirtemplates}

For the fit of  the NIR LCs we adopted the template fitting approach coupled with a procedure to compute NIR  $K$-corrections that we developed (see Appendix~\ref{sec:kcor} for details). The $K$-corrections were computed from the different observed filter systems to the  Mauna Kea Observatories $JHKs$ photometric system  \citep{irfilt} as realized by the filter - detector  combination at the NOTCam instrument at the NOT. Although there are currently several NIR LC templates available, with the excellent new NIR photometry of many nearby SNe~Ia published recently \citep[e.g.,][]{2010AJ....139..519C,2011AJ....142..156S} we also derived new templates. 

To derive the NIR template LCs, only SNe that had (i) good optical LCs and (ii) several NIR observations obtained before phase +10 days and well-distributed along the LC were used. Known peculiar objects such as SN 2002cx-likes, SNe 2006bt, 2006ot, and SN 1991bg-likes, were discarded. As a starting point, the $J$ and $H$-band LCs were simultaneously fitted with the \cite{2008ApJ...689..377W} templates and our new NIR $K$-correction procedure. In the fit, the templates were stretched with the optical stretch $s_\mathrm{opt}$ and the zero time was fixed to $t_{B_\mathrm{max}}$ as in \cite{kri_04,kri_ir_temp}. Examination of the well-sampled LCs revealed that some fits were not very good. The reason appeared to be that the optical stretch $s_\mathrm{opt}$ did not provide adequate stretching to fit the NIR LCs. To further test this we selected a subsample of SNe with well-sampled LCs, which we refer to as the "GOLD" sample. For a SN to enter this sample it has to have (i) at least four points in $J$ and/or $H$ band obtained before phase +10 days with small uncertainty (in general $<0.05$ mag); (ii) the points are evenly-distributed along the LC; (iii) has observations within two days from the time of NIR maximum and at least one observation between phases $5-10$ days. Such SNe allow to perform the fit of the first NIR LC maximum with the NIR stretch $s_\mathrm{NIR}$  and the time of maximum as free fitting parameters. The 40 objects in the GOLD sample are: 
\object{SN\,1998bu}, \object{SN\,1999ee}, \object{SN\,1999ek}, \object{SN\,2000E},  \object{SN\,2001cz}, \object{SN\,2001el}, \object{SN\,2002bo}, \object{SN\,2002dj},
\object{SN\,2003du}, \object{SN\,2004eo}, \object{SN\,2004ey}, \object{SN\,2005M}, \object{SN\,2005cf}, \object{SN\,2005el}, \object{SN\,2005eq},
\object{SN\,2005hc}, 
\object{SN\,2005iq}, \object{SN\,2005kc}, \object{SN\,2005ki}, \object{SN\,2006D},  \object{SN\,2006X}, \object{SN\,2006ax}, \object{SN\,2006et}, \object{SN\,2006kf}, 
\object{SN\,2006le}, \object{SN\,2006lf}, \object{SN\,2006os}, \object{SN\,2007A}, \object{SN\,2007S},  \object{SN\,2007af}, \object{SN\,2007bc}, \object{SN\,2007bm}, 
\object{SN\,2007ca}, \object{SN\,2007le}, \object{SN\,2008Q},  \object{SN\,2008bc}, \object{SN\,2008fp}, \object{SN\,2008gp}, \object{SN\,2008hv}, and \object{SN\,2011fe}. SNe \object{2005el}, \object{2005eq}, \object{2005iq}, \object{2006X,} and \object{2006ax} have LCs from both CSP and CfA, and the LCs were analyzed separately. The fit of the CfA LC of \object{SN\,2005iq} and \object{SN\,2006ac} provided very weak constraints on the stretch and the time of maximum and was moved from the GOLD sample to the SILVER one (see Sect.~\ref{sec:undersampled}). 
The differences between the peak magnitudes estimated from the  CSP and CfA photometry are at most 0.08 mag. The only exception is $H$ band of \object{2006ax}, for which the $H$-band photometry from CfA was 0.22 mag fainter than CSP. Even though the analysis of the maximum colors indicated that those obtained from the CSP photometry were more likely to be correct, we also kept the CfA photometry of this SN.

To derive the final templates, the observed LCs were $K$-corrected to rest frame magnitudes, normalized to zero magnitude at $t_{B_\mathrm{max}}$, converted to phases using Eq.~\ref{eq:ph} with $s=s_\mathrm{NIR}$, and stacked together. The data for phases before $+10$ days were fitted with smoothing-spline function, manually adjusting the smoothing parameter to achieve a smooth fit that leaves no correlated residuals and again taking care to avoid overfitting. A few points that deviated by more than $3\sigma$ were removed from the fits. The same was repeated for the later data and the two fits were stitched together. Because our NIR $K$-correction are reliable up to phases $\sim+40-50$ days, the data past $+50$ days was not used. We noticed that the NIR stretch $s_\mathrm{NIR}$ of SNe with $s_\mathrm{opt}\sim1$ was not 1. For this reason the template time axis was additionally stretched  so that the SNe with $s_\mathrm{opt}\sim1$ also had on average $s_\mathrm{NIR}\sim1$.

\begin{figure}[!t]
\includegraphics*[width=8.8cm]{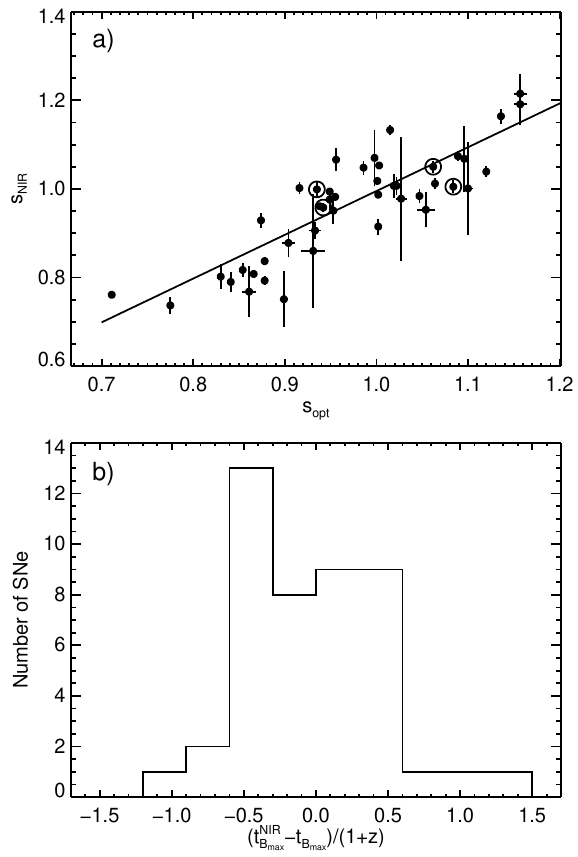}
\caption{ (a) - NIR $vs.$ optical stretch parameter. The solid line is the best-line fit. The four encircled points are the SNe shown in Fig.~\ref{f:nir_fits}; (b) - histogram of the difference between time of $B$ band maximum estimated from the NIR and optical LC fits from the GOLD sample.}\label{f:g_fit}
\end{figure} 

 Armed with the new templates the above steps were repeated to derive our final $J$ and $H$-band LC templates shown in Fig.~\ref{f:templ}. The templates were additionally normalized to have zero magnitude at the first NIR LC peak so that the fitted magnitude offsets are the rest-frame NIR peak magnitudes.
 The weighted rms scatter of the points around templates\footnote{weighted by their photometric errors}, $\sigma_\mathrm{w}$, is $\sigma_\mathrm{w}\simeq0.02$ mag and  $\sigma_\mathrm{w}\simeq0.03$ mags for $J$ and $H$, respectively. The small scatter shows that the templates describe the LCs of the GOLD sample very well. Between +10 and +50 days the scatter is considerably larger, $\sim$0.13 mag and 0.09 mag for $J$ and $H$, respectively.

\paragraph{GOLD SN sample:}

\begin{figure}[!t]
\includegraphics*[width=8.8cm]{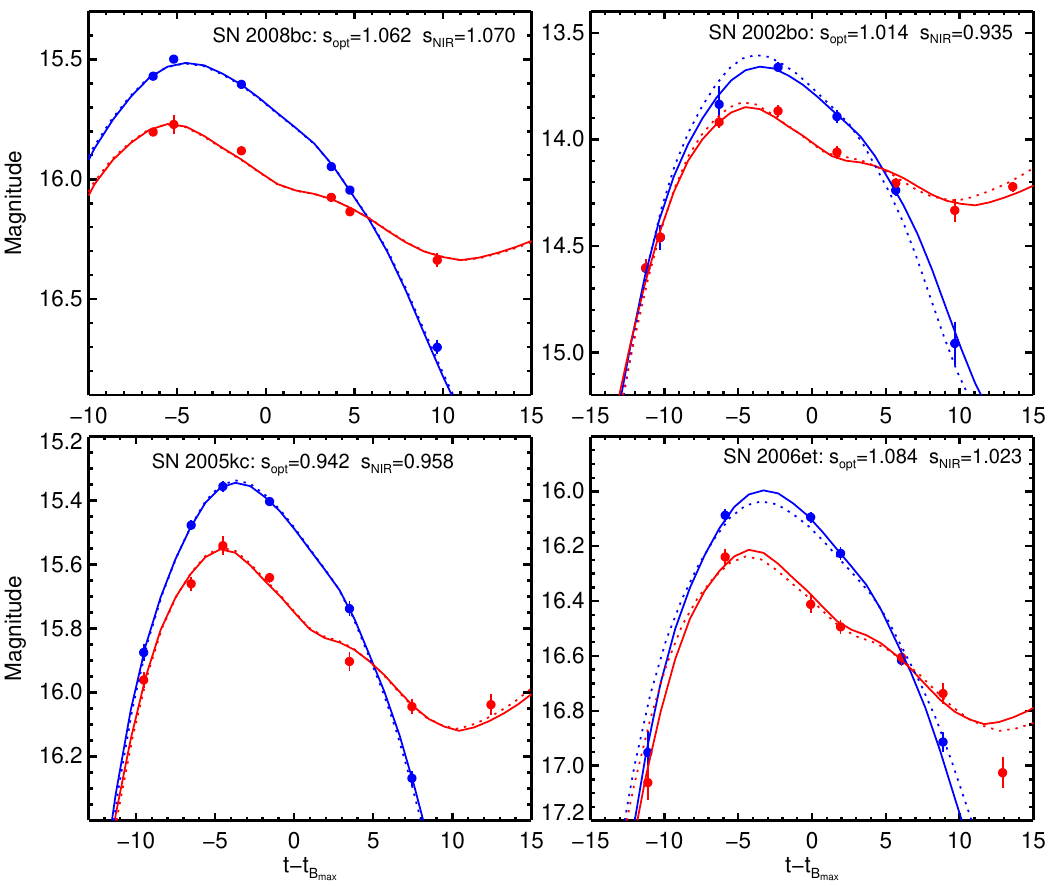}
\caption{Example of NIR LC fit of SNe from the GOLD sample. The solid line is the best fit, and dotted lines are fits with $s_\mathrm{NIR}=s_\mathrm{opt}$. The $J$ band is shown in blue and $H$ in red.}
\label{f:nir_fits}
\end{figure}

The LCs of the GOLD SN sample were fitted with the new templates with the NIR stretch $s_\mathrm{NIR}$  and the time of maximum as free fitting parameters. The NIR and the optical LC parameters are shown in Table~\ref{t:sne_g}. Figure~\ref{f:g_fit}a shows the NIR $vs.$ the optical stretch parameters. The best weighted least-squares linear fit to the points is shown with the solid line and it essentially coincides with the one-to-one correspondence $s_\mathrm{NIR}=s_\mathrm{opt}$. An inspection of the NIR LC fits suggests that the scatter in Fig.~\ref{f:g_fit}a most likely arises from real differences between optical and NIR stretch. In Fig.~\ref{f:nir_fits} are shown the fits of four SNe. The solid lines are the fits with  $s_\mathrm{NIR}$ as a free parameter and the dotted lines with  $s_\mathrm{NIR}=s_\mathrm{opt}$. It is evident that the LCs of \object{SN\,2002bo} and \object{SN\,2006et} are better fitted with  $s_\mathrm{NIR}$ as a free parameter. In order to have  $\chi^2/DoF=1$  ($DoF -$ degrees of freedom) with respect to the line defined by the $s_\mathrm{NIR}=s_\mathrm{opt}$ relation it was necessary to add the quantity $\sigma_s=0.05$ in quadrature to the uncertainties of $s_\mathrm{NIR}$.  This can be interpreted as intrinsic scatter of the  $s_\mathrm{NIR}-s_\mathrm{opt}$ relation.

The zero time of the NIR LC templates is set at $t_{B_\mathrm{max}}$. Thus, the NIR LC fits with the stretch and time of "maximum" as free parameters provide an independent estimate of $t_{B_\mathrm{max}}$. In Fig.~\ref{f:g_fit}b we show the histogram of the difference $\Delta t_{B_\mathrm{max}}$ between the estimates of $t_{B_\mathrm{max}}$ from the NIR and the optical LC fits. The distribution peaks at zero and has a FWHM around one day. The width is larger than what can be expected from the uncertainties of $\Delta t_{B_\mathrm{max}}$ alone and indicates possible presence of small random time-offsets between the optical and NIR LCs. Thus, an additional term $\sigma_t$ was added in quadrature to the uncertainties of $\Delta t_{B_\mathrm{max}}$ and its value $\sigma_t=0.44$ day was estimated so that $\chi^2/DoF$ with respect to the weighted mean of $\Delta t_{B_\mathrm{max}}$ was equal to one. 

\subsubsection{Fitting under-sampled NIR light curves}

\label{sec:undersampled}

\begin{sidewaystable*}\small
\caption{Parameters of the SNe in GOLD sample. The 1$\sigma$ uncertainties are given in parentheses.}
\label{t:sne_g}
\setlength{\tabcolsep}{3pt}
\begin{tabular}{@{}lcrccccccccc@{}}
\hline
\hline
\noalign{\smallskip}
SN   	& $z_\mathrm{CMB}$ & $t^\mathrm{NIR}_{B_\mathrm{max}}-t^\mathrm{opt}_{B_\mathrm{max}}$ & $s_\mathrm{NIR}$ & $J_\mathrm{max}$\tablefootmark{b}  & $H_\mathrm{max}$\tablefootmark{b} & $t_{B_\mathrm{max}}$ & $s_\mathrm{opt}$ &  $\Delta M_{15}$\tablefootmark{c} & $B_\mathrm{max}$ & $V$ at $t_{B_\mathrm{max}}$ & Ref.\tablefootmark{a} \\
\hline
\noalign{\smallskip}
1998bu & 0.00416 &  $-$0.53 (0.15)  & 1.066 (0.026)  & 11.544 (0.014/0.006)  & 11.636 (0.012/0.004)  & 50952.53 (0.02)  & 0.956 (0.002)  & 1.117 (21) & 12.109 (0.002)  & 11.827 (0.001)  & Var \\
1999ee & 0.01053 &     0.00 (0.07)  & 1.074 (0.012)  & 14.734 (0.007/0.008)  & 14.982 (0.007/0.010)  & 51469.39 (0.01)  & 1.089 (0.002)  & 0.903 (20) & 14.873 (0.002)  & 14.610 (0.002)  & Kri \\
1999ek & 0.01759 &  $-$0.21 (0.16)  & 0.906 (0.019)  & 15.742 (0.011/0.010)  & 15.968 (0.009/0.014)  & 51482.24 (0.07)  & 0.933 (0.007)  & 1.162 (25) & 15.893 (0.005)  & 15.714 (0.004)  & Kri \\
2000E  & 0.00422 &  $-$0.48 (0.09)  & 0.984 (0.016)  & 13.206 (0.012/0.007)  & 13.574 (0.023/0.006)  & 52576.77 (0.03)  & 1.047 (0.003)  & 0.962 (21) & 13.010 (0.004)  & 12.844 (0.003)  & Var \\
2001cz & 0.01570 &  $-$0.43 (2.16)  & 0.978 (0.140)  & 15.262 (0.125/0.010)  & 15.615 (0.086/0.012)  & 52103.53 (0.08)  & 1.027 (0.006)  & 0.993 (22) & 15.092 (0.006)  & 15.016 (0.004)  & Kri \\
2001el & 0.00368 &     0.57 (0.06)  & 1.018 (0.009)  & 12.794 (0.012/0.006)  & 12.954 (0.015/0.005)  & 52182.14 (0.02)  & 1.001 (0.002)  & 1.035 (21) & 12.820 (0.003)  & 12.655 (0.002)  & Kri \\
2002bo & 0.00529 &  $-$0.07 (0.16)  & 0.999 (0.019)  & 13.648 (0.019/0.006)  & 13.848 (0.013/0.005)  & 53356.95 (0.03)  & 0.935 (0.003)  & 1.158 (21) & 13.974 (0.006)  & 13.545 (0.004)  & Kri \\
2002dj & 0.01045 &  $-$0.43 (0.27)  & 0.951 (0.031)  & 14.426 (0.020/0.007)  & 14.673 (0.017/0.009)  & 52450.50 (0.02)  & 0.953 (0.002)  & 1.122 (21) & 13.975 (0.004)  & 13.866 (0.003)  & Var \\
2003du & 0.00665 &  $-$0.05 (0.14)  & 1.007 (0.020)  & 14.127 (0.028/0.007)  & 14.383 (0.021/0.007)  & 52766.15 (0.02)  & 1.022 (0.002)  & 1.001 (21) & 13.453 (0.002)  & 13.569 (0.002)  & Var \\
2004eo & 0.01472 &     0.36 (0.13)  & 0.929 (0.017)  & 15.453 (0.007/0.010)  & 15.725 (0.007/0.012)  & 53278.10 (0.02)  & 0.874 (0.002)  & 1.293 (21) & 15.095 (0.004)  & 15.005 (0.004)  & CSP \\
2004ey & 0.01461 &  $-$0.14 (0.25)  & 1.007 (0.027)  & 15.419 (0.010/0.010)  & 15.685 (0.010/0.013)  & 53304.32 (0.01)  & 1.019 (0.001)  & 1.006 (20) & 14.781 (0.001)  & 14.867 (0.002)  & CSP \\
2005M  & 0.02561 &     0.89 (0.09)  & 1.164 (0.017)  & 16.412 (0.009/0.013)  & 16.634 (0.007/0.018)  & 53405.64 (0.02)  & 1.136 (0.002)  & 0.844 (20) & 15.910 (0.002)  & 15.895 (0.002)  & CSP \\
2005cf & 0.00705 &     0.33 (0.13)  & 1.048 (0.016)  & 13.732 (0.012/0.007)  & 13.868 (0.010/0.007)  & 53533.57 (0.01)  & 0.986 (0.001)  & 1.061 (20) & 13.313 (0.003)  & 13.259 (0.002)  & Var \\
2005el & 0.01488 &  $-$0.37 (0.07)  & 0.837 (0.009)  & 15.404 (0.006/0.009)  & 15.643 (0.005/0.012)  & 53646.08 (0.02)  & 0.878 (0.003)  & 1.284 (22) & 14.882 (0.004)  & 14.984 (0.004)  & CSP \\
2005el & 0.01488 &     0.00 (0.07)  & 0.793 (0.010)  & 15.341 (0.007/0.013)  & 15.587 (0.010/0.005)  & 53646.08 (0.02)  & 0.878 (0.003)  & 1.284 (22) & 14.882 (0.004)  & 14.984 (0.004)  & CfA \\
2005eq & 0.02833 &     0.60 (0.28)  & 1.215 (0.045)  & 16.781 (0.015/0.015)  & 17.027 (0.017/0.020)  & 53653.57 (0.06)  & 1.157 (0.007)  & 0.820 (22) & 16.265 (0.005)  & 16.236 (0.005)  & CSP \\
2005eq & 0.02833 &  $-$0.40 (0.52)  & 1.191 (0.047)  & 16.704 (0.024/0.017)  & 17.099 (0.031/0.013)  & 53653.57 (0.06)  & 1.157 (0.007)  & 0.820 (22) & 16.265 (0.005)  & 16.236 (0.005)  & CfA \\
2005hc & 0.04445 &  $-$0.52 (0.67)  & 1.001 (0.106)  & 17.830 (0.039/0.022)  & 17.926 (0.061/0.029)  & 53667.13 (0.06)  & 1.100 (0.005)  & 0.889 (21) & 17.299 (0.003)  & 17.287 (0.003)  & CSP \\
2005iq & 0.03289 &     0.57 (0.61)  & 0.751 (0.063)  & 17.294 (0.040/0.017)  & 17.500 (0.055/0.023)  & 53687.69 (0.03)  & 0.899 (0.004)  & 1.235 (22) & 16.782 (0.004)  & 16.798 (0.003)  & CSP \\
2005kc & 0.01387 &     0.13 (0.07)  & 0.958 (0.011)  & 15.355 (0.009/0.010)  & 15.577 (0.010/0.012)  & 53697.57 (0.02)  & 0.942 (0.003)  & 1.144 (21) & 15.565 (0.004)  & 15.354 (0.003)  & CSP \\
2005ki & 0.02039 &  $-$0.44 (0.10)  & 0.817 (0.015)  & 16.083 (0.015/0.011)  & 16.275 (0.019/0.015)  & 53705.28 (0.02)  & 0.854 (0.002)  & 1.342 (21) & 15.540 (0.004)  & 15.580 (0.003)  & CSP \\
2006D  & 0.00965 &  $-$0.31 (0.22)  & 0.802 (0.028)  & 14.276 (0.013/0.010)  & 14.514 (0.014/0.004)  & 53757.50 (0.01)  & 0.830 (0.001)  & 1.405 (21) & 14.160 (0.002)  & 14.042 (0.002)  & CfA \\
2006X  & 0.00429 &  $-$0.38 (0.06)  & 0.994 (0.009)  & 12.813 (0.005/0.007)  & 12.934 (0.003/0.006)  & 53786.16 (0.01)  & 0.949 (0.001)  & 1.130 (20) & 15.246 (0.002)  & 13.882 (0.002)  & CSP \\
2006X  & 0.00429 &  $-$0.46 (0.13)  & 0.976 (0.023)  & 12.873 (0.017/0.011)  & 12.924 (0.009/0.005)  & 53786.16 (0.01)  & 0.949 (0.001)  & 1.130 (20) & 15.246 (0.002)  & 13.882 (0.002)  & CfA \\
2006ax & 0.01798 &  $-$0.14 (0.05)  & 0.987 (0.007)  & 15.663 (0.007/0.010)  & 15.935 (0.006/0.013)  & 53827.35 (0.01)  & 1.002 (0.001)  & 1.034 (20) & 14.990 (0.002)  & 15.075 (0.001)  & CSP \\
2006ax & 0.01798 &  $-$0.38 (0.13)  & 0.915 (0.018)  & 15.661 (0.017/0.014)  & 16.157 (0.033/0.006)  & 53827.35 (0.01)  & 1.002 (0.001)  & 1.034 (20) & 14.990 (0.002)  & 15.075 (0.001)  & CfA \\
2006et & 0.02162 &     0.26 (0.10)  & 1.005 (0.015)  & 16.023 (0.014/0.012)  & 16.256 (0.015/0.017)  & 53993.74 (0.03)  & 1.084 (0.002)  & 0.910 (20) & 15.959 (0.003)  & 15.783 (0.002)  & CSP \\
2006kf & 0.02079 &  $-$0.26 (0.11)  & 0.737 (0.020)  & 16.203 (0.007/0.012)  & 16.414 (0.028/0.016)  & 54041.32 (0.02)  & 0.775 (0.003)  & 1.564 (23) & 15.929 (0.004)  & 15.898 (0.004)  & CSP \\
2006le & 0.01726 &  $-$0.91 (0.43)  & 1.068 (0.074)  & 15.857 (0.036/0.014)  & 16.305 (0.037/0.006)  & 54048.21 (0.03)  & 1.096 (0.004)  & 0.894 (21) & 14.977 (0.005)  & 15.015 (0.004)  & CfA \\
2006lf & 0.01296 &     0.18 (0.41)  & 0.768 (0.058)  & 14.906 (0.033/0.012)  & 15.213 (0.047/0.004)  & 54044.69 (0.06)  & 0.861 (0.008)  & 1.325 (29) & 14.238 (0.017)  & 14.235 (0.010)  & CfA \\
2006os & 0.03206 &     0.51 (1.91)  & 0.860 (0.128)  & 17.314 (0.131/0.016)  & 17.357 (0.078/0.023)  & 54064.95 (0.31)  & 0.931 (0.013)  & 1.166 (34) & 17.642 (0.013)  & 17.273 (0.008)  & CSP \\
2007A  & 0.01593 &     0.20 (0.21)  & 0.953 (0.040)  & 15.613 (0.020/0.010)  & 15.926 (0.050/0.013)  & 54112.88 (0.06)  & 1.054 (0.010)  & 0.952 (25) & 15.697 (0.005)  & 15.510 (0.004)  & CSP \\
2007S  & 0.01504 &     0.29 (0.11)  & 1.039 (0.012)  & 15.312 (0.008/0.009)  & 15.523 (0.008/0.011)  & 54144.19 (0.02)  & 1.120 (0.003)  & 0.864 (20) & 15.779 (0.004)  & 15.380 (0.003)  & CSP \\
2007af & 0.00629 &     0.03 (0.02)  & 0.982 (0.003)  & 13.424 (0.002/0.007)  & 13.596 (0.003/0.006)  & 54174.24 (0.01)  & 0.955 (0.001)  & 1.118 (20) & 13.152 (0.002)  & 13.098 (0.002)  & CSP \\
2007bc & 0.02187 &     1.00 (0.25)  & 0.878 (0.032)  & 16.298 (0.013/0.011)  & 16.488 (0.046/0.016)  & 54199.57 (0.07)  & 0.904 (0.007)  & 1.224 (26) & 15.868 (0.005)  & 15.894 (0.005)  & CSP \\
2007bm & 0.00745 &     0.07 (0.07)  & 1.002 (0.013)  & 13.872 (0.003/0.007)  & 14.080 (0.009/0.007)  & 54224.38 (0.02)  & 0.916 (0.001)  & 1.198 (21) & 14.488 (0.002)  & 13.947 (0.002)  & CSP \\
2007ca & 0.01509 &  $-$0.16 (0.11)  & 1.012 (0.013)  & 15.525 (0.006/0.009)  & 15.657 (0.011/0.011)  & 54227.48 (0.02)  & 1.064 (0.003)  & 0.937 (21) & 15.906 (0.003)  & 15.635 (0.002)  & CSP \\
2007le & 0.00551 &     0.46 (0.07)  & 1.053 (0.006)  & 13.734 (0.004/0.007)  & 13.930 (0.003/0.007)  & 54398.64 (0.05)  & 1.003 (0.004)  & 1.032 (21) & 13.904 (0.005)  & 13.571 (0.005)  & CSP \\
2007on & 0.00618 &     1.26 (0.02)  & 0.761 (0.003)  & 13.042 (0.005/0.007)  & 13.116 (0.003/0.007)  & 54419.30 (0.01)  & 0.711 (0.001)  & 1.779 (22) & 13.020 (0.003)  & 12.902 (0.002)  & CSP \\
2008Q  & 0.00689 &  $-$0.21 (0.12)  & 0.790 (0.023)  & 13.798 (0.024/0.007)  & 13.902 (0.032/0.008)  & 54505.81 (0.01)  & 0.841 (0.002)  & 1.376 (21) & 13.481 (0.003)  & 13.480 (0.003)  & Var \\
2008bc & 0.01572 &  $-$0.82 (0.06)  & 1.050 (0.014)  & 15.525 (0.005/0.010)  & 15.794 (0.008/0.012)  & 54549.93 (0.01)  & 1.062 (0.001)  & 0.940 (20) & 14.714 (0.002)  & 14.793 (0.002)  & CSP \\
2008fp & 0.00629 &     0.45 (0.05)  & 1.133 (0.012)  & 13.344 (0.002/0.007)  & 13.537 (0.002/0.006)  & 54730.58 (0.02)  & 1.015 (0.002)  & 1.012 (21) & 13.867 (0.003)  & 13.350 (0.002)  & CSP \\
2008gp & 0.03280 &     0.34 (0.21)  & 1.070 (0.063)  & 17.202 (0.046/0.016)  & 17.441 (0.052/0.023)  & 54779.04 (0.02)  & 0.998 (0.002)  & 1.040 (21) & 16.460 (0.004)  & 16.430 (0.003)  & CSP \\
2008hv & 0.01359 &  $-$0.75 (0.05)  & 0.808 (0.005)  & 15.184 (0.015/0.009)  & 15.468 (0.009/0.010)  & 54816.88 (0.01)  & 0.866 (0.001)  & 1.313 (21) & 14.745 (0.003)  & 14.715 (0.002)  & CSP \\
2011fe & 0.00121 &  $-$0.34 (0.02)  & 0.961 (0.003)  & 10.443 (0.005/0.006)  & 10.703 (0.003/0.003)  & 55814.97 (0.01)  & 0.937 (0.002)  & 1.154 (21) &  9.962 (0.004)  &  9.963 (0.002)  & Var \\

\hline
\end{tabular}\\
\tablefoot{
\tablefoottext{a}{See text for the references for the NIR photometry.}
\tablefoottext{b}{The two numbers are the uncertainties of the peak magnitude and the $K$-corrections.}
\tablefoottext{c}{Estimated from our $s_\mathrm{opt}-\Delta M_{15}$ relation. The uncertanties are in units of 0.001 mag.} 
}
\end{sidewaystable*}

\begin{table*}[!t]
\caption{Parameters of the SNe in our new sample. The 1$\sigma$ uncertainties are given in parentheses.}
\label{t:sne_o}
\setlength{\tabcolsep}{3pt}
\begin{tabular}{@{}lccccccc@{}}
\hline
\hline\noalign{\smallskip}
SN &  $J_\mathrm{max}$\tablefootmark{c,d} & $H_\mathrm{max}$\tablefootmark{c,d}  & $t_{B_\mathrm{max}}$\tablefootmark{a} & $s_\mathrm{opt}$ &  $\Delta M_{15}$\tablefootmark{b,c}  & $B_\mathrm{max}$\tablefootmark{c} & $V$ at $t_{B_\mathrm{max}}$\tablefootmark{c} \\
\hline\noalign{\smallskip}

SN 2007hz       & 20.273 (072/49)  & 20.678 (126/080)  & 54350.86 (0.25)  & 1.069 (0.020)  & 0.930 (35) & 19.552 (018)  & 19.639 (014) \\
SN 2007ih       & 21.198 (092/65)  & 21.659 (196/100)  & 54352.61 (0.42)  & 1.184 (0.046)  & 0.791 (52) & 20.767 (025)  & 20.752 (027) \\
SN 2007ik       & 21.319 (232/47)  &     $-$           & 54356.78 (0.28)  & 1.090 (0.030)  & 0.902 (45) & 20.617 (027)  & 20.660 (028) \\
SN 2008bz       & 18.482 (271/27)  & 18.064 (211/033)  & 54579.43 (0.56)  & 0.935 (0.017)  & 1.158 (40) & 17.911 (053)  & 18.006 (030) \\
SNF20080510-005 & 19.742 (121/34)  & 19.569 (115/045)  & 54604.39 (0.25)  & 1.177 (0.029)  & 0.798 (37) & 18.826 (022)  & 18.860 (018) \\
SNF20080512-008 & 19.217 (098/34)  & 19.267 (102/045)  & 54591.19 (0.99)  & 1.237 (0.027)  & 0.738 (33) & 18.179 (087)  & 18.302 (050) \\
SNF20080512-010 & 18.706 (128/28)  & 19.014 (096/035)  & 54602.88 (0.35)  & 0.743 (0.012)  & 1.668 (45) & 18.038 (049)  & 18.115 (030) \\
SNF20080516-000 & 18.865 (067/31)  & 19.101 (071/039)  & 54606.33 (0.51)  & 1.091 (0.015)  & 0.900 (28) & 18.165 (038)  & 18.275 (022) \\
SNF20080516-022 & 19.150 (058/31)  & 19.214 (090/040)  & 54609.79 (0.31)  & 0.947 (0.015)  & 1.134 (36) & 18.379 (021)  & 18.431 (013) \\
SNF20080517-010 & 20.348 (125/42)  & 20.907 (276/068)  & 54605.97 (0.81)  & 1.003 (0.029)  & 1.032 (53) & 19.692 (052)  & 19.625 (028) \\
SNF20080522-000 & 17.684 (033/21)  & 17.887 (030/026)  & 54621.17 (0.13)  & 1.092 (0.007)  & 0.899 (22) & 17.027 (007)  & 17.023 (006) \\
SNF20080522-001 & 17.926 (131/23)  & 18.119 (099/028)  & 54602.83 (0.45)  & 1.138 (0.021)  & 0.842 (32) & 17.422 (049)  & 17.456 (027) \\
SNF20080522-004 & 19.286 (239/40)  & 19.756 (338/058)  & 54606.33 (2.88)  & 1.260 (0.067)  & 0.717 (65) & 18.991 (185)  & 19.317 (100) \\
SNF20080522-011 & 17.284 (025/17)  & 17.521 (043/022)  & 54616.40 (0.15)  & 1.052 (0.009)  & 0.955 (24) & 16.653 (009)  & 16.724 (009) \\
SNF20080606-012 & 18.983 (057/33)  & 19.369 (091/042)  & 54631.93 (0.50)  & 0.954 (0.017)  & 1.120 (39) & 18.684 (036)  & 18.736 (022) \\
SNF20080610-003 & 19.794 (077/37)  & 19.811 (092/051)  & 54633.43 (0.75)  & 1.047 (0.019)  & 0.962 (35) & 18.865 (051)  & 18.957 (030) \\

\hline
\end{tabular} \\
\tablefoot{
\tablefoottext{a}{JD$-$2454000.}
\tablefoottext{b}{Estimated from our $s_\mathrm{opt}-\Delta M_{15}$ relation.}
\tablefoottext{c}{The uncertainties are in units of 0.001 mag.}
\tablefoottext{d}{The two numbers are the uncertainties of the peak magnitude and the $K$-corrections.}
}
\end{table*}

The analysis of the GOLD sample further reinforced the results of \cite{kri_04,kri_ir_temp} and \cite{2008ApJ...689..377W} who showed that smooth LC templates stretched according to the optical stretch accurately describe the rest-frame $J$ and $H$-band LCs within ten days from $B$-band maximum. Thus, if $s_\mathrm{opt}$ and  $t_{B_\mathrm{max}}$ are estimated from optical observations, by fixing the zero time of the NIR template to $t_{B_\mathrm{max}}$  and setting  $s_\mathrm{NIR}=s_\mathrm{opt}$, one could in principle measure the SN~Ia NIR peak magnitudes with a single observation. We used this approach to estimate the peak magnitudes of the SNe in our sample and all published SNe with under-sampled LCs with observations before +10 days. This also includes the \citetalias{2012MNRAS.425.1007B} sample. However, we note that most SNe in this sample have several observations around maximum and the LC quality is closer to that of the GOLD sample; often the only difference is in the photometric uncertainties, which are smaller in the GOLD sample. The parameters derived from the LC fits are given in Tables~\ref{t:sne_o}--\ref{t:sne_lid}.

\paragraph{Expected uncertainties:}

With the adopted fitting procedure the uncertainties of the time of maximum and the stretch, both from the optical LC fitting and the nonzero scatter $\sigma_t$ and $\sigma_s$, will propagate to the uncertainties of the estimated peak magnitudes. This additional source of uncertainty is especially important when only a single NIR observation is available. In this case the magnitude of the uncertainty will also depend of the phase of the observation.

\begin{figure}[!t]
\includegraphics*[width=8.8cm]{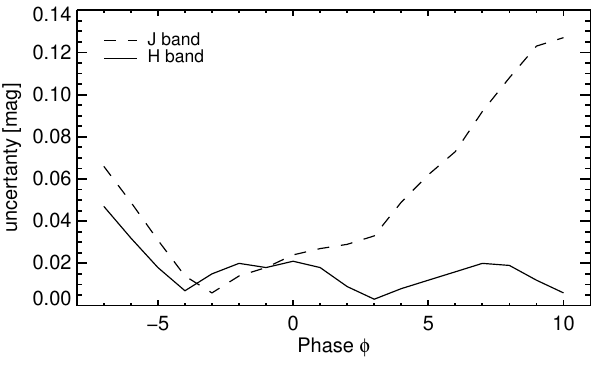}
\caption{Peak magnitude uncertainties of a SN with $s_\mathrm{NIR}=1.0$ $vs.$ the phase of the single observation. The 1$\sigma$ uncertainties of the stretch and the time of maximum were assumed to be  0.05 and 0.44 days, respectively.}
\label{f:sim_templ}
\end{figure}

To estimate the expected magnitude of this uncertainty when only a single observation is available a series of simple Monte Carlo simulations were performed. For a given stretch $s_0$, 10000 random pairs of stretch and time offset ($s_{0,i},t_{0,i}$) were generated with the relations $s_{0,i}=s_0+N(0,\sigma^2_s)$ and $t_{0,i}=N(0,\sigma^2_t)$, where $N(\mu,\sigma^2)$ is Gaussian distribution with mean $\mu$ and variance $\sigma^2$, and $\sigma_t=0.44$ day and $\sigma_s=0.05$. The $J$ and $H$ templates were stretched and shifted with each pair ($s_{0,i},t_{0,i}$), and the scatter with respect to the unperturbed template was computed for each phase from $-7$ to +10 days. The results for $s_\mathrm{NIR}=1.0$ are shown in Fig.~\ref{f:sim_templ}. As expected, when the observation is obtained with a few days from the NIR maximum the peak magnitudes  in both bands are fairly insensitive to the uncertainties of the stretch and the time of maximum. After maximum, the uncertainty in the $H$ band remains less than 0.02 mag, while in $J$ it increases reaching 0.12 mag at +10 days. These results can be understood by noting that the $J$-band flux changes much more rapidly than $H$. 

The previous simulations do not include the uncertainties from the optical LC fitting. Thus, a set of more realistic simulations based on the LCs of the GOLD SN sample were also performed. Two points were randomly drawn from each LC, one from the data within 3 days from the NIR maximum, and another one from the data taken later that 4 days after $t_{B_\mathrm{max}}$. The selected points were fitted with the method for under-sampled LCs, and the weighted mean and RMS of the difference from the peak magnitudes estimated from the full LC fits was computed. The mean values were in all cases close to zero indicating that the LC under-sampling does not lead to bias of the estimated peak magnitudes. The weighted scatter for the points close to maximum was $\sim0.025$ mag in both bands. The scatter increases considerably with the points obtained after phase +4 days: 0.11 mag in $J$ and 0.07 mag in $H$.  
Based on the results of the simulations the literature sample with under-sampled LCs was split into two groups (i) SILVER with the SNe that have observations within two to three days from the expected time of the NIR maximum light, and (ii) BRONZE with the remaining objects. Table~\ref{t:sample_stat} shows statistics of the NIR SNe samples.

\citet{2012MNRAS.425.1007B} and \citet{2014ApJ...784..105W} also performed simulations to estimate the uncertainties from under-sampling the NIR LCs. Although these authors used different approaches from ours, the conclusions was similar -  with prior information from optical LCs, under-sampled NIR LCs can provide accurate enough peak magnitudes to be useful in cosmology.

\longtab{
\begin{landscape}
\begin{longtable}{@{}lccccccccc@{}}
\caption{Parameters of the SNe in the SILVER and BRONZE samples. The 1$\sigma$ uncertainties are given in parentheses.}
\label{t:sne_sb}\\
\hline
\hline
\noalign{\smallskip}
SN & $z_\mathrm{CMB}$ & $J_\mathrm{max}$\tablefootmark{d} & $H_\mathrm{max}$\tablefootmark{d} &  $t_{B_\mathrm{max}}$\tablefootmark{a} & $s_\mathrm{opt}$ &  $\Delta M_{15}$\tablefootmark{b} & $B_\mathrm{max}$ & $V$ at $t_{B_\mathrm{max}}$ &  Ref.\tablefootmark{c} \\
\hline \noalign{\smallskip}
\endfirsthead
\caption{continued.}\\
\hline
\hline
\noalign{\smallskip}
SN & $z_\mathrm{CMB}$ & $J_\mathrm{max}$\tablefootmark{d} & $H_\mathrm{max}$\tablefootmark{d} &  $t_{B_\mathrm{max}}$\tablefootmark{a} & $s_\mathrm{opt}$ &  $\Delta M_{15}$\tablefootmark{b} & $B_\mathrm{max}$ & $V$ at $t_{B_\mathrm{max}}$ &  Ref.\tablefootmark{c} \\
\hline \noalign{\smallskip}
\endhead
\hline
\endfoot

\multicolumn{10}{c}{SILVER sample}\\
\hline \noalign{\smallskip}

2000ca  & 0.02453 &  16.302 (0.028/0.012)  & 16.632 (0.025/0.018)  & 51666.29 (0.12)  & 1.054 (0.008)  & 0.952 (0.023) & 15.581 (0.008)  & 15.699 (0.006)  & Kri \\
2001ba  & 0.03068 &  16.976 (0.014/0.015)  & 17.275 (0.024/0.021)  & 52034.65 (0.14)  & 1.016 (0.009)  & 1.010 (0.025) & 16.235 (0.007)  & 16.329 (0.008)  & Kri \\
2001bt  & 0.01444 &  15.420 (0.029/0.009)  & 15.627 (0.027/0.012)  & 52064.43 (0.03)  & 0.935 (0.004)  & 1.158 (0.022) & 15.343 (0.006)  & 15.126 (0.004)  & Kri \\
2003cg  & 0.00531 &  13.565 (0.054/0.006)  & 13.680 (0.054/0.005)  & 52729.56 (0.03)  & 0.969 (0.005)  & 1.092 (0.022) & 15.813 (0.006)  & 14.609 (0.004)  & Var \\
2005am  & 0.00898 &  13.954 (0.023/0.007)  & 14.200 (0.011/0.008)  & 53436.96 (0.03)  & 0.774 (0.002)  & 1.567 (0.022) & 13.649 (0.002)  & 13.582 (0.002)  & CSP \\
2005eu  & 0.03407 &  17.014 (0.061/0.018)  & 17.339 (0.061/0.016)  & 53659.84 (0.04)  & 1.049 (0.007)  & 0.959 (0.023) & 16.437 (0.007)  & 16.571 (0.007)  & CfA \\
2005iq  & 0.03289 &  17.471 (0.058/0.018)  & 17.627 (0.132/0.016)  & 53687.69 (0.03)  & 0.899 (0.004)  & 1.235 (0.022) & 16.782 (0.004)  & 16.798 (0.003)  & CfA \\
2006D   & 0.00965 &  14.347 (0.017/0.007)  & 14.530 (0.013/0.008)  & 53757.50 (0.01)  & 0.830 (0.001)  & 1.405 (0.021) & 14.160 (0.002)  & 14.042 (0.002)  & CSP \\
2006ac  & 0.02395 &  16.528 (0.077/0.016)  & 16.739 (0.102/0.009)  & 53781.83 (0.08)  & 0.892 (0.007)  & 1.251 (0.026) & 16.193 (0.010)  & 16.043 (0.007)  & CfA \\
2006bh  & 0.01049 &  14.778 (0.012/0.008)  & 14.970 (0.014/0.009)  & 53833.42 (0.01)  & 0.833 (0.001)  & 1.397 (0.021) & 14.362 (0.002)  & 14.365 (0.001)  & CSP \\
2006gt  & 0.04359 &  17.816 (0.069/0.022)  & 18.155 (0.112/0.029)  & 54003.32 (0.10)  & 0.711 (0.006)  & 1.779 (0.030) & 18.235 (0.008)  & 17.912 (0.007)  & CSP \\
2007as  & 0.01790 &  15.812 (0.016/0.010)  & 16.057 (0.028/0.014)  & 54181.61 (0.08)  & 0.909 (0.004)  & 1.213 (0.022) & 15.464 (0.003)  & 15.388 (0.003)  & CSP \\
2007bd  & 0.03186 &  17.022 (0.037/0.015)  & 17.301 (0.093/0.022)  & 54206.78 (0.02)  & 0.910 (0.003)  & 1.211 (0.021) & 16.543 (0.005)  & 16.553 (0.004)  & CSP \\
2007cq  & 0.02377 &  16.254 (0.039/0.016)  & 16.921 (0.132/0.010)  & 54280.27 (0.08)  & 0.985 (0.007)  & 1.063 (0.024) & 15.901 (0.006)  & 15.887 (0.005)  & CfA \\
2007jg  & 0.03656 &  17.578 (0.018/0.018)  & 17.880 (0.074/0.025)  & 54366.51 (0.09)  & 0.942 (0.007)  & 1.144 (0.025) & 17.286 (0.005)  & 17.196 (0.004)  & CSP \\
2008ia  & 0.02255 &  16.403 (0.043/0.012)  & 16.331 (0.031/0.017)  & 54813.04 (0.05)  & 0.872 (0.004)  & 1.298 (0.023) & 15.885 (0.004)  & 15.867 (0.004)  & CSP \\

\hline \noalign{\smallskip}
\multicolumn{10}{c}{BRONZE sample}\\
\hline \noalign{\smallskip}

2000bh  & 0.02396 &  16.369 (0.083/0.012)  & 16.533 (0.032/0.017)  & 51636.19 (0.20)  & 1.002 (0.005)  & 1.034 (0.022) & 15.969 (0.013)  & 15.919 (0.011)  & Kri \\
2000bk  & 0.02662 &  16.595 (0.153/0.013)  & 17.042 (0.055/0.019)  & 51647.92 (0.09)  & 0.703 (0.003)  & 1.808 (0.024) & 16.784 (0.014)  & 16.611 (0.010)  & Kri \\
2000ce  & 0.01649 &  15.521 (0.128/0.010)  & 15.900 (0.034/0.013)  & 51665.51 (0.38)  & 1.060 (0.005)  & 0.943 (0.021) & 16.949 (0.036)  & 16.459 (0.020)  & Kri \\
2001cn  & 0.01508 &  15.640 (0.056/0.010)  & 15.769 (0.043/0.012)  & 52073.46 (0.11)  & 0.939 (0.004)  & 1.150 (0.022) & 15.345 (0.008)  & 15.127 (0.004)  & Kri \\
2003hv  & 0.00510 &  12.979 (0.066/0.007)  & 13.060 (0.025/0.006)  & 52891.79 (0.08)  & 0.773 (0.003)  & 1.571 (0.023) & 12.477 (0.007)  & 12.545 (0.005)  & Var \\
2004S   & 0.00985 &  14.605 (0.070/0.007)  & 14.744 (0.022/0.008)  & 53040.87 (0.33)  & 0.992 (0.011)  & 1.051 (0.028) & 14.208 (0.020)  & 14.153 (0.014)  & Kri \\
2004ef  & 0.02973 &  17.317 (0.094/0.015)  & 17.612 (0.108/0.022)  & 53264.07 (0.01)  & 0.861 (0.001)  & 1.325 (0.021) & 16.873 (0.001)  & 16.721 (0.001)  & CSP \\
2004gs  & 0.02830 &  16.925 (0.133/0.014)  & 17.067 (0.048/0.020)  & 53356.20 (0.04)  & 0.769 (0.004)  & 1.583 (0.024) & 17.170 (0.003)  & 16.925 (0.004)  & CSP \\
2005A   & 0.01833 &  16.310 (0.095/0.011)  & 16.342 (0.028/0.015)  & 53379.88 (0.07)  & 0.948 (0.005)  & 1.132 (0.023) & 18.158 (0.005)  & 17.038 (0.004)  & CSP \\
2005ag  & 0.08032 &  19.169 (0.081/0.032)  & 18.970 (0.057/0.045)  & 53414.26 (0.14)  & 1.046 (0.009)  & 0.964 (0.024) & 18.432 (0.005)  & 18.449 (0.004)  & CSP \\
2005al  & 0.01330 &  15.561 (0.051/0.008)  & 15.747 (0.015/0.010)  & 53430.57 (0.05)  & 0.879 (0.002)  & 1.281 (0.021) & 14.860 (0.003)  & 14.958 (0.002)  & CSP \\
2005na  & 0.02683 &  16.512 (0.078/0.016)  & 17.027 (0.143/0.011)  & 53740.98 (0.06)  & 0.941 (0.003)  & 1.146 (0.021) & 15.966 (0.003)  & 16.046 (0.003)  & CfA \\
2006N   & 0.01426 &  15.581 (0.119/0.013)  & 15.778 (0.129/0.005)  & 53761.62 (0.12)  & 0.777 (0.005)  & 1.558 (0.026) & 15.176 (0.008)  & 15.148 (0.005)  & CfA \\
2006br  & 0.02555 &  17.099 (0.086/0.013)  & 17.181 (0.098/0.018)  & 53852.16 (0.40)  & 0.921 (0.023)  & 1.187 (0.053) & 19.045 (0.023)  & 18.135 (0.012)  & CSP \\
2006cp  & 0.02333 &  16.656 (0.069/0.015)  & 16.720 (0.036/0.009)  & 53896.88 (0.07)  & 0.988 (0.008)  & 1.058 (0.025) & 15.922 (0.008)  & 15.829 (0.007)  & CfA \\
2006ej  & 0.01931 &  16.109 (0.077/0.011)  & 16.459 (0.069/0.016)  & 53975.82 (0.10)  & 0.842 (0.008)  & 1.373 (0.029) & 15.770 (0.009)  & 15.783 (0.009)  & CSP \\
2006eq  & 0.04831 &  18.145 (0.107/0.024)  & 18.157 (0.135/0.031)  & 53978.15 (0.44)  & 0.724 (0.013)  & 1.733 (0.050) & 18.506 (0.035)  & 18.156 (0.022)  & CSP \\
2006ev  & 0.02757 &  17.145 (0.090/0.014)  & 17.204 (0.073/0.020)  & 53989.70 (0.27)  & 0.865 (0.008)  & 1.315 (0.028) & 17.112 (0.021)  & 16.925 (0.012)  & CSP \\
2006gj  & 0.02767 &  17.015 (0.064/0.014)  & 17.209 (0.066/0.020)  & 54000.48 (0.11)  & 0.733 (0.015)  & 1.702 (0.056) & 17.653 (0.008)  & 17.294 (0.007)  & CSP \\
2006hb  & 0.01534 &  15.654 (0.109/0.010)  & 15.818 (0.049/0.012)  & 54003.69 (0.29)  & 0.729 (0.005)  & 1.715 (0.027) & 15.588 (0.030)  & 15.385 (0.018)  & CSP \\
2006hx  & 0.04435 &  17.656 (0.071/0.022)  & 17.885 (0.101/0.029)  & 54021.93 (0.09)  & 0.984 (0.014)  & 1.065 (0.032) & 17.543 (0.021)  & 17.793 (0.032)  & CSP \\
2006is  & 0.03149 &  16.596 (0.083/0.016)  & 17.022 (0.061/0.022)  & 54003.87 (0.15)  & 1.243 (0.004)  & 0.733 (0.020) & 15.934 (0.010)  & 16.111 (0.007)  & CSP \\
2006lu  & 0.05449 &  18.702 (0.256/0.025)  & 18.427 (0.340/0.032)  & 54036.75 (0.37)  & 0.990 (0.011)  & 1.054 (0.028) & 17.556 (0.019)  & 17.593 (0.012)  & CSP \\
2006ob  & 0.05820 &  18.460 (0.064/0.028)  & 18.685 (0.101/0.035)  & 54063.54 (0.11)  & 0.776 (0.009)  & 1.561 (0.035) & 18.210 (0.010)  & 18.139 (0.009)  & CSP \\
2007ai  & 0.03200 &  16.876 (0.029/0.016)  & 16.948 (0.036/0.022)  & 54172.43 (0.21)  & 1.157 (0.013)  & 0.820 (0.025) & 16.968 (0.009)  & 16.852 (0.006)  & CSP \\
2007nq  & 0.04385 &  17.803 (0.088/0.022)  & 17.747 (0.127/0.029)  & 54399.19 (0.11)  & 0.766 (0.006)  & 1.593 (0.028) & 17.420 (0.008)  & 17.329 (0.006)  & CSP \\
2008C   & 0.01708 &  15.949 (0.085/0.010)  & 16.150 (0.043/0.013)  & 54467.29 (0.42)  & 0.918 (0.018)  & 1.193 (0.043) & 15.681 (0.027)  & 15.476 (0.015)  & CSP \\
2008R   & 0.01288 &  15.290 (0.041/0.009)  & 15.368 (0.030/0.011)  & 54494.81 (0.05)  & 0.686 (0.003)  & 1.872 (0.024) & 15.292 (0.003)  & 15.122 (0.003)  & CSP \\
2008bq  & 0.03446 &  17.152 (0.062/0.017)  & 17.609 (0.158/0.023)  & 54563.00 (0.09)  & 1.064 (0.005)  & 0.937 (0.021) & 16.704 (0.005)  & 16.655 (0.003)  & CSP \\

\hline
\end{longtable}
\tablefoot{
\tablefoottext{a}{JD$-$2455000.}
\tablefoottext{b}{Estimated from our $s_\mathrm{opt}-\Delta M_{15}$ relation.}
\tablefoottext{c}{See text for the references for the NIR photometry.}
\tablefoottext{d}{The two numbers are the uncertainties of the peak magnitude and the $K$-corrections.}
}
\end{landscape}
}

\begin{table*}[!t]
\caption{Parameters of the SNe in the BN12 sample. The 1$\sigma$ uncertainties given in parentheses.}
\label{t:sne_lid}
\setlength{\tabcolsep}{3pt}
\begin{tabular}{@{}lcccccccc@{}}
\hline
\hline\noalign{\smallskip}
SN & $z_\mathrm{CMB}$ &  $J_\mathrm{max}$\tablefootmark{c,d} & $H_\mathrm{max}$\tablefootmark{c,d}  & $t_{B_\mathrm{max}}$\tablefootmark{a} & $s_\mathrm{opt}$ &  $\Delta M_{15}$\tablefootmark{b,c}  & $B_\mathrm{max}$\tablefootmark{c} & $V$ at $t_{B_\mathrm{max}}$\tablefootmark{c} \\
\hline\noalign{\smallskip}

PTF09dlc & 0.06656 &  18.675 (24/29)  & 18.948 (26/37)  & 073.79 (0.14)  & 0.991 (0.019)  & 1.052 (39) & 17.911 (52)  & 18.288 (31)  \\
PTF10hdv & 0.05424 &  18.326 (69/24)  & 18.520 (52/30)  & 342.41 (0.82)  & 1.125 (0.035)  & 0.857 (47) & 17.492 (21)  & 17.552 (22)  \\
PTF10hmv & 0.03272 &           $-$    & 17.507 (22/20)  & 351.69 (0.03)  & 1.130 (0.002)  & 0.851 (20) & 17.278 (46)  & 17.116 (24)  \\
PTF10mwb & 0.03117 &  17.137 (80/14)  & 17.401 (47/20)  & 390.32 (0.11)  & 0.885 (0.010)  & 1.267 (31) & 16.709 (32)  & 16.727 (11)  \\
PTF10ndc & 0.08168 &  19.283 (46/34)  & 19.430 (52/44)  & 389.99 (0.16)  & 1.091 (0.010)  & 0.900 (24) & 18.427 (08)  & 18.487 (06)  \\
PTF10nlg & 0.05592 &  18.674 (36/25)  & 18.656 (22/31)  & 391.43 (0.27)  & 0.968 (0.020)  & 1.094 (42) & 18.561 (26)  & 18.445 (19)  \\
PTF10qyx & 0.06515 &  19.028 (40/29)  & 19.200 (35/36)  & 426.36 (0.15)  & 0.913 (0.012)  & 1.204 (33) & 18.214 (16)  & 18.318 (11)  \\
PTF10tce & 0.03971 &  17.871 (25/19)  & 18.016 (20/24)  & 441.92 (0.27)  & 1.088 (0.012)  & 0.904 (26) & 17.177 (14)  & 17.096 (11)  \\
PTF10ufj & 0.07613 &  19.266 (61/32)  & 19.354 (40/41)  & 456.89 (0.13)  & 1.070 (0.021)  & 0.929 (36) & 18.404 (48)  & 18.529 (20)  \\
PTF10wnm & 0.06440 &  18.731 (30/29)  & 18.997 (29/36)  & 476.68 (0.08)  & 1.090 (0.008)  & 0.902 (23) & 18.152 (09)  & 18.129 (06)  \\
PTF10wof & 0.05131 &  18.391 (32/24)  & 18.581 (30/30)  & 473.81 (0.22)  & 1.038 (0.011)  & 0.976 (26) & 17.943 (35)  & 17.826 (08)  \\
PTF10xyt & 0.04831 &  18.406 (37/23)  & 18.416 (86/29)  & 490.85 (0.11)  & 1.117 (0.021)  & 0.867 (33) & 18.452 (24)  & 18.228 (24)  \\

\hline
\end{tabular} \\
\tablefoot{
\tablefoottext{a}{JD$-$2455000.}
\tablefoottext{b}{estimated from our $s_\mathrm{opt}-\Delta M_{15}$ relation.}
\tablefoottext{c}{The uncertainties are in units of 0.001 mag.}
\tablefoottext{d}{The two numbers are the uncertainties of the peak magnitude and the $K$-corrections.}}

\end{table*}

\begin{table}
\centering
\caption{Statistics of the NIR SN samples used in this work: number of SNe, mean redshift, standard deviation $\sigma_{z_\mathrm{CMB}}$ around the mean, and the median redshift.}
\label{t:sample_stat}
\begin{tabular}{@{}lcccc@{}}
\hline
\hline\noalign{\smallskip}
Sample & $N_\mathrm{LCs}$ & $<z_\mathrm{CMB}>$  & $\sigma_{z_\mathrm{CMB}}$ & median \\
\hline
 GOLD   &  45 & 0.0148 & 0.0095 & 0.0147 \\
 SILVER &  16 & 0.0232 & 0.0113 & 0.0239 \\
 BRONZE &  29 & 0.0284 & 0.0165 & 0.0266 \\
\hline
 GOLD ($z\geq0.008$)    &  30 & 0.0196 & 0.0081 & 0.0173 \\
 SILVER  ($z\geq0.008$) &  15 & 0.0244 & 0.0106 & 0.0240 \\
 BRONZE  ($z\geq0.008$) &  28 & 0.0293 & 0.0162 & 0.0268 \\
\hline
 BN12            & 12 & 0.0556 & 0.0161 & 0.0559 \\
 Our new sample  & 14 & 0.0942 & 0.0448 & 0.0850 \\
\hline 
\end{tabular}
\end{table}

\paragraph{Propagating the uncertainty:}

During the fitting the uncertainties of the optical LC stretch and time of $B$-band maximum were propagated to the uncertainty of the peak magnitudes with the following Monte Carlo procedure. The stretch and the time of maximum  were perturbed  according to the relations $s=s_\mathrm{opt}+N(0,\sigma^2_s+ ds^2_\mathrm{opt})$ and $t=t_{B_\mathrm{max}}+ N(0,\sigma^2_t+dt^2_{B_\mathrm{max}})$ and the LC was fitted with the perturbed pair ($s,t$). Here $ds_\mathrm{opt}$ and $dt_{B_\mathrm{max}}$ are the uncertainties of the stretch and time of maximum, and  $\sigma_t=0.44$ days and  $\sigma_s=0.05$ as determined previously. For each SN this was repeated 10000 times and the standard deviation of the resulting peak magnitudes was added in quadrature to the uncertainty derived by the LC fitter. The covariances between $s_\mathrm{opt}$, $t_{B_\mathrm{max}}$, and the peak magnitudes were also computed from the simulations.

\section{Results}

\label{sec:res}

\subsection{$J_\mathrm{max}-H_\mathrm{max}$, $V_{B_\mathrm{max}}-J_\mathrm{max}$ and $V_{B_\mathrm{max}}-H_\mathrm{max}$ colors}

\label{sec:maxcol}

\begin{figure}
\includegraphics*[width=8.8cm]{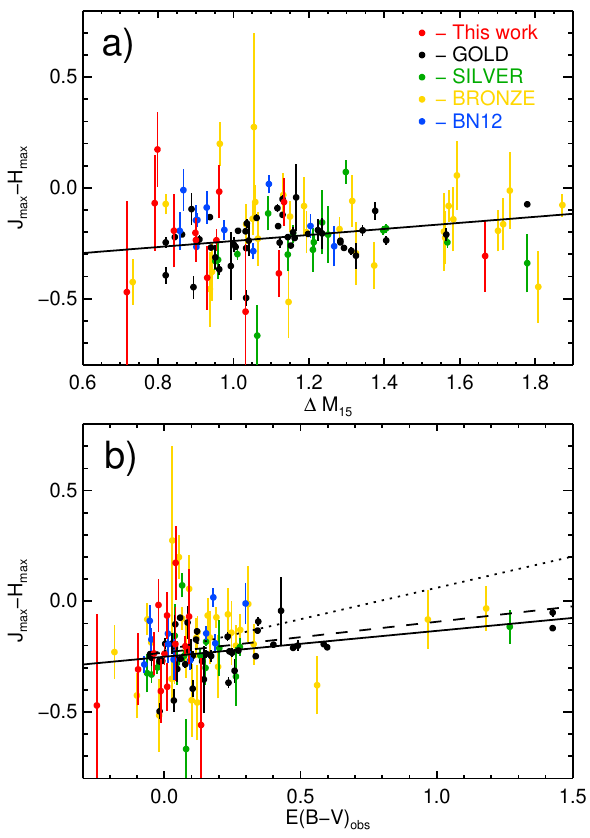}
\caption{(a) $J_\mathrm{max}-H_\mathrm{max}$ color $vs.$ $\Delta M_{15}$. The solid line shows the least-squares linear fit to the GOLD sample; (b)  $J_\mathrm{max}-H_\mathrm{max}$ $vs.$ $E(B-V)_\mathrm{obs}$. The solid line is the least-squares linear fit to the GOLD sample, the dashed and the dotted lines are the expected relation for Milky Way-type dust and $R_V=1.7$ and 3.1, respectively.}
\label{f:nir_col}
\end{figure} 

Figure~\ref{f:nir_col} shows the $J_\mathrm{max}-H_\mathrm{max}$ color $vs.$ $\Delta M_{15}$ and $E(B-V)_\mathrm{obs}$. The $J_\mathrm{max}-H_\mathrm{max}$ color shows a fair amount of scatter, largely due to our sample and the BRONZE sample. However, the scatter in the GOLD sample is fairly small. The weighted least-squares linear fit to $J_\mathrm{max}-H_\mathrm{max}$ \emph{vs.} $\Delta M_{15}$ in the GOLD sample only indicates a possible weak dependence on $\Delta M_{15}$ with a slope of $0.14\,\pm0.06$.  The fit also implies that a typical SN~Ia with $\Delta M_{15}=1.1$ has $J_\mathrm{max}-H_\mathrm{max}=-0.23\pm0.02$ mag. For $\chi^2/DoF$ of the fit to be one it is necessary to add 0.08 mag in quadrature to the errors, which can be interpreted as the intrinsic scatter of the $J_\mathrm{max}-H_\mathrm{max}$ color. Similar intrinsic scatter is also inferred from the BN12 sample, however, $J_\mathrm{max}-H_\mathrm{max}$ color in this sample is systematically redder by about 0.1 mag.

The GOLD sample also reveals a weak linear relation between $J_\mathrm{max}-H_\mathrm{max}$ and $E(B-V)_\mathrm{obs}$. The relation is shown with a solid line in Fig.~\ref{f:nir_col}b. The expected relations for Milky Way-type dust with $R_V=1.7$ and 3.1 are shown with the dashed and the dotted lines, respectively. The slope of the best-fit line, $0.12\pm0.03$, is consistent with a low $R_V$ value around 1.5. The fit indicates that an un-reddened SN should have $J_\mathrm{max}-H_\mathrm{max}=-0.25\pm0.03$. We also note that there is  no apparent dependence of $J_\mathrm{max}-H_\mathrm{max}$ of the redshift. For both the GOLD sample with mean redshift $z\simeq0.02$ and for our sample with  mean redshift $z\simeq0.09$  $J_\mathrm{max}-H_\mathrm{max}$ color is $-0.24$ mag.

Our estimate of the intrinsic $J_\mathrm{max}-H_\mathrm{max}$ color of a SN with $\Delta M_{15}=1.1$ mag is similar to the results of 
\cite{kri_04} and \cite{2008ApJ...689..377W}. It is also consistent with \cite{2014ApJ...789...32B} who, using a different approach,  found that a SN with stretch equal to one has $J_\mathrm{max}-H_\mathrm{max}=-0.23$ mag with intrinsic dispersion of $0.09$ mag. It is interesting to note, however, that \cite{2010AJ....139..120F} based on part of the data we used in our analysis find it to be close to zero. From Fig.~\ref{f:nir_col} one can see that this is inconsistent with our results. We also note that the observed variation of the $J_\mathrm{max}-H_\mathrm{max}$ color is far too big to be caused by dust extinction of any reasonable law.

\begin{figure}
\includegraphics*[width=8.8cm]{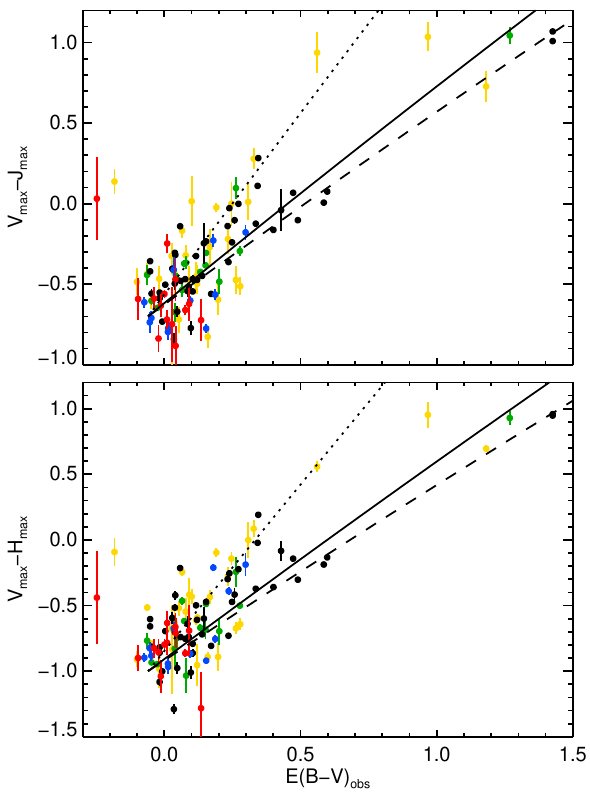}
\caption{$V_{B_\mathrm{max}}-J_\mathrm{max}$ (left) and $V_{B_\mathrm{max}}-H_\mathrm{max}$ (right) colors $vs.$ $E(B-V)_\mathrm{obs}$. The dashed, the solid and the dotted lines show the expected relations for Milky Way-type dust and $R_V=1.7$, 1.9 and 3.1, respectively. Color codes are as in Fig.~\ref{f:nir_col}. }
\label{f:nir_col2}
\end{figure} 

Finally, Fig.~\ref{f:nir_col2} shows $V_{B_\mathrm{max}}-J_\mathrm{max}$ and $V_{B_\mathrm{max}}-H_\mathrm{max}$ colors $vs.$ $E(B-V)_\mathrm{obs}$. Most of the SNe that have $E(B-V)_\mathrm{obs}\geq0.4$  mag align along the line tracing Milky Way-type extinction with $R_V=1.7-1.9$. There is also a set of SNe with $E(B-V)_\mathrm{obs}$ between 0.1 mag  and 0.6 mag that seem to follow the line with $R_V=3.1$. The two SNe with $E(B-V)_\mathrm{obs}<-0.15$ mag appear to be outliers. However, one of them SNF20080522-004 has large uncertainty of the maximum $B-V$ colors of $\sim0.2$ mag. The other object is SN~2006hx, which is also an $\sim1$ mag outlier on the optical Hubble diagram, hence its anomalous $V-$NIR colors.

\subsection{Hubble diagram}

The main goal of SNe~Ia Hubble diagram analysis is to derive the cosmological parameters, such as the mass density of the Universe. This requires a sample of SNe spanning from low- to high-redshifts \citep{1995ApJ...450...14G}. The nearby SNe Ia sample alone cannot constrain the cosmological parameters but instead can be used to derive the calibration  parameters  $a_X$ and $R_X$  in Eq.~\ref{eq:b_std}, the absolute SN~Ia magnitudes and to assess the precision of the SN~Ia standard candle. In our analysis the $a_X$ and $R_X$ parameters, and the absolute magnitude in the band $X$, $M_X$, of a fiducial SN~Ia with $\Delta M_{15}=1.1$ mag were determined by $\chi^2$-minimization of:
\begin{equation}
\chi^2=\sum_{i=1}^{N_\mathrm{set}}\sum_{j=1}^{N_\mathrm{SN}} \frac{\left(m^\ast_{X,i,j}-M^i_X-\mu_{i,j}\right)^2}{\sigma^2_{\mathrm{lc},i,j}+\sigma^2_{m_X,\mathrm{intr},i}+w\,\sigma^2_{\mathrm{ecorr},i,j}+\sigma^2_{\mathrm{ext},i,j}}.
\label{eq:chi2}
\end{equation}
with iterative $3\sigma$ outlier rejection and where the distance modulus $\mu$ was computed with the adopted cosmological parameters. In Eq.~\ref{eq:chi2} we allowed for the different data sets to have different absolute magnitudes $M^i_X$. Here  $\sigma_{\mathrm{ext}}$   account for the random host galaxy peculiar velocities with dispersion $\sigma_\mathrm{pec}=150$~km\,s$^{-1}$ and $\sigma_{m_X,\mathrm{intr}}$ accounts for possible intrinsic SN~Ia luminosity scatter, which cannot be corrected for by the model defined by Eq.~\ref{eq:b_std}. In principle, $\sigma_{m_X,\mathrm{intr}}$ should also include contribution from any other sources of uncertainty such as calibration uncertainties, possible mismatch between the templates and the actual fitted LCs, etc., which could be different in different data sets. For this reason, the values of the terms $\sigma_{m_X,\mathrm{intr}}$ were obtained so that the reduced $\chi^2$ is one for each data set. This approach helps to prevent samples with poorer-quality data from decreasing the accuracy of the derived parameters \citep{2010ApJ...716..712A}. For low-redshift samples the inferred value of  $\sigma_{m_X,\mathrm{intr}}$ is also dependent on the adopted value of the dispersion of the galaxies peculiar velocity $\sigma_\mathrm{ext}$, which is somehow uncertain. The term $\sigma_{\mathrm{lc}}$
\begin{equation}
\sigma^2_{\mathrm{lc}}=\sigma^2_{m_X}+a_X^2\sigma^2_{\Delta M_{15}}-2\,a_X\,\sigma_{[\Delta M_{15},m_X]}.
\label{eq:huberr}
\end{equation}
accounts for the uncertainties of the peak magnitude and $\Delta M_{15}$ and their covariance derived from the LC fit. Hereafter $\sigma_{[X,Y]}$ denotes co-variance of the variables $X$ and $Y$. The term $\sigma_{\mathrm{ecorr}}$
\begin{eqnarray}
\sigma^2_{\mathrm{ecorr}} &=& R_X^2(\sigma^2_{B_\mathrm{max}}+\sigma^2_{V_\mathrm{max}}-2\sigma_{[B_\mathrm{max},V_\mathrm{max}]})  \nonumber \\
    & & -2a_XR_X(\sigma_{[\Delta M_{15},V_\mathrm{max}]}-\sigma_{[\Delta M_{15},B_\mathrm{max}]})+R_X^2\sigma^2_{\mathrm{col,intr}}.
\label{eq:huberr1}
\end{eqnarray}
is the uncertainty of the extinction correction and it is asumed that $\sigma_{[m_X,B_\mathrm{max}]}=\sigma_{[m_X,V_\mathrm{max}]}=0$. Following \cite{2011ApJ...740...72M} we included a nonzero intrinsic color scatter $\sigma_{\mathrm{col,intr}}$, which does not correlate with luminosity. In general, the intrinsic color scatter could be different for each data set, but in our analysis we assumed that it was the same for all data sets $\sigma_\mathrm{col,intr}\simeq0.04$ mag. This value was estimated from analysis of the distribution of $(B-V)_{B_\mathrm{max}}$ as a function of $\Delta M_{15}$ (see Appendix~\ref{sec:colbv0} for the details).

The weight factor $w$ in Eq.~\ref{eq:chi2} accounts for the fact that the term  $\sigma_{\mathrm{ecorr}}$ arises only when an extinction correction is applied. Ideally, $w$ is zero for $E(B-V)_\mathrm{obs}\leq0$ and one for $E(B-V)_\mathrm{obs}>0$. However, the nonzero intrinsic color scatter makes the situation more complicated. For given stretch factor, among the SNe with $E(B-V)_\mathrm{obs}<0$ there should be a fraction of intrinsically blue objects that suffered only small reddening. In our analysis these objects will not be extinction corrected, but they should. In an attempt to treat these objects in a more correct way, $w$ was set to linearly decrease from 1 at $E(B-V)_\mathrm{obs}$ to 0 at $E(B-V)_\mathrm{obs}=-\sigma_\mathrm{col,intr}=-0.04$ mag,  meaning that an extra uncertainty was assigned to the SNe with small negative $E(B-V)_\mathrm{obs}$ even though no extinction corrections was applied.

\subsubsection{NIR Hubble diagram}

\begin{sidewaystable*}
\caption{Fit results for the whole NIR sample. In addition to $A$, $R$, $\gamma$, and the global weighted $1\sigma$ RMS scatter around the Hubble line, for each sample is also shown the mean absolute magnitude $M$,
the weighted $1\sigma$ scatter and the intrinsic scatter. The numbers in the parentheses are the uncertainty of the number directly above it.} 
\label{t:fitpars}
\centering
\setlength{\tabcolsep}{3pt}
\begin{tabular}{@{}rrr|c|rcc|rcc|rcc|rcc|rcc@{}} 
\hline\hline             
 & & & weighted & \multicolumn{3}{c|}{GOLD} & \multicolumn{3}{c|}{SILVER} & \multicolumn{3}{c|}{BRONZE} & \multicolumn{3}{c|}{BN12} & \multicolumn{3}{c}{{\bf Our sample}}  \\
\cline{5-7}\cline{8-10}\cline{11-13}\cline{14-16}\cline{17-19}
 $a_X$ & $R_X$ & $\gamma$ &  rms & $M$ & rms & $\sigma_\mathrm{intr}$ & $M$ & rms & $\sigma_\mathrm{intr}$ &  $M$ &rms & $\sigma_\mathrm{intr}$ & $M$ & rms & $\sigma_\mathrm{intr}$ &  $M$ & rms & $\sigma_\mathrm{intr}$ \\ 
\hline
\multicolumn{19}{c}{$J$ band} \\ \hline

   $-$    &    $-$    &    $-$    &   0.154 &  $-$18.645 &   0.116 &   0.090 & $-$18.629 &   0.171 &   0.154 & $-$18.513 &   0.238 &   0.213 & $-$18.432 &   0.149 &   0.138 & $-$18.604 &   0.246 &   0.218  \\
   $-$    &    $-$    &    $-$    & (0.011) &    (0.021) & (0.015) & (0.012) &   (0.044) & (0.032) & (0.029) &   (0.045) & (0.032) & (0.029) &   (0.045) & (0.033) & (0.031) &   (0.066) & (0.048) & (0.043)  \\
 \hline    
    0.223 &     0.421 &    $-$    &   0.146 &  $-$18.692 &   0.118 &   0.096 & $-$18.699 &   0.170 &   0.152 & $-$18.643 &   0.188 &   0.152 & $-$18.445 &   0.125 &   0.109 & $-$18.590 &   0.240 &   0.212  \\
  (0.071) &   (0.100) &    $-$    & (0.010) &    (0.025) & (0.015) & (0.013) &   (0.047) & (0.032) & (0.029) &   (0.044) & (0.026) & (0.021) &   (0.040) & (0.028) & (0.024) &   (0.065) & (0.047) & (0.042)  \\
  \hline
  0.187 &     0.314 &     0.569 &   0.134 &  $-$18.675 &   0.119 &   0.099 & $-$18.672 &   0.154 &   0.134 & $-$18.652 &   0.178 &   0.152 & $-$18.487 &   0.094 &   0.081 & $-$18.604 &   0.207 &   0.182  \\
  (0.068) &   (0.096) &   (0.120) & (0.010) &    (0.025) & (0.016) & (0.013) &   (0.043) & (0.029) & (0.025) &   (0.042) & (0.024) & (0.021) &   (0.032) & (0.021) & (0.018) &   (0.059) & (0.042) & (0.037)  \\

\hline
\multicolumn{19}{c}{$H$ band} \\ \hline 

   $-$    &    $-$    &    $-$    &   0.141 & $-$18.394 &   0.123 &   0.098 & $-$18.383 &   0.168 &   0.139 & $-$18.336 &   0.225 &   0.200 & $-$18.274 &   0.093 &   0.074 & $-$18.405 &   0.206 &   0.161 \\
   $-$    &    $-$    &    $-$    & (0.010) &   (0.022) & (0.016) & (0.013) &   (0.043) & (0.032) & (0.026) &   (0.042) & (0.031) & (0.027) &   (0.027) & (0.020) & (0.016) &   (0.057) & (0.042) & (0.033) \\
   \hline
   0.164 &     0.226 &    $-$    &   0.136 & $-$18.421 &   0.131 &   0.111 & $-$18.424 &   0.174 &   0.147 & $-$18.417 &   0.198 &   0.171 & $-$18.277 &   0.083 &   0.060 & $-$18.399 &   0.191 &   0.145 \\
  (0.070) &   (0.098) &    $-$    & (0.010) &   (0.027) & (0.017) & (0.015) &   (0.047) & (0.033) & (0.028) &   (0.046) & (0.027) & (0.023) &   (0.027) & (0.018) & (0.013) &   (0.054) & (0.039) & (0.030) \\
 \hline    
    0.187 &     0.314 &    -0.431 &   0.134 & $-$18.435 &   0.119 &   0.099 & $-$18.432 &   0.154 &   0.134 & $-$18.412 &   0.178 &   0.152 & $-$18.247 &   0.094 &   0.081 & $-$18.364 &   0.207 &   0.182 \\
  (0.068) &   (0.096) &   (0.120) & (0.010) &   (0.025) & (0.016) & (0.013) &   (0.043) & (0.029) & (0.025) &   (0.042) & (0.024) & (0.021) &   (0.032) & (0.021) & (0.018) &   (0.059) & (0.042) & (0.037) \\

\hline
\end{tabular}
\end{sidewaystable*}

\begin{figure*}[!th]
\includegraphics*[width=18cm]{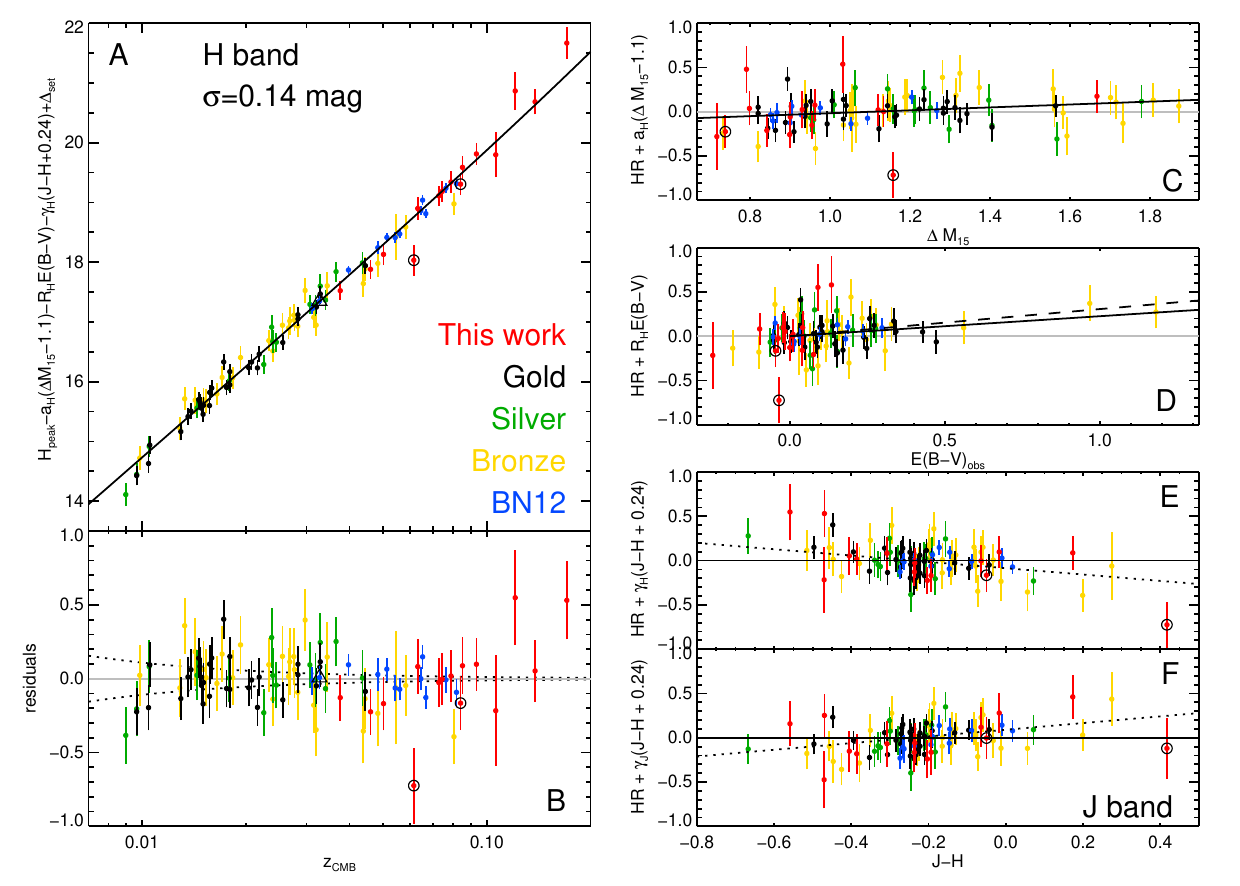}
\caption{$H$-band Hubble diagram for the full sample with the $a_H$ and $R_H$  correction terms included. Panels (c-d) show the Hubble residuals when only  $a_H$ or $R_H$ terms are applied $vs.$ the quantity multiplied by the other, for example, in panel (d) the Hubble residuals $vs.$ $E(B-V)_\mathrm{obs}$ when only the  $a_H$ correction term is applied. Panels (e) and (f) show $H$ and $J$ band residuals $vs.$ $J_\mathrm{max}-H_\mathrm{max}$ color . In panel (d) the dashed lines are the extinction corrections for Milky Way extinction law with $R_V=1.7$. The two red points surrounded by black circles are SN~2008bz and SNF20080512-008, which have been excluded from the fit (see text). The blue point surrounded a black triangle  is PTF10hmv, which was observed only in $H$. }
\label{f:hub_nir}
\end{figure*}

The model defined by Eqs.~\ref{eq:b_std}-\ref{eq:huberr1} was used to analyze the NIR $J$ and $H$ peak magnitudes of the SNe with $z\geq0.008$.
The data was considered consisting of five SN samples: the new SN observations presented in this paper, the \citetalias{2012MNRAS.425.1007B} sample, and the published SN sample, which is split into three subsamples - GOLD, SILVER and BRONZE. Two SNe from our sample, SN~2008bz and SNF20080512-008, were observed in the NIR at stretch-corrected phase $\sim+13$ days. This is beyond the range where our fitting method could be used and these two SNe were excluded from the analysis. Further, two SNe from the CSP sample, SNe~2005ku and 2007hx, were found to be significantly brighter than the best-fit Hubble line in $H$ band. CSP (C.Contreras, private communication) confirmed that most of the $H$ band photometry of these two SNe was unreliable. Only one $H$ observation of SN~2005ku was deemed good and the updated photometry was even brighter than the published one.  For this reason  SNe~2005ku and 2007hx were also excluded. The results of the analysis are shown in Table~\ref{t:fitpars}. 

The data was first analyzed assuming that SNe~Ia are natural standard candles in the NIR and no corrections to the peak magnitudes are needed. The weighted scatter around the Hubble line was 0.154 mag in $J$ and 0.141 mag in $H$, which is comparable to the best values reported for $B$ band. Inspection of the Hubble residuals revealed a weak dependence on the color excess $E(B-V)_\mathrm{obs}$ and possibly on $\Delta M_{15}$. When the $a_{J/H}$ and $R_{J/H}$ correction terms were included in the analysis the scatter of the Hubble residuals only marginally decreased to 0.146 mag in $J$ and 0.136 mag in $H$. The corresponding $H$-band Hubble diagram is shown in Fig.~\ref{f:hub_nir} with panels (c) and (d) showing the residuals versus $\Delta M_{15}$  and $E(B-V)_\mathrm{obs}$. The weak dependence of the peak magnitudes on $\Delta M_{15}$  and $E(B-V)_\mathrm{obs}$, combined with the lack of many highly reddened SNe, is the main reason for the rather small decrease of the scatter. Furthermore, the low values of $R_{J/H}=0.42/0.22$ are consistent with Milky Way type extinction  with $R_V\simeq1.5-2$ as commonly found in the optical studies of SNe~Ia. 
The $a_X$ coefficient is $\approx0.20$ in both bands and is smaller than  the estimates of \cite{2010AJ....139..120F} and \cite{2012PASP..124..114K}. The \cite{2012PASP..124..114K} analysis shows that the estimate of $a_X$ is highly dependent on the sample used. Thus, the fact that we used  a different sample, together with the different approaches to estimate $\Delta M_{15}$  and $E(B-V)_\mathrm{obs}$, is the most likely the reason for the  difference with the previous studies. We also note that the $a_X$ coefficient is much smaller than in the $B$ band \citep[see next section and, e.g.,][]{2010AJ....139..120F},  which further supports the conclusion that SNe~Ia are better natural standard candles in the NIR than in the optical.

The overall rms around the Hubble line is similar in $J$ and $H$ band, $0.146\,\pm0.010$ mag and $0.136\,\pm0.010$ mag, respectively. This is somehow surprising because several previous studies have claimed that $H$-band Hubble diagram has considerably smaller scatter than $J$ \citep[e.g.,][]{2008ApJ...689..377W}, but we find that SNe~Ia are nearly equally good standard candles in both bands. However, we note that there is no big difference between the  $J$ and $H$ band rms' in the literature samples, but in our and BN12 samples the $H$ band rms is noticeably smaller than in $J$.

The value of $\sigma_{\mathrm{intr}}$ is a very important quantity because it tells us how precisely SNe~Ia peak magnitudes could be standardized. The uncertainty of the value of $\sigma_\mathrm{ext}$ will affect the estimation of  $\sigma_{\mathrm{intr}}$ from the GOLD, SILVER, and BRONZE samples more compared to BN12 and our samples because the latter have higher mean redshift. The BN12 sample has $\sigma_{\mathrm{intr}}\simeq0.11$ mag in $J$ and even smaller $\sigma_{\mathrm{intr}}\simeq0.06$ mag in $H$. Even though we assumed rather small $\sigma_\mathrm{ext}=150$~km\,s$^{-1}$ (300~km\,s$^{-1}$ and even 450~km\,s$^{-1}$ are sometimes used), the  GOLD sample also has $\sigma_{\mathrm{intr}}\simeq0.1$ in both NIR bands. This suggest that the intrinsic luminosity scatter in $J$ and $H$ is most likely around  $\sigma_{\mathrm{intr}}\simeq0.10$ mag.
  
The GOLD, SILVER, and BRONZE have similar redshift distribution and should be equally affected by the uncertainty of $\sigma_\mathrm{ext}$. Given this, if all sources of uncertainties of the peak magnitudes have been properly accounted for, these three samples should have similar $\sigma_{\mathrm{intr}}$. Our analysis shows that $\sigma_{\mathrm{intr}}$ in the SILVER and BRONZE samples is larger than in the GOLD sample by a factor of about 1.5. Our sample has larger $\sigma_{\mathrm{intr}}$ in $J$ than any of the other samples, but in $H$ $\sigma_{\mathrm{intr}}$ is similar to the SILVER sample. This suggest that in spite of our efforts to identify and include all uncertainties that affect the peak magnitudes estimated from incomplete LCs, some
additional scatter is still present. One possible source of such scatter is systematic deviation from the $s_\mathrm{NIR}=s_\mathrm{opt}$ relation. Careful inspection of Fig.~\ref{f:g_fit}a reveals that the points in the range  $s_\mathrm{opt}=0.92-1.02$ are systematically above the line, while those in the ranges  $s_\mathrm{opt}=0.8-0.9$ and $s_\mathrm{opt}=1.02-1.15$ are systematically below it. This raises the interesting possibility that $s_\mathrm{NIR}$ has a bi-modal distribution with mean values of $s_\mathrm{NIR}\simeq0.8$ and $s_\mathrm{NIR}\simeq1.05$ for the SNe with $s_\mathrm{opt}$ below and above $\sim0.91$, respectively, but clearly more observations are needed to confirm  whether these systematic trends are real or not. If real, these systematic deviations may affect our and the BRONZE samples differently because the two samples have different stretch distributions. In our sample 4 SNe are in the range $s_\mathrm{opt}=0.92-1.04$, 11 have $s_\mathrm{opt}>1.04$ and only one has $s_\mathrm{opt}<0.92$. In the BRONZE sample on the other hand 8 are in the range $s_\mathrm{opt}=0.92-1.04$, only 5 have $s_\mathrm{opt}>1.04$ and the majority of 15 SNe have $s_\mathrm{opt}<0.92$. The SILVER sample has stretch distribution similar to that of the BRONZE and BN12 similar to ours. However, because the SILVER and BN12  SNe have observations close the NIR maximum, these samples should be less sensitive to the uncertainties of the stretch.

\begin{figure*}[!t]
\includegraphics*[width=18cm]{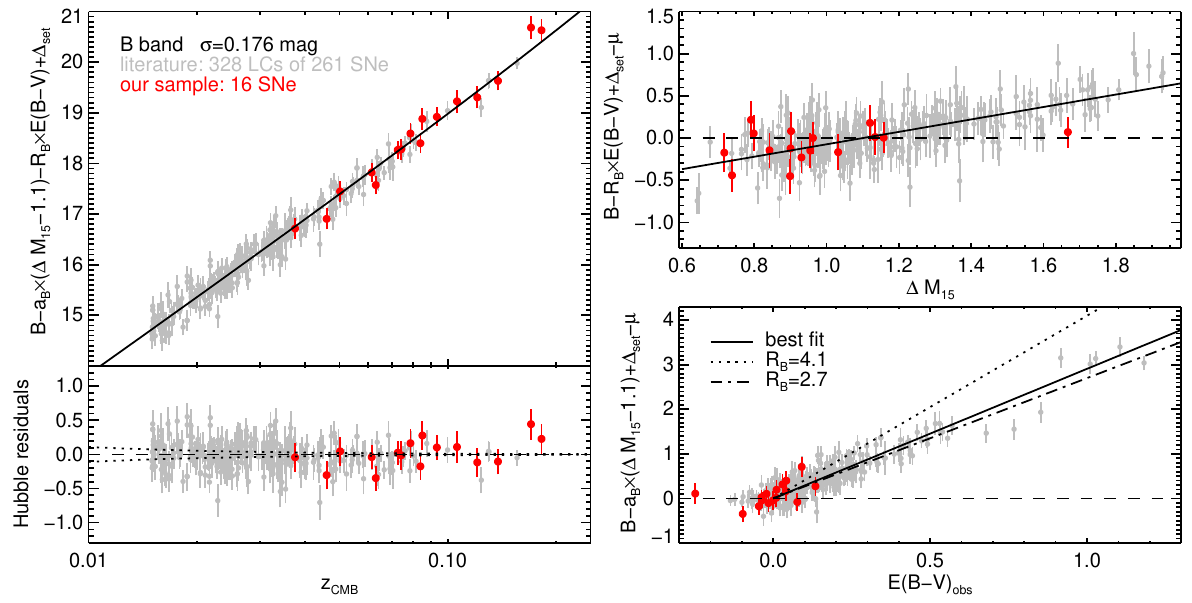}
\caption{$B$-band Hubble diagram. The right panels show the Hubble residual when only the LC shape or the extinction correction is applied.}
\label{f:hub_opt}
\end{figure*} 

Despite the differences of the inferred values of $\sigma_{\mathrm{intr}}$ and the observed scatter, the SN~Ia absolute magnitudes estimated from the GOLD, SILVER, BRONZE and our sample are in a good agreement and differ by less than $1\sigma$. This indicates that using under-sampled LCs does not lead to significant bias in estimating the NIR peak magnitudes. However, the  \cite{2012MNRAS.425.1007B} sample is on average 0.15-0.20 mag fainter than the rest. This difference is not negligible and appears to be too large to be explained with differences between the photometric systems used. It is also likely not related to our analysis procedures because \citet{2014ApJ...784..105W} also noted that the absolute magnitudes inferred from BN12 sample differ from the other samples.

Upon closer examination, the Hubble residuals hinted at possible dependence on the  $J_\mathrm{max}-H_\mathrm{max}$ color ($V-[J/H]$ were also tested, but the highest degree of correlation was found with the $J_\mathrm{max}-H_\mathrm{max}$ color). Linear least-squares fit produced nonzero slopes of $0.38\pm0.11$ and $-0.36\pm0.11$ in $J$ and $H$, respectively. These fits are shown in Fig.~\ref{f:hub_nir}(e) and (f). The uncertainties were estimated from fitting 10000 bootstrap replications of the  original sample, which with or without preserving the number of SNe in the samples, gave uncertainties of $\sim0.1$. Thus the $\gamma$ term: 
\begin{equation}
 -\gamma_X\times\left[(J_\mathrm{max}-H_\mathrm{max})-(J-H)_0\right],
 \label{eq:gamma}
\end{equation}
and the relevant error terms were added to Eqs.~\ref{eq:b_std} and \ref{eq:huberr}, respectively. Here $(J-H)_0=-0.24$ mag is set to the mean SN~Ia $J_\mathrm{max}-H_\mathrm{max}$ color. The results of the fits with the updates equation also indicate the presence of a significant nonzero $\gamma$ term. 
We note that including the $\gamma$ terms makes the $J$- and $H$-band residuals highly correlated and forces $\gamma_J=\gamma_H+1$. As a results all derived parameters, except for the absolute magnitudes, become the same for $J$ and $H$ (Table~\ref{t:fitpars}).

From Table~\ref{t:fitpars}, one can see that including  the $\gamma$ term does not lead to decrease of the inferred $\sigma_{\mathrm{intr}}$ in all samples and in few an increase is observed. However, in all cases the change is within the estimated uncertainties. It can also be seen that in the best observed GOLD and BN12 samples $\sigma_{\mathrm{intr}}$ does not show a clear decrease and the nonzero $\gamma$ term is likely not driven by these samples. All this casts doubt of whether the correlation of the Hubble residuals with the  $J_\mathrm{max}-H_\mathrm{max}$ color is real or an artifact of the small sample sizes. However, we note that the range of $J_\mathrm{max}-H_\mathrm{max}$ color in the GOLD and BN12 samples is smaller than in the others. Thus, the question of whether the Hubble residuals depend on the $J_\mathrm{max}-H_\mathrm{max}$ color should be answered in the future when a larger sample of well-observed SNe~Ia in the NIR becomes available. For this reason we base our further discussion of the results without the $\gamma$ term.

\subsubsection{$B$ and $V$ bands}

\begin{table*}            
\caption{Results from the analysis of the optical observations assuming intrisic color scatter $\sigma_{\mathrm{col,intr}}=0.04$ mag. For each sample are listed the mean absolute magnitude, the systematic luminosity scatter, the weighted observed scatter around the Hubble line, and the number of used and total SNe.} 
\label{t:hub_opt}
\begin{tabular}{@{}lccccccccc@{}}
\hline
\hline\noalign{\smallskip}
Data set   &   $M_B$        & $\sigma_{B,\mathrm{intr}}$ &  $\sigma_{\mathrm{obs}}$\tablefootmark{b} & $N_\mathrm{used}$($N_\mathrm{tot}$) & &   $M_V$        & $\sigma_{V,\mathrm{intr}}$ &  $\sigma_{\mathrm{obs}}$\tablefootmark{b} & $N_\mathrm{used}$($N_\mathrm{tot}$) \\
\hline\noalign{\smallskip}
Berkeley        & $-$19.264 (0.022) & 0.142 & 0.188 & 87 (88)  & &  $-$19.183  (0.022) &  0.165 &  0.188 &  87   (88)   \\  
CSP             & $-$19.387 (0.025) & 0.095 & 0.154 & 50 (51)  & &  $-$19.298  (0.023) &  0.114 &  0.141 &  49   (51)   \\  
CfA1            & $-$19.376 (0.063) & 0.137 & 0.189 &  9  (9)  & &  $-$19.290  (0.065) &  0.167 &  0.194 &   9    (9)   \\  
CfA2            & $-$19.255 (0.052) & 0.143 & 0.193 & 14 (14)  & &  $-$19.181  (0.052) &  0.170 &  0.194 &  14   (14)   \\  
CfA3            & $-$19.342 (0.025) & 0.153 & 0.205 & 85 (89)  & &  $-$19.254  (0.025) &  0.192 &  0.216 &  87   (89)   \\  
CfA4            & $-$19.304 (0.041) & 0.210 & 0.249 & 39 (39)  & &  $-$19.227  (0.040) &  0.226 &  0.247 &  39   (39)   \\  
Kowalski        & $-$19.274 (0.061) & 0.115 & 0.149 &  6  (6)  & &  $-$19.193  (0.070) &  0.152 &  0.170 &   6    (6)   \\  
Krisciunas      & $-$19.384 (0.030) & 0.0\tablefootmark{a}   & 0.081 &  8  (8)  & &  $-$19.297  (0.026) &  0.017 &  0.076 &   8    (8)   \\  
Cal\`an-Tololo  & $-$19.313 (0.041) & 0.117 & 0.172 & 18 (18)  & &  $-$19.245  (0.039) &  0.137 &  0.165 &  18   (18)   \\  
BN12            & $-$19.189 (0.047) & 0.114 & 0.160 & 12 (12)  & &  $-$19.083  (0.044) &  0.126 &  0.150 &  12   (12)   \\  
{\bf This work} & $-$19.271 (0.051) & 0.170 & 0.202 & 16 (16)  & &  $-$19.183  (0.050) &  0.186 &  0.200 &  16   (16)   \\  
\hline\noalign{\smallskip}
Mean $M_X$      & $-$19.318 (0.062) &       & 0.176\tablefootmark{c} &          & & $-$19.236 (0.064) &       & 0.172\tablefootmark{c} &            \\ 
$a_X$           &     2.901 (0.062) &       &       &          & &      1.812 (0.060) &       &       &            \\ 
$R_X$           &     0.737 (0.038) &       &       &          & &      0.587 (0.037) &       &       &            \\ 
\hline
\end{tabular}\\
\tablefoot{
\tablefoottext{a}{$\chi^2/DoF<1$ even with $\sigma_{B,\mathrm{intr}}=0.0$ mag.}
\tablefoottext{b}{weighted standard deviation.}
\tablefoottext{c}{weighted standard deviation of all points around the Hubble line. Its uncertainty is 0.007 mag.}
}
\end{table*}

The model defined by Eqs.~\ref{eq:b_std}-\ref{eq:huberr1} was also used to analyze the $B$- and $V$-band peak magnitudes of the whole nearby SN~Ia sample\footnote{We note that no attempt has been made to select only SNe with "good" optical LCs as in other studies. All SNe that allow stretch fitting were used.} with redshift above 0.015 plus the 16 SNe from this work -- 348 peak magnitudes in total. The term $2R_X(\sigma_{[B_\mathrm{max},V_\mathrm{max}]}-\sigma^2_{B_\mathrm{max}})$ was added to Eq.~\ref{eq:huberr1} because in this case $m_X=B_\mathrm{max}$. The results of the analysis are shown in Table~\ref{t:hub_opt} and the $B$-band Hubble diagram in Fig.~\ref{f:hub_opt}. First we note that the parameters of our sample fall well in the ranges of the other samples. The overall weighted $1\sigma$ scatter around the best-fit Hubble line is $0.176\,\pm0.007$ mag and $0.172\,\pm0.007$ mag and $B$ and $V$ bands, respectively. In the individual samples the observed scatter varies between 0.15 and 0.25 mag. The intrinsic luminosity  scatter $\sigma_{B,\mathrm{intr}}$ is between 0.10 and 0.15 mag, and larger by about 0.02 mag in $V$. \cite{2014A&A...568A..22B}, \cite{2014ApJ...795...45S} and \cite{2017arXiv171000845S} estimate similar $B$-band intrinsic luminosity scatter for the nearby SN samples, even though using a different LC fitting approaches and applying different cuts to the samples used. From Table~\ref{t:hub_opt} one can also see that the absolute magnitudes for the different data sets span a range of $\sim0.16$ mag. Tensions between the  nearby SN~Ia samples have been reported previously \citep[e.g.,][]{2009ApJ...700.1097H,2012ApJ...746...85S,2013MNRAS.433.2240G,2014ApJ...795...45S} and possibly indicate systematic differences between the data sets, and the peak magnitudes of the SNe in common between the four largest data sets indeed show offsets up to 0.05 mag (Table~\ref{t:diff}). The nearby SN sample does suffer from selection effects, which can be seen from plots of  $(B-V)_{B_\mathrm{max}}$ and $\Delta M_{15}$ $vs.$ redshift $z$. Above $z\simeq0.035-0.040$ there are no SNe with $(B-V)_{B_\mathrm{max}}>0.5$ mag, while there are many at the lower redshifts. Similarly, all SNe with $z>0.06$ have $\Delta M_{15}<1.4$ mag, but at the lower redshifts many SNe with $\Delta M_{15}>1.4$ mag are present.  However, the distributions of the redshift, $\Delta M_{15}$ and  $(B-V)_{B_\mathrm{max}}$ in the largest samples are very similar and the selection biases are unlikely to be the main cause for the tensions between the sets. Another potential bias is related the differences of the stretch distributions and possibly the luminosity distributions of \emph{normal}  SNe~Ia in passive and active host galaxies \citep[see, e.g.,][]{2010MNRAS.406..782S}. The nearby samples contain the same mixture of SNe in passive and active hosts and this bias is probably not important in our case.

Similar to other recent studies we also find low values of $R_V=1.81\pm0.060$ and $R_B=2.90\pm0.062$ even tough intrinsic color scatter of 0.04 mag was included in our analysis. \cite{2014ApJ...780...37S} have claimed that  erroneous treatment of the color scatter as luminosity scatter will bias the estimates of $a_B$ and $R_B$ coefficients. $R_B$ in particular can be biased by as much as 30\% toward lower values and the authors claim that when the color scatter is treated properly the usual $R_B\simeq4.1$ should be recovered. However, from the lower right panel of Fig.~\ref{f:hub_opt} one can see that $R_B=4.1$, (shown with the dotted line) is inconsistent with the data. Moreover, from analysis of simulated data sets with luminosity
and color scatter, and with the same number of SNe and distributions of $(B-V)_{B_\mathrm{max}}$, $\Delta M_{15}$ and redshift as the full data set, we could only find small bias in $R_B$ at the level of $\leq10$\% and even less in $a_B$. The differences between our results and those of \cite{2014ApJ...780...37S} are likely due to the different color distributions of the SN samples  analyzed. The majority of the SNe in the sample simulated by \cite{2014ApJ...780...37S} have $(B-V)_{B_\mathrm{max}}<+0.2$ mag and none with $(B-V)_{B_\mathrm{max}}>+0.3$ mag, while the full nearby sample has many SNe with $(B-V)_{B_\mathrm{max}}=0.2-1.5$ mag. Repeating the analysis only with the SNe with $E(B-V)_\mathrm{obs}<+0.3$ we find $R_B=3.57\pm0.13$ and $R_B=2.76\pm0.13$ with and without intrinsic color scatter, respectively. Simulated data sets were also analyzed and the results showed that the bias is larger in samples with lower color range.  It should also be emphasized that the assumption of universal $R_B$ value for all SNe  most likely is not valid for any real SN sample.  Indeed,  large  variations of $R_V$ from event to event have been found by \cite{2014ApJ...789...32B} and  \citet{2015MNRAS.453.3300A}, with the latter work also including UV measurements, where the differences in extinction laws  are the largest.

The slopes of the $B_\mathrm{max}-\Delta M_{15}$ and $V_\mathrm{max}-\Delta M_{15}$ relations are $a_B=0.737\pm0.037$ and $a_V=0.587\pm0.037$. 
These values are again lower than those found by \cite{2010AJ....139..120F} and similarly to the NIR, we attribute this to the different sample, and the different approaches we used to estimate $\Delta M_{15}$  and $E(B-V)_\mathrm{obs}$.

 The analysis of the color indexes at maximum light suggested that
 several SNe may be reddened by dust with $R_B\simeq4.1$ and not 2.9 (see Fig.~\ref{f:nir_col2}).  The mean color excess of these SNe is
 $\sim+0.3$ mag. If these SNe had to be corrected with $R_B=4.1$ but
 we used $R_B=2.9$ instead, then their Hubble residuals should be on average $\sim+0.3$ mag. However, this is not the case and the
 Hubble residuals are on average close to zero.  Thus, most likely the
 behavior of the maximum colors of these SNe is an intrinsic property
 and not caused by dust extinction with $R_B=4.1$. We also find that it is 
  likely to be unrelated to the expansion velocity dichotomy noted by \cite{2009ApJ...699L.139W}.

\subsection{$V-J$ and $V-H$ color indexes evolution}

\begin{figure}[!t]
\includegraphics*[width=8.8cm]{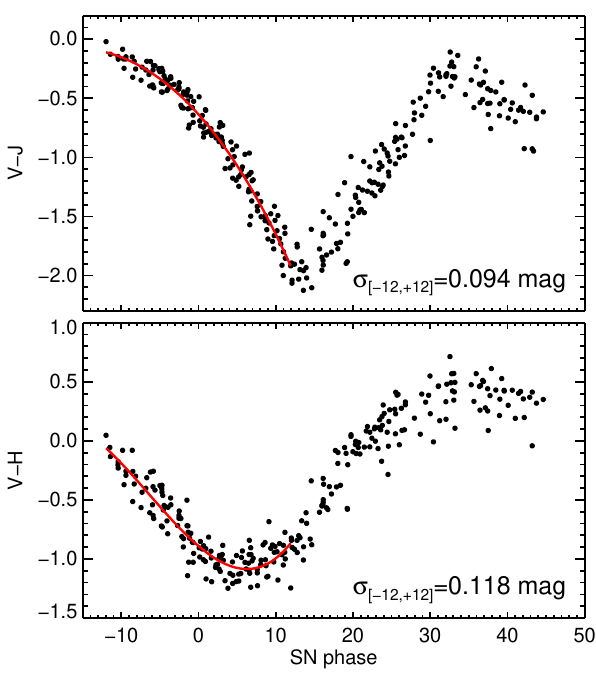}
\caption{$V-J$ and $V-H$ un-reddened loci for a SN with $\Delta M_{15}=1.1$. The third-order polynomial fit of the data between phases $-12$ and +12 days is also shown. For clarity the error bars are not plotted.}
\label{f:vnir}
\end{figure}

\cite{kri_ir_temp} have studied the $V-$NIR colors of SNe~Ia and found that their evolution between phases $-10$ and $+10$ days is fairly uniform. They also showed that SNe with $\Delta M_{15}<1.1$ mag and  $\Delta M_{15}>1.1$ follow different $V-H$ loci. We studied the evolution of the $V-J$ and $V-H$ colors with the SNe of the GOLD sample with low reddening $E(B-V)<0.2$ mag. The observed $V$, $J$ and $H$ magnitudes, corrected for the Milky Way dust extinction,  were converted to rest-frame and the observed epochs were converted to stretch-corrected phases using Eq.~\ref{eq:ph} with $s_\mathrm{opt}$. The $V$ magnitudes were interpolated at the phases of the NIR observations. Most SNe have well-sampled $V$ band LCs and the interpolation was done with smoothing splines. Only in a few cases interpolation from the templates fits was used. 

We assumed that the evolution of the intrinsic colors indexes with the stretch-corrected phase is the same for all SNe. Under this assumption  the offsets between color curves of the different SNe arise from different dust extinction in the host galaxies and possibly from intrinsic color scatter. The offset should also be a function of the decline rate parameter $\Delta M_{15}$ because the slope of the "luminosity -- LC shape" relation in $V$ and the NIR bands is different. Thus, the observed $V-X$  color curves of each SN were corrected by
\begin{equation}
 (V-X)^\ast=(V-X)-R_{(V-X)}\,E(B-V)-a_{(V-X)}(\Delta M_{15}-1.1),
 \label{eq:vnirmodel}
\end{equation}
where $X$ is either $J$ or $H$, and $R_{(V-X)}=R_V-R_X$ and $a_{(V-X)}=a_V-a_X$ were computed from the $R$ and $a$ coefficients determined from the Hubble diagram fits.  The corrected $(V-X)^\ast$ curves between phases $-12$ and $+12$ days were stacked together and weighted least-squares fit with third-order polynomial function was performed.  Intrinsic scatter term $\sigma_\mathrm{intr}$ was also added in quadrature to all uncertainties so that the reduced $\chi^2$ of the fit was one.

After the first iteration it was noted that four SNe had color curves significantly deviating from the rest. SNe 2002dj, 2006le, 2006lf and 2008gp were all too blue at all epochs and were excluded from the analysis. The final sample consisted of 22 LCs of 20 SNe with $\Delta M_{15}$ and $E(B-V)_\mathrm{obs}$ in the ranges $0.82\div1.56$ mag and $-0.05\div+0.16$ mag, respectively. The coefficients of the best-fit polynomial function, the inferred intrinsic scatter and the weighted standard deviation around the fit are shown in Table~\ref{t:vnir}. The corrected color curves up to phases +45 days and the best fits are shown in Fig.~\ref{f:vnir}.

The derived intrinsic scatter of $\sigma_\mathrm{intr}\simeq0.09-0.11$ mag is comparable to that reported by \cite{2014ApJ...789...32B}, even tough we  used one mean value of $R_{(V-X)}$, while \cite{2014ApJ...789...32B} derived individual values for each SN. We note also that part of the scatter of the $V-$NIR colors is not statistical. A closer examination of the data shows that most SNe are either bluer or redder than the mean loci at all epochs. This behavior may be related to the use of average $R_{(V-X)}$, but it may also result from other sources such as systematic differences in photometry and intrinsic color scatter. The fact that four SNe from the low-reddened sample had systematically too blue colors demonstrates this point. We note, however, that the offsets of SNe 2006le and 2006lf are too big to be explained by a different extinction law.

\begin{table}[!t]
\centering
\caption{Parameters of the third-order polynomials describing the $V-J$ and $V-H$ colors evolution, the intrinsic scatter and the RMS around the best fit.} 
\label{t:vnir}
\begin{tabular}{l|c|c}
\hline
\hline
 Parameter &    $V-J$   & $V-H$  \\
\hline
$c_0$      & $-$0.63581        & $-$0.89079   \\
$c_1$      & $-$0.07315        & $-$0.05582    \\
$c_2$      & $-$2.65367E$-$03  & 2.98528E$-$03 \\
$c_3$      & $-$1.78536E$-$05  & 1.55660E$-$04 \\
\hline
$\sigma_\mathrm{intr}$ & 0.089 & 0.108 \\ 
RMS [$-$12,+12]       & 0.094            & 0.118         \\
\hline
\end{tabular}
\end{table}

\section{Discussion}

\label{sec:disc_concl}

\subsection{SNe~Ia in the optical}

We analyzed the $B$ and $V$ LCs of our new SNa~Ia sample together with the largest nearby SN~Ia compilation to date -- 458 LCs from 10 different samples.  The Hubble diagram of all nearby SNe with redshift above $z=0.015$ has weighted scatter around the best-fit line of $\sim0.176$ mag in $B$ and $\sim0.172$ mag in $V$ band. The $B$ band scatter in the individual samples varies (Table~\ref{t:hub_opt}) and for example it is between 0.15 and 0.25 mag for the four largest samples CSP, Berkeley, CfA3 and CfA4, and our sample has 0.2 mag. Similarly,  the corresponding $B$ band intrinsic scatter $\sigma_{\mathrm{intr}}$  varies in the range 0.1--0.2 mag with our sample having 0.17 mag. These results show that our data fit very well with findings in the literature. The four largest samples have similar redshift distributions and photometric quality, and the differences in $\sigma_{\mathrm{intr}}$ indicate the likely presence of additional sample-dependent source of uncertainty beyond the intrinsic SN~Ia luminosity scatter because the latter should be the same for all samples. Several of the samples that we analized have intrinsic scatter $\sigma_{\mathrm{intr}}=0.10-0.12$ mag. From a similarly heterogeneous low-redshift samples \cite{2014A&A...568A..22B}, \cite{2014ApJ...795...45S} and \cite{2017arXiv171000845S} find an intrinsic luminosity scatter  $\sim0.10-0.12$ mag, a value which agrees with our estimate.

Analyzing the distribution of the maximum $B-V$ color of 393 SNe~Ia we conclude that the intrinsic $B-V$ color of normal SNe~Ia at maximum light is a nonlinear function of $\Delta M_{15}$ and has intrinsic scatter $\sim0.04$ mag. Including this in the analysis, we found that the host galaxy dust extinction is best described with Milky way-type dust with low $R_V\simeq1.8-1.9$ in accordance with many other studies. Because of the large span of the colors this estimate is robust and it is easy to see from Fig.~\ref{f:hub_opt} that the typical Milky Way value $R_V=4.1$ is inconsistent with the data.

Further, restricting the $B$ band Hubble diagam analysis only to the 78 SNe in common with \cite{2014A&A...568A..22B}, which were treated as a single sample,  weighted scatter of the Hubble residuals of 0.162 mag and $\sigma_{\mathrm{intr}}=0.11$ mag were obtained. \cite{2014A&A...568A..22B} selected the best-observed low-reddening normal SNe and our results confirm that such selection criteria provide improvement over the use of all SNe. However, repeating the analysis using only the SNe with  $E(B-V)_\mathrm{obs}<0.3$ mag we find overall scatter of 0.173 mag, thus suggesting that LC quality is a more important contributor to the scatter than the reddening. In general, the similarity of our results to those of  \cite{2014A&A...568A..22B}, despite the different analysis approaches, is encouraging given that we only used $B$ and $V$ band LCs.

\subsection{SNe~Ia in the NIR}

The analysis of our new sample of 16 SNe together with the largest compilation of 102 SN~Ia NIR LCs to date confirms previous conclusions that SNe~Ia are as good, if not better, standard candles in the NIR as in the optical. The overall scatter of the Hubble diagram is $\sim0.14-0.15$ mag in both $J$ and $H$, which is similar to the best published value of 0.14 mag in $B$ band \citep{2014A&A...568A..22B}. The analysis yields $R_J\simeq0.4$ and $R_H\simeq0.2$, which are consistent with the low value of $R_V\simeq1.8-1.9$ derived from the optical observations. This implies that the extinction corrections in the NIR are by a factor of $\sim7-15$ smaller than in the optical $B$ band. The same holds true for the  LC shape correction too. We find $a_{J/H}\simeq0.20$, which is by a factor of 3.5 smaller than in $B$ band. This demonstrates one of the advantages of the NIR observations, namely, that the effect of possible systematic uncertainties associated with the LC shape and the extinction corrections are greatly reduced compared to the optical. 

The best-observed nearby GOLD and the BN12 samples indicate rather small intrinsic luminosity scatter $\sigma_\mathrm{intr}\simeq0.10$ mag, which  is comparable to the latest estimates in the optical $B$ band \citep{2014A&A...568A..22B,2014ApJ...795...45S,2017arXiv171000845S}.  However, one should be cautious about the derived low intrinsic scatter.  Estimating this quantity is difficult because it requires identifying all other  sources of uncertainty and correctly accounting for them. In addition, low-redshift samples are sensitive to the assumed value of the galaxy velocity dispersion, the exact value of which is rather uncertain. One possible source of uncertainty is the inclusion of the $K$-correction uncertainties in the error budget. We have assumed that the scatter of the individual $K$-corrections reflects real difference on the energy distribution in the NIR spectra of different SNe, hence this scatter directly adds to the uncertainty of the peak magnitudes. However, this may not necessarily be the case. It is possible that least part of the observed scatter of the $K$-corrections is introduced by our procedure to prepare the spectra in the form suitable for computing $K$-corrections (Appendix~\ref{sec:kcor}). For example, the linear interpolation across the telluric bands is only a crude approximation. Both the spectra and the photometry to which they were matched may potentially contain calibration errors, and the host galaxy dust extinction estimation may not be accurate enough. All this certainly introduces extra scatter in the individual $K$-corrections. Adding this extra scatter will lead to deriving smaller  $\sigma_\mathrm{intr}$. Therefore, for accurate determination of the NIR intrinsic luminosity scatter it will be necessary to estimate correctly the uncertainty of the $K$-corrections. To achieve that, more high-quality photometry and spectroscopy of nearby SNe~Ia is clearly needed. Considering this in Tables~\ref{t:sne_g}-\ref{t:sne_lid} the uncertainties of the peak magnitudes estimated by the LC fitter and the uncertainties of the $K$-corrections are given separately. 

Our findings are also expected from theoretical radiative transfer models. \cite{2006ApJ...649..939K} found that the NIR peak magnitudes are much less sensitive to the amount of $^{56}$Ni generated in the explosion (the quantity that to a large extent controls the "luminosity -- LC shape" relation) than the optical peak magnitudes. Furthermore, \cite{2006ApJ...649..939K} found that significant variation of other physical parameters, such as progenitor metallicity, electron capture elements, and production of intermediate mass elements led to only a 
small scatter of $\sim0.1$~mag of the NIR peak magnitudes, a value which is consistent with the intrinsic scatter found in both optical and NIR.

After applying the LC shape and the extinction corrections it was noticed that the Hubble residuals showed a possible correlation with the $J_\mathrm{max}-H_\mathrm{max}$ color in both $J$ and $H$. Adding the $J_\mathrm{max}-H_\mathrm{max}$ term in the regression analysis leads to further small reduction of the scatter of the whole sample. Even though our analysis indicated that this term is statistically significant, further observations are needed to confirm this finding and whether this correction term will improve the NIR SN~Ia standard candle. Recently, the CSP, CfA and SweetSpot programs published updates on their previous data releases and added newly observed objects \citep{2017AJ....154..211K,2017arXiv170302402W,2015ApJS..220....9F}. Most of the new objects are in the Hubble flow and will provide further opportunities to verify the NIR SN~Ia standard candle and look for additional correlations such as with the $J_\mathrm{max}-H_\mathrm{max}$ color.  

Recently, the BAO technique has emerged as a very accurate probe for studying the expansion history of the universe, a very welcome addition to the cosmology toolkit. However, this technique also has certain limitations, most notably the  small cosmic volume probed at low redshift \citep[see, e.g.,][]{2017MNRAS.470.2617A}. At low redshift SNe~Ia provide significantly better measurements of relative distances and hence finding ways to reduce the systematic uncertainties of the SN~Ia technique is particularly relevant. We have seen that rest-frame NIR observations can provide a way to achieve that. Furthermore, SN~Ia appear indispensable to constrain models of the dark energy involving time-dependent equation of state parameter (see Figs.~29 and 17 in \cite{2014MNRAS.441...24A} and \cite{2017MNRAS.470.2617A}, respectively).  For this, as systematic-free SN~Ia observations as possible over a wide redshift range will be needed and in the next section we discuss the prospects for rest-frame NIR observations of SNe~Ia at high redshift.

\subsection{NIR observations at high redshift}

In order to provide constraints of the cosmological parameters with rest-frame NIR observations of SNe~Ia, a large sample of events out to redshift $z=1$ needs to be obtained. NIR observations of SNe~Ia beyond $z\sim0.5$ will be possible with the next generation of 40-m class telescopes (ESO Extremely Large Telescope, The Thirty Meter Telescope) and James Webb Space Telescope (JWST). For example, in the redshift range $z=0.6-1.2$ the rest-frame $J$ and $H$ bands are shifted to the JWST F200W, F277W and F356W bands, respectively\footnote{We note, however, that the JWST bands are twice broader than the red-shifted $J$ and $H$.}. The observed peak (Vega) magnitudes of a typical SN~Ia in these bands will be  $\sim23-24$ mag and the JWST ETC estimates that with 15 min exposure it should be possible to obtain photometry with S/N$\simeq20-80$ at NIR maximum light and with S/N$\simeq10-45$ at 10 days past $t_{B_\mathrm{max}}$. Observations out to $z\sim1.8$ will also be possible with exposure times of 40--60 min, but only at maximum light. We have seen that the GOLD sample - the SNe with well-sampled NIR LCs - provides the best estimate of the peak magnitudes and the smallest scatter on the Hubble diagram. However, obtaining LCs with at least four points (plus one more if galaxy reference image is needed) is a considerable investment in terms of observing time at the world's most advanced and expensive observatory. Thus, the main goal of this study has been to test the accuracy of estimating the NIR peak magnitudes of SNe~Ia with only a single observation coupled with good optical data, which is the most efficient approach to use rest-frame NIR observations of SNe~Ia in cosmology.

To measure the peak magnitudes from under-sampled LCs, we emplyed a method orgnally used by \cite{kri_04,kri_ir_temp}. We have further verified the applicability of this method by analyzing the optical and NIR LCs of the GOLD sample and discussed in detail the associated uncertainties when under-sampled LCs are used. We have performed series of simulations based on the NIR templates and re-sampling the NIR LCs of the GOLD sample. The results showed that it is  possible to estimate the NIR peak magnitudes from a single observation with accuracy better than 0.03 mag if the observation is obtained within a few days from the NIR maximum. If the observation is obtained further away from the maximum the accuracy decreases but is still better than $\sim0.10$ mag.

To test the method on real data we analyzed published LCs together with our new sample of 16 SNe. The published low-redshift sample was divided into three groups GOLD, SILVER, and BRONZE according to the sampling of the NIR LCs, plus the BN12 sample, which is at higher redshift. By design our new sample contains SNe with single NIR observations and only two SNe have two observations. Out of 28 SNe in the BRONZE sample, 16 have at most two observations per band, and all the others have more observations in at least one band. Out of the 15 SNe in the SILVER sample, only four have two or fewer observations. Thus, along with our new sample, the BRONZE one is also well-suited to test the methodology.

The intrinsic luminosity scatter $\sigma_\mathrm{intr}$ of SNe~Ia in the NIR was estimated from the GOLD and BN12 samples to be around 0.10 mag in both $J$ and $H$ bands. The scatter inferred from the other samples is larger, most likely because of unaccounted additional uncertainties arising from fitting under-sampled NIR LCs. The intrinsic scatter inferred from our and the BRONZE samples is in the range $\sim0.15-0.21$ mag and $\sim0.15-0.17$ mag in $J$ and $H$, respectively, i.e. larger by a factor of $\sim2$ and $\sim1.5$ than that inferred from the GOLD and BN12 samples.  This difference might be related to the much larger flux variation in $J$ than in $H$ after maximum, $\sim1.3$ mag $vs.$ $\sim0.4$ mag, but it may also be an intrinsic property of SNe~Ia. Regardless of the reason, our results suggest that when working with under-sampled LCs $H$ band should be preferred over $J$. 

One may then ask the question what is the best observing strategy when limited observing time is available. Should one obtain only well-sampled LCs with four or five points plus a reference image or three times larger sample of SNe with only one observation plus a reference image? Considering the values of the intrinsic luminosity scatter that we estimate in the different samples and the results of the simulations, the latter approach may be advantageous,  especially if $H$-band observations can be obtained within two to three days from the NIR maximum.

In the future most SN discoveries at high redshift will likely come from LSST, WFIRST and the DESIRE surveys,  which will also have the capability to provide $B$ and $V$ LCs out to redshift $z\sim1.5$. Taking into account that the minimum technical time to trigger JWST will likely be around two days, to observe a SN within a few days from NIR maximum, the SN candidates need to be spectroscopically classified as SNe~Ia at least one week before $t_{B_\mathrm{max}}$. Very effective selection of SNe~Ia for spectroscopic screening (90\% success) can be made by studying the early multicolor LCs of the candidates, as shown by \cite{2008AJ....135..348S}. Thus, together with a dedicated program to classify SNe~Ia on 8-m class telescopes, LSST and space missions should provide enough targets for NIR follow-up to build NIR Hubble diagram out to redshift $z=1$.

\section{Conclusions}

We obtained multi-epoch optical $UBVRI$ and single-epoch NIR  $J$ and $H$ photometry of 16 SNe~Ia in the
redshift range $z=0.037-0.183$. This sample doubles the leverage of the current SN~Ia NIR Hubble diagram. 
We analyzed this new data set together with 102 NIR and 458 optical LCs of normal SNe~Ia from the literature. 
Our findings can be summarized as follows.

 A subsample of SNe with well-sampled NIR LCs (GOLD sample) allows template fitting with the stretch and the time of maximum as free parameters. We find that the optical and NIR stretches are nearly one-to-one related and that the time of $B$-band maximum light estimated from the NIR and optical LC fits are very close.  This further reinforces the results of \cite{kri_04,kri_ir_temp} and \cite{2008ApJ...689..377W} who showed that smooth LC templates, the time axis of which is stretched according to the optical stretch parameter, accurately describe the rest-frame $J$ and $H$-band LCs within 10 days from $B$-band maximum. This allowed us to estimate the NIR peak magnitudes of the SNe with under-sampled LCs, including our new sample.

 Based on simulations, we find that  it is possible to estimate the NIR peak magnitudes with accuracy better that 0.025~mag with even a single observation if obtained within two to three days from the time of NIR maximum light. The accuracy decreases to $\sim0.12$ mag if the observation is obtained ten or more days after the time of NIR maximum light.

 Our analysis confirms previous findings that SNe~Ia are at least as good standard candles in the NIR as in the optical, now with good statistics of $>100$ objects. We find evidence that the NIR Hubble residuals weakly correlate with the LC shape parameter $\Delta M_{15}$ and the color excess $E(B-V)_\mathrm{obs}$,  and for the first time we report a possible dependence  on the  $J_\mathrm{max}-H_\mathrm{max}$ color. The intrinsic luminosity scatter of SNe~Ia in the NIR is estimated  from the best-observed SNe to be around 0.10 mag. The weighted rms scatter of the Hubble diagram is $0.14\,\pm0.01$ mag in both $J$ and $H$, and SNe~Ia appear to be equally good standard candles in both bands.

 The intrinsic luminosity scatter derived from the SNe with under-sampled LCs is larger than that estimated   from the best-observed SNe by a factor of $\sim1.5$ in $H$ band and by $\sim2$ in $J$ band. Together with the results of the simulations, this suggests that obtaining three times more SNe with single observation may be advantageous compared to obtaining one SN with well-sampled LC,  especially if the observation can be obtained close to the NIR maximum.

Compared to the optical, the corrections for the LC shape in the NIR are by a factor of 3.5 smaller. The extinction corrections are also smaller by an order of magnitude and consistent with low $R_V\simeq1.9$. Thus, possible systematic uncertainties associated with the LC shape and the extinction corrections are greatly reduced in the NIR. 

We find that the intrinsic $B-V$ color of SNe~Ia at $B$-band maximum light, $(B-V)_{B_\mathrm{max},0}$, can be described by the following  nonlinear function of $\Delta M_{15}$:
\begin{equation}
 (B-V)_{B_\mathrm{max},0}=0.036\,\exp\left[\frac{\Delta M_{15}-1.3}{0.3455}\right]-0.084 (\pm0.01). \nonumber
\end{equation}
The intrinsic scatter of $(B-V)_{B_\mathrm{max},0}$ is estimated to be $\sim0.04$ mag.

 The weighted scatter of the $B$ band Hubble residuals is $0.176\,\pm0.007$ mag. The inferred intrinsic luminosity scatter was found to vary between the samples indicating possible systematic uncertainties. The dust extinction in the host galaxies is on average described with $R_V\sim1.8-19$.  Because of the large span of the maximum $B-V$ color of the nearby SN~Ia sample our estimation of $R_V$ is robust and $R_V=3.1$ is inconsistent with the data.

In conclusion, we have now extended the NIR SN~Ia Hubble diagram to its nonlinear part at $z\sim0.2$, that is, where the lever arm in redshift makes it possible to start probing the cosmic constituents, most notably the properties of dark energy. Interestingly, our study confirms that it is feasible to accomplish this result with very modest sampling of the NIR LCs, if complemented by well-sampled optical LCs to determine the SN phase and the optical stretch parameter. This bodes well for the future, in particular when JWST will be operational. JWST will have the capabilities to observe the rest-frame $J$ and $H$ out to redshift $z\simeq1.8$, and LSST, WFIRST and DESIRE surveys will provide SNe with well-sampled rest-frame $B$ and $V$ LCs out to redshift $z\simeq1.5$. In the view of our results, we suggest that the most efficient way to extend the SN~Ia NIR Hubble diagram to high redshifts will be to observe a large samples of SNe~Ia with a single observation close to the NIR maximum.

\begin{acknowledgements}

This work is based on observations collected at the Nordic Optical Telescope (NOT) located at the
Spanish Observatorio del Roque de los Muchachos of the Instituto de
Astrofisica de Canarias (La Palma, Spain). Part of the The data presented here were obtained in part with ALFOSC, which is provided by the Instituto de Astrofisica de Andalucia (IAA) under a joint agreement with the University of Copenhagen and NOTSA. This paper also uses observations made at the South African Astronomical Observatory (SAAO).
Part of the observations were also obtained at the NOT during a student training course in
Observational Astronomy provided by Stockholm Observatory and the NorFA Summer School in Observational Astronomy. We are grateful to the NOT astronomers Errki Kankade, Amanda Diupvik, Thomas Augusteijn, Carolin Villforth, Sami Niemi, Tapio Pursimo, and Helena Uthas who performed Service observations,  Lison Malo and Celine Reyle for giving up part of their time to observe two of our targets, and Kate Maguire for providing us with optical LCs.
We thank the SNfactory team, Greg Aldering, Peter Nugent, and Rollin Thomas in particular, and the SDSS-II team for making their SN discoveries known to us.

V.S. acknowledges financial support from Funda\c{c}\~{a}o para a Ci\^{e}ncia
e a Tecnologia (FCT) under program Ci\^{e}ncia 2008. This work was partly funded by FCT with the research grant 
PTDC/CTE-AST/112582/2009. AG and RA acknowledge generous support from the Swedish Research Council.

This work has made use of the NASA/IPAC Extragalactic Database (NED),
NASA's Astrophysics Data System, and data products from SDSS and SDSS-II surveys.
Funding for the SDSS and SDSS-II has been provided by the Alfred P. Sloan Foundation, 
the Participating Institutions, the National Science Foundation, the
 U.S. Department of Energy, the National Aeronautics and Space Administration, 
 the Japanese Monbukagakusho, the Max Planck Society, and the Higher Education Funding Council 
 for England. The SDSS Web Site is \url{http://www.sdss.org/}.

\end{acknowledgements}


\begin{appendix}

\section{Local sequences of stars}

\begin{figure*}[!h]
\includegraphics*[width=18cm]{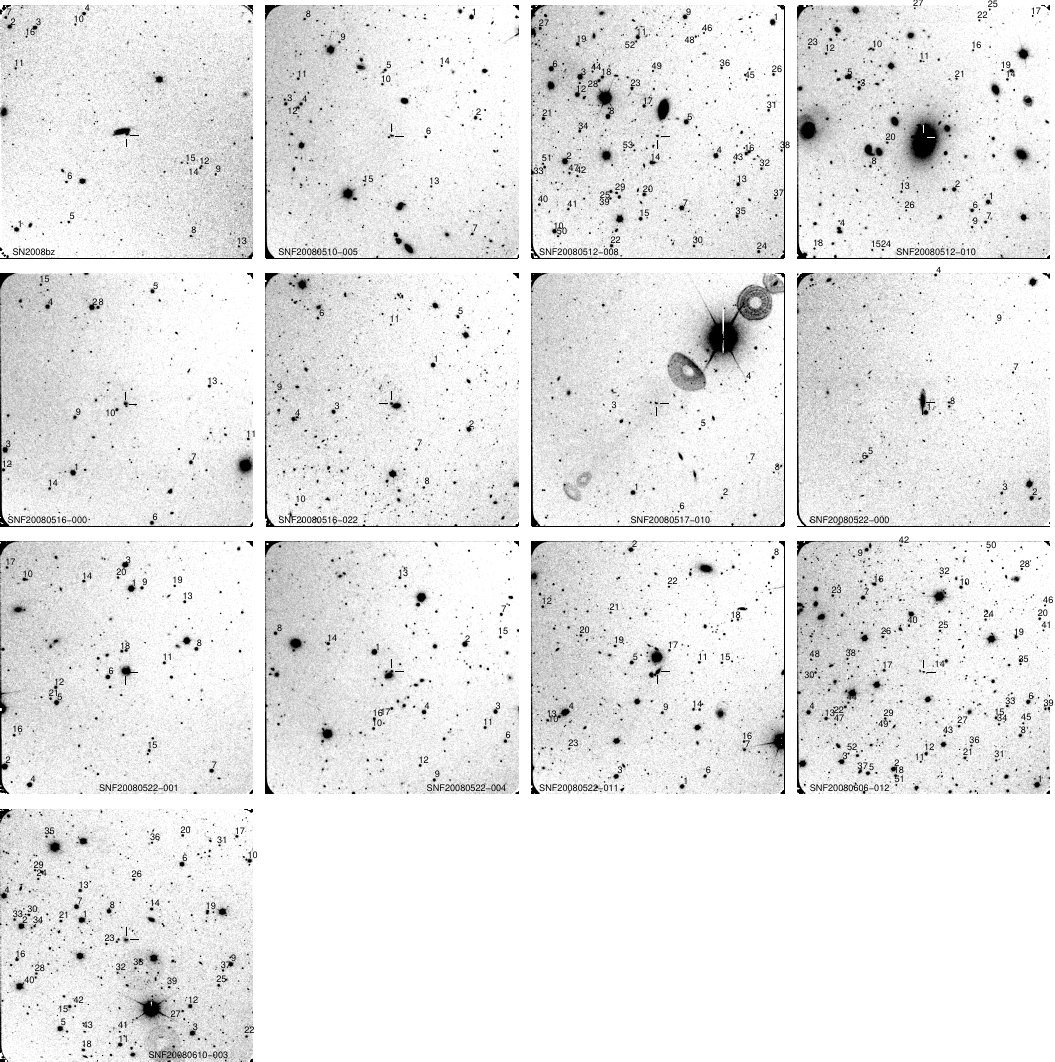}
\caption{Numbers indicate the local sequence stars. In most cases the number is right-up of the star and in a few cases only it is left-down. The SNs are located at the intersection of the two ticks.}
\label{f:fc1}
\end{figure*} 

The calibrated magnitudes of the local sequences of stars are given in Table~\ref{t:stds} and the stars are indicated on the finding charts in Fig.~\ref{f:fc1}.

\longtab[1]{
\begin{longtable}{@{}rcccccccccc@{}}
\caption{Calibrated magnitudes of the local sequences of stars around the SNe. The 1$\sigma$ uncertainties are given in parentheses. $N_\mathrm{obs}$ shows the number of individual calibrations that were averaged.}
\label{t:stds}\\
\hline\hline
No. &  $U$ & $N_\mathrm{obs}$ & $B$ & $N_\mathrm{obs}$ & $V$ & $N_\mathrm{obs}$ & $R$ & $N_\mathrm{obs}$ &  $I$ & $N_\mathrm{obs}$  \\
\hline\noalign{\smallskip}
\endfirsthead
\caption{continued.}\\
\hline\hline
No. &  $U$ & $N_\mathrm{obs}$ & $B$ & $N_\mathrm{obs}$ & $V$ & $N_\mathrm{obs}$ & $R$ & $N_\mathrm{obs}$ &  $I$ & $N_\mathrm{obs}$  \\
\hline\noalign{\smallskip}
\endhead
\hline
\endfoot
\hline
\endlastfoot
\multicolumn{11}{c}{SN~2008bz}\\
\hline\noalign{\smallskip}
    1 & 15.755 (0.013)  &        1 & 15.814 (0.014)  &        1 & 15.220 (0.011)  &        1 & 14.895 (0.012)  &        1 & 14.552 (0.012)  &        1 \\
    2 & 16.029 (0.013)  &        1 & 16.101 (0.013)  &        1 & 15.587 (0.011)  &        1 & 15.278 (0.012)  &        1 & 14.973 (0.012)  &        1 \\
    3 & 17.657 (0.018)  &        1 & 16.858 (0.016)  &        1 & 15.935 (0.012)  &        1 & 15.393 (0.013)  &        1 & 14.933 (0.012)  &        1 \\
    4 & 19.522 (0.036)  &        1 & 18.324 (0.020)  &        1 & 17.159 (0.012)  &        1 & 16.398 (0.014)  &        1 & 15.766 (0.013)  &        1 \\
    5 & 20.519 (0.077)  &        1 & 19.187 (0.020)  &        1 & 17.945 (0.012)  &        1 & 17.143 (0.013)  &        1 & 16.433 (0.015)  &        1 \\
    6 & 19.023 (0.027)  &        1 & 18.719 (0.020)  &        1 & 17.973 (0.017)  &        1 & 17.531 (0.018)  &        1 & 17.115 (0.015)  &        1 \\
    7 & 20.961 (0.108)  &        1 & 19.816 (0.033)  &        1 & 18.233 (0.015)  &        1 & 17.169 (0.016)  &        1 & 15.672 (0.016)  &        1 \\
    8 & 18.942 (0.027)  &        1 & 19.028 (0.025)  &        1 & 18.479 (0.012)  &        1 & 18.095 (0.024)  &        1 & 17.750 (0.013)  &        1 \\
    9 & 19.111 (0.030)  &        1 & 19.300 (0.013)  &        1 & 18.733 (0.017)  &        1 & 18.408 (0.029)  &        1 & 18.047 (0.039)  &        1 \\
   10 & 19.065 (0.028)  &        1 & 19.364 (0.039)  &        1 & 18.972 (0.019)  &        1 & 18.701 (0.017)  &        1 & 18.415 (0.017)  &        1 \\
   11 & 21.601 (0.191)  &        1 & 20.547 (0.023)  &        1 & 19.112 (0.014)  &        1 & 18.176 (0.015)  &        1 & 16.968 (0.014)  &        1 \\
   12 & 21.224 (0.150)  &        1 & 20.482 (0.033)  &        1 & 19.175 (0.014)  &        1 & 18.349 (0.018)  &        1 & 17.533 (0.033)  &        1 \\
   13 & 20.605 (0.095)  &        1 & 20.133 (0.021)  &        1 & 19.228 (0.015)  &        1 & 18.692 (0.014)  &        1 & 18.177 (0.031)  &        1 \\
   14 &  $-$            & $-$      & 20.778 (0.053)  &        1 & 19.382 (0.021)  &        1 & 18.341 (0.016)  &        1 & 16.808 (0.016)  &        1 \\
   15 & 19.956 (0.058)  &        1 & 20.164 (0.012)  &        1 & 19.695 (0.049)  &        1 & 19.335 (0.026)  &        1 & 18.963 (0.037)  &        1 \\
   16 & 20.148 (0.058)  &        1 & 20.305 (0.015)  &        1 & 19.737 (0.027)  &        1 & 19.366 (0.015)  &        1 & 19.071 (0.020)  &        1 \\
\hline\noalign{\smallskip}
\multicolumn{11}{c}{SNF20080510-005}\\
\hline\noalign{\smallskip}
    1 & 18.666 (0.023)  &        3 & 17.572 (0.028)  &        3 & 16.046 (0.010)  &        3 & 15.065 (0.015)  &        2 &  $-$            & $-$      \\
    2 & 17.311 (0.015)  &        3 & 17.316 (0.031)  &        3 & 16.691 (0.015)  &        3 & 16.325 (0.022)  &        3 & 15.957 (0.023)  &        3 \\
    3 & 17.619 (0.021)  &        3 & 17.931 (0.010)  &        3 & 17.450 (0.013)  &        3 & 17.148 (0.015)  &        3 & 16.786 (0.015)  &        3 \\
    4 & 19.090 (0.100)  &        3 & 18.482 (0.022)  &        3 & 17.591 (0.017)  &        3 & 17.030 (0.018)  &        3 & 16.530 (0.014)  &        3 \\
    5 & 17.808 (0.038)  &        3 & 18.155 (0.017)  &        3 & 17.693 (0.014)  &        3 & 17.390 (0.011)  &        3 & 17.055 (0.013)  &        3 \\
    6 & 18.845 (0.078)  &        3 & 18.945 (0.028)  &        3 & 18.371 (0.016)  &        3 & 18.045 (0.013)  &        3 & 17.688 (0.022)  &        3 \\
    7 &  $-$            & $-$      & 20.019 (0.016)  &        3 & 18.674 (0.018)  &        3 & 17.780 (0.019)  &        3 & 16.989 (0.016)  &        3 \\
    8 & 19.740 (0.031)  &        3 & 19.542 (0.045)  &        3 & 18.810 (0.003)  &        3 & 18.352 (0.009)  &        3 & 17.910 (0.019)  &        3 \\
    9 & 19.483 (0.034)  &        3 & 19.623 (0.011)  &        3 & 19.028 (0.022)  &        3 & 18.689 (0.009)  &        3 & 18.385 (0.010)  &        3 \\
   10 & 20.913 (0.196)  &        3 & 20.417 (0.030)  &        3 & 19.033 (0.012)  &        3 & 18.093 (0.007)  &        3 & 16.999 (0.020)  &        3 \\
   11 &  $-$            & $-$      & 20.831 (0.050)  &        3 & 19.140 (0.012)  &        3 & 18.049 (0.028)  &        3 & 16.410 (0.014)  &        3 \\
   12 & 18.823 (0.046)  &        3 & 19.460 (0.022)  &        3 & 19.241 (0.038)  &        3 & 18.917 (0.022)  &        3 & 18.346 (0.019)  &        3 \\
   13 &  $-$            & $-$      & 21.079 (0.065)  &        3 & 19.630 (0.045)  &        3 & 18.634 (0.018)  &        3 & 17.238 (0.015)  &        3 \\
   14 &  $-$            & $-$      & 21.175 (0.048)  &        3 & 19.692 (0.018)  &        3 & 18.725 (0.021)  &        3 & 17.438 (0.013)  &        3 \\
   15 & 20.126 (0.062)  &        3 & 20.510 (0.023)  &        3 & 20.048 (0.002)  &        3 & 19.723 (0.039)  &        3 & 19.461 (0.027)  &        3 \\
\hline\noalign{\smallskip}
\multicolumn{11}{c}{SNF20080512-008}\\
\hline\noalign{\smallskip}
    1 & 15.949 (0.015)  &        1 & 15.909 (0.008)  &        2 & 15.208 (0.004)  &        2 & 14.836 (0.009)  &        2 &  $-$            & $-$      \\
    2 & 16.568 (0.017)  &        1 & 16.042 (0.011)  &        2 & 15.211 (0.005)  &        2 & 14.750 (0.012)  &        1 &  $-$            & $-$      \\
    3 & 16.984 (0.018)  &        1 & 16.403 (0.009)  &        2 & 15.528 (0.003)  &        2 & 15.026 (0.002)  &        2 &  $-$            & $-$      \\
    4 & 16.215 (0.015)  &        1 & 16.223 (0.002)  &        2 & 15.608 (0.007)  &        2 & 15.255 (0.006)  &        2 &  $-$            & $-$      \\
    5 & 16.944 (0.006)  &        2 & 16.608 (0.005)  &        3 & 15.845 (0.005)  &        3 & 15.428 (0.015)  &        3 & 15.073 (0.069)  &        3 \\
    6 & 16.539 (0.016)  &        1 & 16.570 (0.033)  &        2 & 15.940 (0.017)  &        2 & 15.555 (0.015)  &        2 & 15.160 (0.011)  &        1 \\
    7 & 17.317 (0.015)  &        2 & 16.854 (0.005)  &        3 & 16.013 (0.009)  &        3 & 15.518 (0.010)  &        3 & 15.087 (0.012)  &        1 \\
    8 & 16.180 (0.003)  &        2 & 16.457 (0.002)  &        3 & 16.069 (0.002)  &        3 & 15.799 (0.008)  &        3 & 15.507 (0.018)  &        3 \\
    9 & 17.265 (0.018)  &        2 & 16.982 (0.010)  &        3 & 16.215 (0.007)  &        3 & 15.789 (0.003)  &        3 & 15.370 (0.007)  &        2 \\
   10 & 17.046 (0.005)  &        2 & 16.953 (0.010)  &        3 & 16.256 (0.003)  &        3 & 15.878 (0.009)  &        3 & 15.505 (0.007)  &        2 \\
   11 & 16.974 (0.013)  &        2 & 17.143 (0.008)  &        3 & 16.620 (0.010)  &        3 & 16.303 (0.000)  &        3 & 15.957 (0.008)  &        3 \\
   12 & 19.299 (0.055)  &        2 & 18.087 (0.004)  &        3 & 16.730 (0.006)  &        3 & 15.856 (0.004)  &        3 & 15.052 (0.012)  &        1 \\
   13 & 17.595 (0.009)  &        2 & 17.617 (0.009)  &        3 & 17.005 (0.009)  &        3 & 16.640 (0.009)  &        3 & 16.270 (0.006)  &        3 \\
   14 & 17.638 (0.019)  &        2 & 17.628 (0.003)  &        3 & 17.017 (0.006)  &        3 & 16.653 (0.019)  &        3 & 16.298 (0.010)  &        3 \\
   15 & 17.311 (0.002)  &        2 & 17.598 (0.004)  &        3 & 17.134 (0.002)  &        3 & 16.820 (0.007)  &        3 & 16.489 (0.003)  &        3 \\
   16 & 17.988 (0.004)  &        2 & 18.057 (0.014)  &        3 & 17.441 (0.008)  &        3 & 17.081 (0.013)  &        3 & 16.699 (0.003)  &        3 \\
   17 & 17.932 (0.006)  &        2 & 18.053 (0.006)  &        3 & 17.500 (0.001)  &        3 & 17.166 (0.008)  &        3 & 16.803 (0.005)  &        3 \\
   18 & 17.931 (0.002)  &        2 & 18.087 (0.007)  &        3 & 17.560 (0.005)  &        3 & 17.247 (0.005)  &        3 & 16.918 (0.004)  &        3 \\
   19 & 18.284 (0.018)  &        2 & 18.276 (0.007)  &        3 & 17.631 (0.005)  &        3 & 17.252 (0.008)  &        3 & 16.875 (0.006)  &        3 \\
   20 & 20.284 (0.064)  &        2 & 19.110 (0.010)  &        3 & 17.690 (0.010)  &        3 & 16.763 (0.014)  &        3 & 15.833 (0.001)  &        3 \\
   21 & 19.759 (0.048)  &        2 & 18.834 (0.014)  &        3 & 17.797 (0.002)  &        3 & 17.173 (0.010)  &        3 & 16.625 (0.006)  &        3 \\
   22 & 19.427 (0.009)  &        2 & 18.724 (0.003)  &        3 & 17.808 (0.007)  &        3 & 17.262 (0.004)  &        3 & 16.785 (0.002)  &        3 \\
   23 & 18.823 (0.051)  &        2 & 18.631 (0.006)  &        3 & 17.885 (0.009)  &        3 & 17.452 (0.005)  &        3 & 17.019 (0.005)  &        3 \\
   24 & 18.259 (0.012)  &        2 & 18.400 (0.014)  &        3 & 17.949 (0.012)  &        3 & 17.634 (0.015)  &        3 & 17.309 (0.008)  &        3 \\
   25 & 19.820 (0.012)  &        2 & 19.054 (0.011)  &        3 & 18.108 (0.008)  &        3 & 17.515 (0.010)  &        3 & 17.006 (0.019)  &        3 \\
   26 & 18.823 (0.021)  &        2 & 18.799 (0.007)  &        3 & 18.141 (0.006)  &        3 & 17.779 (0.005)  &        3 & 17.397 (0.018)  &        3 \\
   27 & 19.193 (0.028)  &        2 & 18.940 (0.005)  &        3 & 18.166 (0.006)  &        3 & 17.724 (0.012)  &        3 & 17.314 (0.008)  &        3 \\
   28 & 20.670 (0.117)  &        2 & 19.735 (0.023)  &        3 & 18.376 (0.004)  &        3 & 17.474 (0.011)  &        3 & 16.480 (0.002)  &        3 \\
   29 & 19.002 (0.018)  &        2 & 19.108 (0.023)  &        3 & 18.498 (0.007)  &        3 & 18.144 (0.006)  &        3 & 17.796 (0.010)  &        3 \\
   30 & 18.764 (0.005)  &        2 & 18.987 (0.009)  &        3 & 18.519 (0.011)  &        3 & 18.212 (0.005)  &        3 & 17.873 (0.011)  &        3 \\
   31 & 20.381 (0.040)  &        2 & 19.609 (0.010)  &        3 & 18.618 (0.009)  &        3 & 17.973 (0.013)  &        3 & 17.373 (0.010)  &        3 \\
   32 & 19.282 (0.035)  &        2 & 19.322 (0.017)  &        3 & 18.672 (0.006)  &        3 & 18.292 (0.016)  &        3 & 17.907 (0.008)  &        3 \\
   33 & 20.689 (0.075)  &        2 & 19.799 (0.022)  &        3 & 18.736 (0.006)  &        3 & 18.126 (0.009)  &        3 & 17.593 (0.008)  &        3 \\
   34 & 19.532 (0.182)  &        2 & 19.480 (0.011)  &        3 & 18.794 (0.016)  &        3 & 18.396 (0.006)  &        3 & 17.990 (0.009)  &        3 \\
   35 &  $-$            & $-$      & 20.321 (0.005)  &        3 & 18.812 (0.006)  &        3 & 17.835 (0.019)  &        3 & 16.710 (0.005)  &        3 \\
   36 &  $-$            & $-$      & 20.064 (0.039)  &        3 & 18.836 (0.007)  &        3 & 18.017 (0.015)  &        3 & 17.295 (0.007)  &        3 \\
   37 & 20.194 (0.039)  &        2 & 19.757 (0.012)  &        3 & 18.916 (0.017)  &        3 & 18.419 (0.013)  &        3 & 17.933 (0.003)  &        3 \\
   38 & 19.465 (0.087)  &        2 & 19.587 (0.008)  &        3 & 18.960 (0.020)  &        3 & 18.577 (0.017)  &        3 & 18.199 (0.014)  &        3 \\
   39 & 19.184 (0.015)  &        2 & 19.455 (0.020)  &        3 & 18.975 (0.001)  &        3 & 18.649 (0.018)  &        3 & 18.332 (0.015)  &        3 \\
   40 & 20.594 (0.035)  &        2 & 19.929 (0.007)  &        3 & 18.993 (0.016)  &        3 & 18.436 (0.021)  &        3 & 17.920 (0.020)  &        3 \\
   41 & 20.297 (0.049)  &        2 & 19.840 (0.015)  &        3 & 19.004 (0.006)  &        3 & 18.518 (0.004)  &        3 & 18.038 (0.013)  &        3 \\
   42 & 20.044 (0.091)  &        2 & 20.048 (0.022)  &        3 & 19.340 (0.004)  &        3 & 18.971 (0.025)  &        3 & 18.573 (0.015)  &        3 \\
   43 &  $-$            & $-$      & 21.181 (0.122)  &        3 & 19.634 (0.028)  &        3 & 18.602 (0.005)  &        3 & 17.321 (0.008)  &        3 \\
   44 & 19.703 (0.113)  &        2 & 20.046 (0.014)  &        3 & 19.663 (0.013)  &        3 & 19.399 (0.012)  &        3 & 19.039 (0.003)  &        3 \\
   45 & 19.792 (0.138)  &        2 & 20.136 (0.020)  &        3 & 19.667 (0.029)  &        3 & 19.380 (0.005)  &        3 & 19.032 (0.014)  &        3 \\
   46 & 20.997 (0.110)  &        2 & 20.570 (0.053)  &        3 & 19.683 (0.014)  &        3 & 19.122 (0.013)  &        3 & 18.644 (0.027)  &        3 \\
   47 & 20.842 (0.004)  &        2 & 20.635 (0.039)  &        3 & 19.724 (0.004)  &        3 & 19.211 (0.027)  &        3 & 18.738 (0.017)  &        3 \\
   48 &  $-$            & $-$      & 21.342 (0.013)  &        3 & 19.797 (0.032)  &        3 & 18.782 (0.006)  &        3 & 17.300 (0.006)  &        3 \\
   49 &  $-$            & $-$      & 21.013 (0.051)  &        3 & 19.813 (0.002)  &        3 & 19.030 (0.017)  &        3 & 18.258 (0.012)  &        3 \\
   50 & 22.579 (0.005)  &        2 & 21.085 (0.115)  &        3 & 19.830 (0.020)  &        3 & 19.030 (0.016)  &        3 & 18.344 (0.015)  &        3 \\
   51 &  $-$            & $-$      & 21.715 (0.085)  &        3 & 20.197 (0.020)  &        3 & 19.215 (0.006)  &        3 & 17.927 (0.011)  &        3 \\
   52 &  $-$            & $-$      & 21.787 (0.122)  &        3 & 20.384 (0.022)  &        3 & 19.367 (0.019)  &        3 & 17.881 (0.008)  &        3 \\
   53 &  $-$            & $-$      & 22.473 (0.089)  &        3 & 20.747 (0.048)  &        3 & 19.526 (0.016)  &        3 & 17.655 (0.016)  &        3 \\
\hline\noalign{\smallskip}
\multicolumn{11}{c}{SNF20080512-010}\\
\hline\noalign{\smallskip}
    1 & 16.985 (0.024)  &        2 & 16.839 (0.009)  &        3 & 16.138 (0.009)  &        3 & 15.756 (0.002)  &        2 & 15.361 (0.013)  &        1 \\
    2 & 19.479 (0.005)  &        2 & 18.426 (0.006)  &        3 & 16.933 (0.006)  &        3 & 15.924 (0.005)  &        2 &  $-$            & $-$      \\
    3 & 17.406 (0.003)  &        2 & 17.607 (0.009)  &        3 & 17.056 (0.008)  &        3 & 16.733 (0.009)  &        3 & 16.367 (0.011)  &        3 \\
    4 & 19.239 (0.032)  &        2 & 18.354 (0.010)  &        3 & 17.351 (0.006)  &        3 & 16.738 (0.005)  &        3 & 16.196 (0.011)  &        3 \\
    5 & 18.066 (0.018)  &        2 & 18.071 (0.036)  &        3 & 17.408 (0.020)  &        3 & 17.037 (0.012)  &        3 & 16.654 (0.008)  &        3 \\
    6 & 20.373 (0.115)  &        2 & 19.086 (0.028)  &        3 & 17.660 (0.006)  &        3 & 16.722 (0.011)  &        3 & 15.795 (0.002)  &        2 \\
    7 & 20.927 (0.078)  &        2 & 19.750 (0.010)  &        3 & 18.286 (0.008)  &        3 & 17.329 (0.014)  &        3 & 16.241 (0.001)  &        3 \\
    8 & 20.458 (0.112)  &        2 & 19.528 (0.004)  &        3 & 18.560 (0.008)  &        3 & 17.934 (0.001)  &        3 & 17.413 (0.012)  &        3 \\
    9 & 18.794 (0.013)  &        2 & 19.087 (0.009)  &        3 & 18.661 (0.014)  &        3 & 18.365 (0.013)  &        3 & 18.040 (0.002)  &        3 \\
   10 & 19.868 (0.016)  &        2 & 19.688 (0.004)  &        3 & 18.902 (0.013)  &        3 & 18.441 (0.008)  &        3 & 17.944 (0.015)  &        3 \\
   11 & 19.051 (0.022)  &        2 & 19.407 (0.015)  &        3 & 18.922 (0.009)  &        3 & 18.612 (0.013)  &        3 & 18.263 (0.014)  &        3 \\
   12 & 19.970 (0.057)  &        2 & 19.881 (0.035)  &        3 & 19.215 (0.014)  &        3 & 18.829 (0.010)  &        3 & 18.437 (0.025)  &        3 \\
   13 & 21.373 (0.147)  &        2 & 20.356 (0.009)  &        3 & 19.218 (0.013)  &        3 & 18.518 (0.010)  &        3 & 17.920 (0.014)  &        3 \\
   14 &  $-$            & $-$      & 20.519 (0.025)  &        3 & 19.261 (0.019)  &        3 & 18.430 (0.011)  &        3 & 17.739 (0.012)  &        3 \\
   15 & 19.461 (0.021)  &        2 & 19.710 (0.020)  &        3 & 19.282 (0.008)  &        3 & 19.005 (0.011)  &        3 & 18.695 (0.027)  &        3 \\
   16 & 19.854 (0.048)  &        2 & 19.989 (0.003)  &        3 & 19.418 (0.017)  &        3 & 19.071 (0.004)  &        3 & 18.697 (0.031)  &        3 \\
   17 & 19.567 (0.008)  &        2 & 19.969 (0.054)  &        3 & 19.460 (0.011)  &        3 & 19.184 (0.022)  &        3 & 18.848 (0.024)  &        3 \\
   18 & 20.183 (0.148)  &        2 & 20.269 (0.015)  &        3 & 19.516 (0.009)  &        3 & 19.119 (0.006)  &        3 & 18.712 (0.009)  &        3 \\
   19 & 19.687 (0.019)  &        2 & 20.030 (0.008)  &        3 & 19.562 (0.007)  &        3 & 19.253 (0.008)  &        3 & 18.910 (0.048)  &        3 \\
   20 & 22.052 (0.158)  &        2 & 21.141 (0.016)  &        3 & 19.622 (0.037)  &        3 & 18.520 (0.010)  &        3 & 16.812 (0.006)  &        3 \\
   21 & 22.098 (0.015)  &        2 & 21.061 (0.011)  &        3 & 19.687 (0.006)  &        3 & 18.800 (0.005)  &        3 & 17.942 (0.010)  &        3 \\
   22 & 20.662 (0.038)  &        2 & 20.535 (0.028)  &        3 & 19.772 (0.002)  &        3 & 19.336 (0.008)  &        3 & 18.918 (0.024)  &        3 \\
   23 & 21.740 (0.140)  &        2 & 21.052 (0.047)  &        3 & 19.891 (0.028)  &        3 & 19.087 (0.017)  &        3 & 18.378 (0.009)  &        3 \\
   24 &  $-$            & $-$      & 21.424 (0.078)  &        3 & 20.036 (0.024)  &        3 & 19.110 (0.012)  &        3 & 18.065 (0.011)  &        3 \\
   25 &  $-$            & $-$      & 21.635 (0.147)  &        3 & 20.276 (0.026)  &        3 & 19.370 (0.008)  &        3 & 18.407 (0.015)  &        3 \\
   26 &  $-$            & $-$      & 21.997 (0.067)  &        3 & 20.508 (0.032)  &        3 & 19.451 (0.010)  &        3 & 17.912 (0.013)  &        3 \\
   27 & 20.278 (0.046)  &        2 & 20.853 (0.042)  &        3 & 20.514 (0.024)  &        3 & 20.111 (0.028)  &        3 & 19.695 (0.028)  &        3 \\
\hline\noalign{\smallskip}
\multicolumn{11}{c}{SNF20080516-000}\\
\hline\noalign{\smallskip}
    1 &  $-$            & $-$      & 15.750 (0.004)  &        2 & 15.145 (0.003)  &        2 &  $-$            & $-$      &  $-$            & $-$      \\
    2 &  $-$            & $-$      & 16.011 (0.000)  &        2 & 15.221 (0.007)  &        2 &  $-$            & $-$      &  $-$            & $-$      \\
    3 & 16.989 (0.026)  &        1 & 16.337 (0.011)  &        3 & 15.405 (0.007)  &        3 & 14.899 (0.016)  &        1 &  $-$            & $-$      \\
    4 & 16.639 (0.022)  &        1 & 16.864 (0.004)  &        4 & 16.312 (0.006)  &        4 & 15.983 (0.008)  &        4 & 15.629 (0.013)  &        3 \\
    5 & 18.985 (0.046)  &        1 & 18.021 (0.004)  &        4 & 16.876 (0.008)  &        4 & 16.160 (0.008)  &        4 & 15.562 (0.012)  &        3 \\
    6 & 17.486 (0.024)  &        1 & 17.495 (0.017)  &        3 & 16.909 (0.007)  &        4 & 16.550 (0.004)  &        4 & 16.202 (0.012)  &        4 \\
    7 & 18.227 (0.031)  &        1 & 17.782 (0.004)  &        4 & 16.914 (0.003)  &        4 & 16.408 (0.008)  &        4 & 15.951 (0.015)  &        4 \\
    8 & 19.533 (0.067)  &        1 & 18.844 (0.019)  &        4 & 17.777 (0.004)  &        4 & 17.112 (0.009)  &        4 & 16.547 (0.011)  &        4 \\
    9 & 20.700 (0.153)  &        1 & 19.701 (0.012)  &        4 & 18.069 (0.014)  &        4 & 17.001 (0.005)  &        4 & 15.506 (0.021)  &        4 \\
   10 & 20.644 (0.146)  &        1 & 19.436 (0.019)  &        4 & 18.106 (0.007)  &        4 & 17.223 (0.011)  &        4 & 16.288 (0.014)  &        4 \\
   11 & 18.760 (0.041)  &        1 & 18.802 (0.035)  &        4 & 18.172 (0.009)  &        4 & 17.828 (0.012)  &        4 & 17.442 (0.022)  &        4 \\
   12 &  $-$            & $-$      & 19.641 (0.025)  &        4 & 18.325 (0.017)  &        4 & 17.482 (0.010)  &        4 & 16.612 (0.017)  &        4 \\
   13 & 20.302 (0.118)  &        1 & 19.659 (0.020)  &        4 & 18.346 (0.008)  &        4 & 17.493 (0.017)  &        4 & 16.729 (0.017)  &        4 \\
   14 & 18.990 (0.044)  &        1 & 19.031 (0.013)  &        4 & 18.347 (0.010)  &        4 & 17.952 (0.017)  &        4 & 17.549 (0.017)  &        4 \\
   15 &  $-$            & $-$      & 21.106 (0.050)  &        4 & 19.621 (0.013)  &        4 & 18.539 (0.020)  &        4 & 16.936 (0.011)  &        4 \\
\hline\noalign{\smallskip}
\multicolumn{11}{c}{SNF20080516-022}\\
\hline\noalign{\smallskip}
    1 & 16.508 (0.006)  &        2 & 16.486 (0.005)  &        4 & 15.844 (0.005)  &        4 & 15.468 (0.008)  &        3 &  $-$            & $-$      \\
    2 & 16.712 (0.002)  &        2 & 16.591 (0.009)  &        4 & 15.884 (0.003)  &        4 & 15.462 (0.007)  &        3 &  $-$            & $-$      \\
    3 & 17.385 (0.001)  &        2 & 17.370 (0.001)  &        4 & 16.780 (0.007)  &        4 & 16.425 (0.007)  &        4 & 16.081 (0.004)  &        4 \\
    4 & 19.424 (0.021)  &        2 & 18.286 (0.003)  &        4 & 17.122 (0.010)  &        4 & 16.388 (0.011)  &        4 & 15.775 (0.005)  &        4 \\
    5 & 18.680 (0.026)  &        2 & 18.456 (0.003)  &        4 & 17.706 (0.008)  &        4 & 17.274 (0.017)  &        4 & 16.884 (0.006)  &        4 \\
    6 & 18.836 (0.057)  &        2 & 19.086 (0.013)  &        4 & 18.582 (0.007)  &        4 & 18.274 (0.011)  &        4 & 17.947 (0.010)  &        4 \\
    7 & 20.297 (0.067)  &        2 & 19.775 (0.008)  &        4 & 18.914 (0.009)  &        4 & 18.367 (0.014)  &        4 & 17.883 (0.017)  &        4 \\
    8 & 19.459 (0.009)  &        2 & 19.758 (0.027)  &        4 & 19.328 (0.030)  &        4 & 19.045 (0.013)  &        4 & 18.781 (0.011)  &        4 \\
    9 & 21.549 (0.089)  &        2 & 20.678 (0.026)  &        4 & 19.608 (0.021)  &        4 & 18.929 (0.021)  &        4 & 18.339 (0.019)  &        4 \\
   10 &  $-$            & $-$      & 21.127 (0.023)  &        4 & 19.656 (0.017)  &        4 & 18.709 (0.016)  &        4 & 17.479 (0.011)  &        4 \\
   11 &  $-$            & $-$      & 21.659 (0.076)  &        4 & 20.318 (0.025)  &        4 & 19.373 (0.018)  &        4 & 18.103 (0.010)  &        4 \\
\hline\noalign{\smallskip}
\multicolumn{11}{c}{SNF20080517-010}\\
\hline\noalign{\smallskip}
    1 & 16.357 (0.062)  &        3 & 16.562 (0.009)  &        4 & 16.058 (0.021)  &        4 & 15.744 (0.009)  &        4 & 15.420 (0.022)  &        3 \\
    2 & 19.073 (0.008)  &        2 & 19.381 (0.023)  &        3 & 18.957 (0.025)  &        3 & 18.677 (0.007)  &        3 & 18.375 (0.026)  &        3 \\
    3 & 19.256 (0.038)  &        3 & 20.335 (0.025)  &        4 & 20.131 (0.012)  &        4 & 19.755 (0.022)  &        4 & 19.376 (0.017)  &        3 \\
    4 &  $-$            & $-$      & 21.247 (0.035)  &        3 & 20.150 (0.025)  &        3 & 19.466 (0.005)  &        3 & 18.874 (0.014)  &        3 \\
    5 &  $-$            & $-$      & 21.679 (0.114)  &        4 & 20.297 (0.026)  &        4 & 19.251 (0.027)  &        4 & 17.663 (0.025)  &        4 \\
    6 & 19.671 (0.039)  &        3 & 20.581 (0.023)  &        4 & 20.416 (0.039)  &        4 & 20.215 (0.035)  &        4 & 19.591 (0.062)  &        4 \\
    7 &  $-$            & $-$      & 22.328 (0.102)  &        3 & 20.957 (0.014)  &        3 & 19.984 (0.018)  &        3 & 18.869 (0.046)  &        3 \\
    8 & 22.416 (0.099)  &        2 & 22.312 (0.098)  &        3 & 20.987 (0.031)  &        3 & 20.103 (0.009)  &        3 & 18.957 (0.028)  &        3 \\
\hline\noalign{\smallskip}
\multicolumn{11}{c}{SNF20080522-000}\\
\hline\noalign{\smallskip}
    1 & 17.567 (0.005)  &        2 & 16.795 (0.015)  &        4 & 15.847 (0.005)  &        4 & 15.274 (0.023)  &        4 & 14.860 (0.096)  &        4 \\
    2 & 16.564 (0.004)  &        2 & 16.595 (0.016)  &        4 & 15.939 (0.010)  &        4 & 15.526 (0.012)  &        3 & 15.086 (0.018)  &        1 \\
    3 & 20.581 (0.059)  &        2 & 19.465 (0.017)  &        4 & 17.981 (0.006)  &        4 & 17.006 (0.007)  &        4 & 15.809 (0.007)  &        4 \\
    4 & 19.870 (0.027)  &        2 & 19.068 (0.011)  &        4 & 18.052 (0.009)  &        4 & 17.388 (0.001)  &        4 & 16.749 (0.013)  &        4 \\
    5 & 19.337 (0.033)  &        2 & 19.125 (0.021)  &        4 & 18.350 (0.010)  &        4 & 17.896 (0.004)  &        4 & 17.449 (0.009)  &        4 \\
    6 & 20.012 (0.098)  &        2 & 19.361 (0.017)  &        4 & 18.450 (0.006)  &        4 & 17.896 (0.014)  &        4 & 17.386 (0.013)  &        4 \\
    7 & 19.654 (0.015)  &        2 & 19.344 (0.034)  &        4 & 18.538 (0.017)  &        4 & 18.069 (0.016)  &        4 & 17.602 (0.012)  &        4 \\
    8 & 21.162 (0.193)  &        2 & 20.192 (0.052)  &        4 & 18.786 (0.014)  &        4 & 17.849 (0.009)  &        4 & 16.752 (0.017)  &        4 \\
    9 &  $-$            & $-$      & 21.984 (0.132)  &        4 & 20.516 (0.051)  &        4 & 19.428 (0.009)  &        4 & 17.775 (0.018)  &        4 \\
\hline\noalign{\smallskip}
\multicolumn{11}{c}{SNF20080522-001}\\
\hline\noalign{\smallskip}
    1 & 15.153 (0.021)  &        1 & 15.024 (0.003)  &        2 & 14.273 (0.001)  &        2 & 13.849 (0.019)  &        1 &  $-$            & $-$      \\
    2 & 15.493 (0.018)  &        2 & 15.434 (0.007)  &        3 & 14.674 (0.011)  &        4 & 14.296 (0.019)  &        1 & 13.886 (0.019)  &        1 \\
    3 & 16.138 (0.012)  &        2 & 15.898 (0.006)  &        3 & 15.111 (0.007)  &        4 & 14.684 (0.008)  &        4 & 14.249 (0.019)  &        1 \\
    4 & 16.268 (0.021)  &        2 & 16.219 (0.010)  &        3 & 15.543 (0.008)  &        4 & 15.189 (0.012)  &        4 & 14.810 (0.020)  &        3 \\
    5 & 18.513 (0.012)  &        2 & 17.384 (0.005)  &        3 & 16.181 (0.002)  &        4 & 15.399 (0.008)  &        4 & 14.611 (0.008)  &        2 \\
    6 & 17.136 (0.019)  &        2 & 16.962 (0.006)  &        3 & 16.227 (0.005)  &        4 & 15.804 (0.012)  &        4 & 15.363 (0.014)  &        4 \\
    7 & 17.728 (0.015)  &        2 & 17.513 (0.012)  &        3 & 16.799 (0.011)  &        4 & 16.397 (0.010)  &        4 & 16.001 (0.012)  &        4 \\
    8 & 18.604 (0.013)  &        2 & 17.796 (0.010)  &        3 & 16.809 (0.012)  &        4 & 16.197 (0.017)  &        4 & 15.654 (0.016)  &        4 \\
    9 & 17.869 (0.062)  &        2 & 18.076 (0.010)  &        3 & 17.562 (0.007)  &        4 & 17.243 (0.011)  &        4 & 16.905 (0.020)  &        4 \\
   10 & 19.175 (0.015)  &        2 & 18.664 (0.010)  &        3 & 17.781 (0.004)  &        4 & 17.225 (0.009)  &        4 & 16.750 (0.014)  &        4 \\
   11 & 18.369 (0.009)  &        2 & 18.439 (0.019)  &        3 & 17.830 (0.012)  &        4 & 17.486 (0.013)  &        4 & 17.119 (0.019)  &        4 \\
   12 & 20.674 (0.039)  &        2 & 19.418 (0.045)  &        3 & 18.094 (0.011)  &        4 & 17.232 (0.009)  &        4 & 16.388 (0.017)  &        4 \\
   13 &  $-$            & $-$      & 20.023 (0.088)  &        3 & 18.456 (0.011)  &        4 & 17.433 (0.019)  &        4 & 16.124 (0.024)  &        4 \\
   14 & 19.527 (0.159)  &        2 & 19.348 (0.018)  &        3 & 18.564 (0.015)  &        4 & 18.143 (0.025)  &        4 & 17.753 (0.029)  &        4 \\
   15 &  $-$            & $-$      & 19.975 (0.070)  &        3 & 18.607 (0.013)  &        4 & 17.694 (0.009)  &        4 & 16.726 (0.022)  &        4 \\
   16 & 18.700 (0.004)  &        2 & 19.107 (0.016)  &        3 & 18.610 (0.016)  &        4 & 18.334 (0.023)  &        4 & 18.005 (0.024)  &        4 \\
   17 & 18.874 (0.038)  &        2 & 19.049 (0.011)  &        3 & 18.754 (0.014)  &        4 & 18.613 (0.035)  &        4 & 18.380 (0.035)  &        4 \\
   18 & 21.040 (0.137)  &        2 & 20.288 (0.083)  &        3 & 18.806 (0.030)  &        4 & 17.882 (0.023)  &        4 & 16.926 (0.023)  &        4 \\
   19 & 19.239 (0.138)  &        2 & 19.498 (0.016)  &        3 & 19.018 (0.013)  &        4 & 18.708 (0.027)  &        4 & 18.269 (0.017)  &        3 \\
   20 & 21.209 (0.026)  &        2 & 20.207 (0.022)  &        3 & 19.055 (0.016)  &        4 & 18.282 (0.009)  &        4 & 17.569 (0.016)  &        4 \\
   21 &  $-$            & $-$      & 21.791 (0.109)  &        3 & 19.832 (0.018)  &        4 & 18.795 (0.015)  &        4 & 17.219 (0.016)  &        4 \\
\hline\noalign{\smallskip}
\multicolumn{11}{c}{SNF20080522-004}\\
\hline\noalign{\smallskip}
    1 & 16.881 (0.001)  &        2 & 16.539 (0.006)  &        4 & 15.707 (0.013)  &        4 & 15.216 (0.010)  &        4 & 14.790 (0.086)  &        4 \\
    2 & 18.114 (0.020)  &        2 & 16.996 (0.008)  &        4 & 15.867 (0.014)  &        4 & 15.181 (0.013)  &        3 & 14.607 (0.019)  &        1 \\
    3 & 17.332 (0.021)  &        2 & 17.437 (0.006)  &        4 & 16.804 (0.016)  &        4 & 16.447 (0.009)  &        4 & 16.044 (0.007)  &        4 \\
    4 & 17.313 (0.013)  &        2 & 17.577 (0.005)  &        4 & 17.044 (0.020)  &        4 & 16.708 (0.011)  &        4 & 16.328 (0.008)  &        4 \\
    5 & 17.874 (0.052)  &        2 & 17.911 (0.018)  &        4 & 17.302 (0.019)  &        4 & 16.951 (0.003)  &        4 & 16.579 (0.015)  &        4 \\
    6 & 18.565 (0.040)  &        2 & 18.648 (0.026)  &        4 & 18.066 (0.022)  &        4 & 17.716 (0.010)  &        4 & 17.321 (0.004)  &        4 \\
    7 & 18.768 (0.041)  &        2 & 18.809 (0.014)  &        3 & 18.096 (0.013)  &        4 & 17.703 (0.008)  &        4 & 17.318 (0.025)  &        4 \\
    8 & 18.497 (0.054)  &        2 & 18.758 (0.038)  &        4 & 18.151 (0.020)  &        4 & 17.782 (0.006)  &        4 & 17.354 (0.021)  &        4 \\
    9 & 19.764 (0.025)  &        2 & 19.278 (0.033)  &        4 & 18.444 (0.012)  &        4 & 17.915 (0.017)  &        4 & 17.405 (0.016)  &        4 \\
   10 & 18.925 (0.002)  &        2 & 19.175 (0.014)  &        4 & 18.668 (0.024)  &        4 & 18.326 (0.020)  &        4 & 17.939 (0.019)  &        4 \\
   11 & 19.639 (0.151)  &        2 & 19.427 (0.053)  &        4 & 18.687 (0.020)  &        4 & 18.219 (0.011)  &        4 & 17.753 (0.026)  &        4 \\
   12 & 18.793 (0.060)  &        2 & 19.114 (0.032)  &        3 & 18.694 (0.012)  &        4 & 18.431 (0.016)  &        4 & 18.175 (0.037)  &        4 \\
   13 & 19.789 (0.092)  &        2 & 19.515 (0.049)  &        4 & 18.737 (0.023)  &        4 & 18.297 (0.019)  &        4 & 17.797 (0.022)  &        4 \\
   14 & 19.862 (0.156)  &        2 & 19.605 (0.046)  &        4 & 18.878 (0.026)  &        4 & 18.424 (0.018)  &        4 & 17.945 (0.006)  &        3 \\
   15 & 21.046 (0.160)  &        2 & 20.208 (0.148)  &        4 & 19.237 (0.043)  &        4 & 18.578 (0.010)  &        4 & 17.982 (0.009)  &        4 \\
   16 &  $-$            & $-$      & 20.974 (0.077)  &        4 & 19.544 (0.027)  &        4 & 18.457 (0.019)  &        4 & 16.839 (0.005)  &        4 \\
   17 &  $-$            & $-$      & 21.699 (0.263)  &        3 & 20.220 (0.034)  &        4 & 19.218 (0.010)  &        4 & 17.857 (0.014)  &        4 \\
\hline\noalign{\smallskip}
\multicolumn{11}{c}{SNF20080522-011}\\
\hline\noalign{\smallskip}
    1 & 16.565 (0.018)  &        1 & 16.624 (0.023)  &        3 & 16.043 (0.017)  &        3 & 15.667 (0.015)  &        3 & 15.277 (0.014)  &        1 \\
    2 & 17.026 (0.019)  &        1 & 16.825 (0.011)  &        3 & 16.065 (0.018)  &        3 & 15.630 (0.017)  &        3 & 15.194 (0.014)  &        1 \\
    3 & 18.297 (0.033)  &        1 & 17.387 (0.006)  &        3 & 16.358 (0.006)  &        3 & 15.729 (0.010)  &        3 & 15.212 (0.014)  &        1 \\
    4 & 19.371 (0.063)  &        1 & 18.119 (0.112)  &        3 & 16.840 (0.020)  &        3 & 15.977 (0.026)  &        3 & 15.157 (0.014)  &        1 \\
    5 & 18.045 (0.027)  &        1 & 17.679 (0.015)  &        3 & 16.860 (0.006)  &        3 & 16.371 (0.014)  &        3 & 15.934 (0.017)  &        3 \\
    6 & 17.794 (0.025)  &        1 & 17.841 (0.024)  &        3 & 17.216 (0.016)  &        3 & 16.840 (0.014)  &        3 & 16.465 (0.026)  &        3 \\
    7 & 17.651 (0.025)  &        1 & 18.017 (0.025)  &        3 & 17.544 (0.019)  &        3 & 17.226 (0.009)  &        3 & 16.881 (0.028)  &        3 \\
    8 & 19.910 (0.102)  &        1 & 19.081 (0.047)  &        3 & 17.782 (0.002)  &        3 & 16.955 (0.005)  &        3 & 16.236 (0.012)  &        3 \\
    9 & 19.422 (0.070)  &        1 & 18.910 (0.011)  &        3 & 17.996 (0.010)  &        3 & 17.450 (0.016)  &        3 & 17.000 (0.025)  &        3 \\
   10 & 18.491 (0.035)  &        1 & 18.761 (0.014)  &        3 & 18.228 (0.018)  &        3 & 17.889 (0.006)  &        3 & 17.520 (0.024)  &        3 \\
   11 & 18.677 (0.041)  &        1 & 18.932 (0.035)  &        3 & 18.454 (0.006)  &        3 & 18.120 (0.008)  &        3 & 17.747 (0.039)  &        3 \\
   12 & 20.003 (0.101)  &        1 & 19.453 (0.012)  &        3 & 18.535 (0.020)  &        3 & 17.957 (0.009)  &        3 & 17.468 (0.030)  &        3 \\
   13 & 19.470 (0.067)  &        1 & 19.395 (0.040)  &        3 & 18.610 (0.005)  &        3 & 18.171 (0.008)  &        3 & 17.723 (0.020)  &        3 \\
   14 & 18.925 (0.052)  &        1 & 19.335 (0.031)  &        3 & 18.778 (0.009)  &        3 & 18.432 (0.014)  &        3 & 18.051 (0.029)  &        3 \\
   15 & 20.407 (0.161)  &        1 & 19.925 (0.037)  &        3 & 18.851 (0.014)  &        3 & 18.111 (0.008)  &        3 & 17.486 (0.022)  &        3 \\
   16 & 20.116 (0.135)  &        1 & 19.907 (0.017)  &        3 & 19.044 (0.031)  &        3 & 18.503 (0.018)  &        3 & 18.002 (0.027)  &        3 \\
   17 &  $-$            & $-$      & 20.108 (0.005)  &        3 & 19.057 (0.017)  &        3 & 18.376 (0.023)  &        3 & 17.779 (0.024)  &        3 \\
   18 & 20.185 (0.130)  &        1 & 19.977 (0.031)  &        3 & 19.101 (0.007)  &        3 & 18.590 (0.020)  &        3 & 18.120 (0.042)  &        3 \\
   19 & 20.489 (0.168)  &        1 & 20.081 (0.055)  &        3 & 19.131 (0.014)  &        3 & 18.529 (0.016)  &        3 & 17.924 (0.018)  &        3 \\
   20 &  $-$            & $-$      & 20.726 (0.065)  &        3 & 19.356 (0.016)  &        3 & 18.488 (0.013)  &        3 & 17.740 (0.030)  &        3 \\
   21 & 20.057 (0.123)  &        1 & 20.161 (0.018)  &        3 & 19.451 (0.020)  &        3 & 19.016 (0.024)  &        3 & 18.545 (0.037)  &        3 \\
   22 &  $-$            & $-$      & 21.437 (0.167)  &        3 & 19.971 (0.029)  &        3 & 18.935 (0.024)  &        3 & 17.584 (0.033)  &        3 \\
   23 &  $-$            & $-$      & 21.387 (0.055)  &        3 & 20.008 (0.029)  &        3 & 19.176 (0.018)  &        3 & 18.328 (0.011)  &        3 \\
\hline\noalign{\smallskip}
\multicolumn{11}{c}{SNF20080606-012}\\
\hline\noalign{\smallskip}
    1 & 16.230 (0.015)  &        1 & 16.028 (0.015)  &        1 & 15.289 (0.012)  &        1 &  $-$            & $-$      &  $-$            & $-$      \\
    2 & 16.481 (0.018)  &        2 & 16.509 (0.009)  &        2 & 15.877 (0.010)  &        2 & 15.521 (0.013)  &        1 &  $-$            & $-$      \\
    3 & 17.477 (0.044)  &        2 & 16.846 (0.013)  &        2 & 15.882 (0.008)  &        2 & 15.384 (0.013)  &        1 &  $-$            & $-$      \\
    4 & 17.292 (0.015)  &        2 & 17.233 (0.009)  &        2 & 16.536 (0.003)  &        2 & 16.170 (0.002)  &        2 & 15.733 (0.013)  &        1 \\
    5 & 17.480 (0.033)  &        2 & 17.412 (0.005)  &        2 & 16.748 (0.009)  &        2 & 16.384 (0.006)  &        2 & 15.973 (0.013)  &        1 \\
    6 & 19.395 (0.122)  &        2 & 18.137 (0.035)  &        2 & 16.773 (0.009)  &        2 & 15.920 (0.003)  &        2 &  $-$            & $-$      \\
    7 & 19.091 (0.057)  &        2 & 17.984 (0.014)  &        2 & 16.832 (0.008)  &        2 & 16.148 (0.004)  &        2 &  $-$            & $-$      \\
    8 & 17.900 (0.012)  &        2 & 17.822 (0.000)  &        2 & 17.173 (0.014)  &        2 & 16.801 (0.011)  &        2 & 16.382 (0.013)  &        1 \\
    9 & 19.780 (0.061)  &        2 & 18.568 (0.001)  &        2 & 17.304 (0.008)  &        2 & 16.480 (0.007)  &        2 & 15.721 (0.013)  &        1 \\
   10 & 18.034 (0.043)  &        2 & 17.977 (0.011)  &        2 & 17.313 (0.011)  &        2 & 16.951 (0.003)  &        2 & 16.525 (0.013)  &        1 \\
   11 & 18.081 (0.022)  &        2 & 18.049 (0.022)  &        2 & 17.360 (0.002)  &        2 & 16.964 (0.010)  &        2 & 16.514 (0.013)  &        1 \\
   12 & 18.081 (0.022)  &        2 & 18.049 (0.022)  &        2 & 17.360 (0.002)  &        2 & 16.964 (0.010)  &        2 & 16.514 (0.013)  &        1 \\
   13 & 18.342 (0.043)  &        2 & 18.226 (0.006)  &        2 & 17.504 (0.016)  &        2 & 17.128 (0.004)  &        2 & 16.706 (0.013)  &        1 \\
   14 & 19.782 (0.099)  &        2 & 18.699 (0.011)  &        2 & 17.654 (0.011)  &        2 & 17.025 (0.006)  &        2 & 16.425 (0.013)  &        1 \\
   15 & 18.405 (0.010)  &        2 & 18.454 (0.025)  &        2 & 17.801 (0.009)  &        2 & 17.422 (0.001)  &        2 & 16.995 (0.013)  &        1 \\
   16 & 19.097 (0.084)  &        2 & 18.716 (0.020)  &        2 & 17.880 (0.004)  &        2 & 17.392 (0.004)  &        2 & 16.882 (0.013)  &        1 \\
   17 & 18.733 (0.088)  &        2 & 18.596 (0.009)  &        2 & 17.887 (0.009)  &        2 & 17.484 (0.002)  &        2 & 17.053 (0.013)  &        1 \\
   18 & 18.779 (0.054)  &        2 & 18.739 (0.008)  &        2 & 18.072 (0.014)  &        2 & 17.707 (0.012)  &        2 & 17.290 (0.013)  &        1 \\
   19 & 18.804 (0.070)  &        2 & 18.842 (0.029)  &        2 & 18.201 (0.005)  &        2 & 17.834 (0.001)  &        2 & 17.426 (0.013)  &        1 \\
   20 & 18.802 (0.078)  &        2 & 18.896 (0.013)  &        2 & 18.261 (0.006)  &        2 & 17.913 (0.008)  &        2 & 17.475 (0.013)  &        1 \\
   21 & 19.810 (0.161)  &        2 & 19.290 (0.052)  &        2 & 18.377 (0.012)  &        2 & 17.849 (0.004)  &        2 & 17.345 (0.013)  &        1 \\
   22 & 18.918 (0.083)  &        2 & 18.998 (0.001)  &        2 & 18.386 (0.015)  &        2 & 18.028 (0.008)  &        2 & 17.625 (0.013)  &        1 \\
   23 & 18.902 (0.001)  &        2 & 19.146 (0.015)  &        2 & 18.663 (0.002)  &        2 & 18.367 (0.000)  &        2 & 17.977 (0.014)  &        1 \\
   24 & 19.064 (0.137)  &        2 & 19.246 (0.017)  &        2 & 18.718 (0.021)  &        2 & 18.399 (0.009)  &        2 & 17.989 (0.014)  &        1 \\
   25 & 18.947 (0.037)  &        2 & 19.233 (0.026)  &        2 & 18.722 (0.000)  &        2 & 18.402 (0.003)  &        2 & 17.974 (0.014)  &        1 \\
   26 & 20.340 (0.195)  &        2 & 19.769 (0.016)  &        2 & 18.838 (0.015)  &        2 & 18.299 (0.000)  &        2 & 17.721 (0.014)  &        1 \\
   27 & 21.102 (0.027)  &        2 & 20.042 (0.007)  &        2 & 18.883 (0.020)  &        2 & 18.103 (0.012)  &        2 & 17.387 (0.014)  &        1 \\
   28 &  $-$            & $-$      & 19.899 (0.021)  &        2 & 19.044 (0.033)  &        2 & 18.533 (0.001)  &        2 & 18.015 (0.014)  &        1 \\
   29 & 20.117 (0.058)  &        2 & 19.866 (0.003)  &        2 & 19.047 (0.003)  &        2 & 18.564 (0.007)  &        2 & 18.075 (0.014)  &        1 \\
   30 &  $-$            & $-$      & 20.015 (0.011)  &        2 & 19.097 (0.024)  &        2 & 18.611 (0.003)  &        2 & 18.079 (0.014)  &        1 \\
   31 & 19.506 (0.119)  &        2 & 19.712 (0.022)  &        2 & 19.160 (0.009)  &        2 & 18.808 (0.010)  &        2 & 18.411 (0.015)  &        1 \\
   32 & 19.473 (0.028)  &        2 & 19.727 (0.013)  &        2 & 19.174 (0.021)  &        2 & 18.841 (0.020)  &        2 & 18.431 (0.015)  &        1 \\
   33 &  $-$            & $-$      & 20.668 (0.035)  &        2 & 19.251 (0.007)  &        2 & 18.349 (0.004)  &        2 & 17.393 (0.014)  &        1 \\
   34 & 19.758 (0.012)  &        2 & 19.967 (0.052)  &        2 & 19.279 (0.018)  &        2 & 18.823 (0.008)  &        2 & 18.325 (0.015)  &        1 \\
   35 &  $-$            & $-$      & 20.053 (0.026)  &        2 & 19.355 (0.024)  &        2 & 18.895 (0.006)  &        2 & 18.384 (0.015)  &        1 \\
   36 & 19.604 (0.063)  &        2 & 19.936 (0.026)  &        2 & 19.388 (0.000)  &        2 & 19.056 (0.012)  &        2 & 18.691 (0.016)  &        1 \\
   37 &  $-$            & $-$      & 20.983 (0.037)  &        2 & 19.440 (0.009)  &        2 & 18.375 (0.001)  &        2 & 16.661 (0.015)  &        1 \\
   38 & 20.071 (0.160)  &        2 & 20.125 (0.007)  &        2 & 19.545 (0.014)  &        2 & 19.255 (0.032)  &        2 & 18.843 (0.017)  &        1 \\
   39 &  $-$            & $-$      & 21.407 (0.019)  &        2 & 19.666 (0.007)  &        2 & 18.516 (0.008)  &        2 & 16.701 (0.015)  &        1 \\
   40 &  $-$            & $-$      & 21.052 (0.047)  &        2 & 19.709 (0.029)  &        2 & 18.901 (0.020)  &        2 & 18.061 (0.015)  &        1 \\
   41 & 20.397 (0.099)  &        2 & 20.544 (0.081)  &        2 & 19.769 (0.003)  &        2 & 19.332 (0.022)  &        2 & 18.865 (0.017)  &        1 \\
   42 & 19.926 (0.079)  &        2 & 20.224 (0.018)  &        2 & 19.772 (0.010)  &        2 & 19.448 (0.013)  &        2 & 19.055 (0.018)  &        1 \\
   43 &  $-$            & $-$      & 21.079 (0.022)  &        2 & 19.799 (0.033)  &        2 & 18.992 (0.015)  &        2 & 18.224 (0.015)  &        1 \\
   44 &  $-$            & $-$      & 21.129 (0.081)  &        2 & 19.903 (0.006)  &        2 & 19.119 (0.001)  &        2 & 18.393 (0.016)  &        1 \\
   45 &  $-$            & $-$      & 21.561 (0.102)  &        2 & 19.992 (0.014)  &        2 & 19.008 (0.026)  &        2 & 17.680 (0.015)  &        1 \\
   46 &  $-$            & $-$      & 21.395 (0.056)  &        2 & 20.017 (0.008)  &        2 & 19.188 (0.000)  &        2 & 18.335 (0.016)  &        1 \\
   47 &  $-$            & $-$      & 20.554 (0.012)  &        2 & 20.039 (0.038)  &        2 & 19.746 (0.018)  &        2 & 19.324 (0.020)  &        1 \\
   48 &  $-$            & $-$      & 21.196 (0.027)  &        2 & 20.238 (0.005)  &        2 & 19.589 (0.004)  &        2 & 18.948 (0.018)  &        1 \\
   49 &  $-$            & $-$      & 21.767 (0.011)  &        2 & 20.489 (0.015)  &        2 & 19.617 (0.025)  &        2 & 18.769 (0.017)  &        1 \\
   50 & 22.084 (0.200)  &        2 & 22.137 (0.034)  &        2 & 20.587 (0.010)  &        2 & 19.597 (0.005)  &        2 & 18.128 (0.016)  &        1 \\
   51 &  $-$            & $-$      & 22.048 (0.051)  &        2 & 20.619 (0.004)  &        2 & 19.661 (0.015)  &        2 & 18.401 (0.016)  &        1 \\
   52 &  $-$            & $-$      & 21.813 (0.243)  &        2 & 20.692 (0.002)  &        2 & 19.993 (0.018)  &        2 & 19.276 (0.021)  &        1 \\
\hline\noalign{\smallskip}
\multicolumn{11}{c}{SNF20080610-003}\\
\hline\noalign{\smallskip}
    1 & 15.941 (0.014)  &        1 & 15.862 (0.014)  &        1 & 15.215 (0.012)  &        1 & 14.846 (0.012)  &        1 & 14.582 (0.012)  &        1 \\
    2 & 17.660 (0.022)  &        1 & 16.605 (0.017)  &        1 & 15.506 (0.012)  &        1 & 14.842 (0.013)  &        1 &  $-$            & $-$      \\
    3 & 16.781 (0.016)  &        1 & 16.330 (0.015)  &        1 & 15.551 (0.012)  &        1 & 15.112 (0.012)  &        1 &  $-$            & $-$      \\
    4 & 16.343 (0.014)  &        1 & 16.225 (0.014)  &        1 & 15.552 (0.012)  &        1 & 15.169 (0.012)  &        1 & 14.753 (0.012)  &        1 \\
    5 & 18.616 (0.030)  &        2 & 17.283 (0.018)  &        2 & 15.975 (0.007)  &        2 & 15.157 (0.013)  &        1 &  $-$            & $-$      \\
    6 & 16.493 (0.037)  &        2 & 16.560 (0.016)  &        2 & 16.002 (0.001)  &        2 & 15.685 (0.012)  &        1 & 15.327 (0.012)  &        1 \\
    7 & 16.549 (0.015)  &        1 & 16.698 (0.013)  &        1 & 16.161 (0.012)  &        1 & 15.817 (0.013)  &        1 & 15.438 (0.012)  &        1 \\
    8 & 16.910 (0.020)  &        2 & 16.946 (0.011)  &        2 & 16.324 (0.008)  &        2 & 15.973 (0.002)  &        2 & 15.608 (0.045)  &        2 \\
    9 & 17.444 (0.040)  &        2 & 17.210 (0.027)  &        2 & 16.460 (0.000)  &        2 & 16.065 (0.002)  &        2 & 15.643 (0.012)  &        1 \\
   10 & 16.742 (0.031)  &        2 & 17.022 (0.031)  &        2 & 16.549 (0.003)  &        2 & 16.261 (0.000)  &        2 & 15.906 (0.013)  &        2 \\
   11 & 17.815 (0.030)  &        2 & 17.476 (0.013)  &        2 & 16.704 (0.007)  &        2 & 16.274 (0.007)  &        2 & 15.870 (0.012)  &        1 \\
   12 & 17.559 (0.019)  &        2 & 17.474 (0.020)  &        2 & 16.781 (0.006)  &        2 & 16.383 (0.013)  &        2 & 15.939 (0.006)  &        2 \\
   13 & 17.987 (0.024)  &        2 & 18.151 (0.006)  &        2 & 17.495 (0.004)  &        2 & 17.086 (0.013)  &        2 & 16.610 (0.012)  &        2 \\
   14 & 17.992 (0.007)  &        2 & 18.166 (0.015)  &        2 & 17.632 (0.003)  &        2 & 17.318 (0.011)  &        2 & 16.953 (0.012)  &        2 \\
   15 & 18.471 (0.030)  &        2 & 18.349 (0.020)  &        2 & 17.658 (0.008)  &        2 & 17.278 (0.005)  &        2 & 16.872 (0.005)  &        2 \\
   16 & 20.765 (0.141)  &        2 & 19.453 (0.011)  &        2 & 18.046 (0.000)  &        2 & 17.170 (0.009)  &        2 & 16.267 (0.010)  &        2 \\
   17 & 20.008 (0.085)  &        2 & 19.169 (0.038)  &        2 & 18.173 (0.011)  &        2 & 17.599 (0.001)  &        2 & 17.090 (0.016)  &        2 \\
   18 & 19.297 (0.065)  &        2 & 19.033 (0.008)  &        2 & 18.275 (0.011)  &        2 & 17.848 (0.003)  &        2 & 17.423 (0.013)  &        2 \\
   19 & 20.263 (0.010)  &        2 & 19.369 (0.070)  &        2 & 18.322 (0.010)  &        2 & 17.637 (0.001)  &        2 & 17.036 (0.013)  &        2 \\
   20 & 19.827 (0.043)  &        2 & 19.348 (0.013)  &        2 & 18.493 (0.001)  &        2 & 18.019 (0.014)  &        2 & 17.553 (0.007)  &        2 \\
   21 &  $-$            & $-$      & 20.202 (0.075)  &        2 & 18.665 (0.020)  &        2 & 17.697 (0.010)  &        2 & 16.532 (0.023)  &        2 \\
   22 & 18.960 (0.005)  &        2 & 19.174 (0.056)  &        2 & 18.704 (0.003)  &        2 & 18.406 (0.016)  &        2 & 18.054 (0.025)  &        2 \\
   23 & 19.255 (0.016)  &        2 & 19.504 (0.016)  &        2 & 19.059 (0.006)  &        2 & 18.781 (0.001)  &        2 & 18.456 (0.008)  &        2 \\
   24 & 20.067 (0.051)  &        2 & 19.904 (0.017)  &        2 & 19.144 (0.006)  &        2 & 18.757 (0.029)  &        2 & 18.283 (0.002)  &        2 \\
   25 &  $-$            & $-$      & 20.663 (0.012)  &        2 & 19.160 (0.012)  &        2 & 18.187 (0.003)  &        2 & 16.992 (0.029)  &        2 \\
   26 & 21.732 (0.032)  &        2 & 20.695 (0.022)  &        2 & 19.174 (0.007)  &        2 & 18.188 (0.001)  &        2 & 16.892 (0.002)  &        2 \\
   27 &  $-$            & $-$      & 20.227 (0.004)  &        2 & 19.347 (0.007)  &        2 & 18.816 (0.009)  &        2 & 18.302 (0.008)  &        2 \\
   28 &  $-$            & $-$      & 20.608 (0.006)  &        2 & 19.347 (0.026)  &        2 & 18.548 (0.007)  &        2 & 17.791 (0.007)  &        2 \\
   29 & 19.510 (0.004)  &        2 & 19.815 (0.041)  &        2 & 19.349 (0.011)  &        2 & 19.026 (0.011)  &        2 & 18.682 (0.001)  &        2 \\
   30 &  $-$            & $-$      & 21.039 (0.027)  &        2 & 19.386 (0.006)  &        2 & 18.351 (0.001)  &        2 & 16.825 (0.012)  &        2 \\
   31 & 18.971 (0.034)  &        2 & 19.954 (0.020)  &        2 & 19.552 (0.010)  &        2 & 19.399 (0.007)  &        2 & 19.195 (0.017)  &        2 \\
   32 & 21.493 (0.188)  &        2 & 20.610 (0.043)  &        2 & 19.595 (0.013)  &        2 & 18.906 (0.006)  &        2 & 18.321 (0.027)  &        2 \\
   33 & 21.461 (0.194)  &        2 & 20.654 (0.069)  &        2 & 19.609 (0.007)  &        2 & 18.961 (0.013)  &        2 & 18.301 (0.009)  &        2 \\
   34 &  $-$            & $-$      & 21.073 (0.057)  &        2 & 19.644 (0.014)  &        2 & 18.707 (0.002)  &        2 & 17.530 (0.026)  &        2 \\
   35 & 20.202 (0.148)  &        2 & 20.334 (0.048)  &        2 & 19.658 (0.001)  &        2 & 19.268 (0.046)  &        2 & 18.798 (0.015)  &        2 \\
   36 &  $-$            & $-$      & 20.694 (0.005)  &        2 & 19.667 (0.015)  &        2 & 19.036 (0.003)  &        2 & 18.472 (0.032)  &        2 \\
   37 &  $-$            & $-$      & 21.194 (0.052)  &        2 & 19.720 (0.021)  &        2 & 18.705 (0.014)  &        2 & 17.205 (0.005)  &        2 \\
   38 &  $-$            & $-$      & 21.016 (0.000)  &        2 & 19.906 (0.001)  &        2 & 19.162 (0.017)  &        2 & 18.502 (0.013)  &        2 \\
   39 &  $-$            & $-$      & 21.765 (0.076)  &        2 & 20.178 (0.030)  &        2 & 19.167 (0.008)  &        2 & 17.691 (0.018)  &        2 \\
   40 & 20.458 (0.038)  &        2 & 20.739 (0.054)  &        2 & 20.264 (0.009)  &        2 & 19.973 (0.006)  &        2 & 19.649 (0.006)  &        2 \\
   41 &  $-$            & $-$      & 22.000 (0.049)  &        2 & 20.326 (0.008)  &        2 & 19.349 (0.003)  &        2 & 17.902 (0.011)  &        2 \\
   42 &  $-$            & $-$      & 21.666 (0.021)  &        2 & 20.369 (0.007)  &        2 & 19.560 (0.039)  &        2 & 18.833 (0.003)  &        2 \\
   43 &  $-$            & $-$      & 22.120 (0.004)  &        2 & 20.610 (0.008)  &        2 & 19.695 (0.023)  &        2 & 18.563 (0.008)  &        2 \\
\end{longtable}
}

\section{Second-order atmospheric extinction}

The second-order extinction term in $U$ and $B$ bands arises from the fact that 
the transmission of Earth's atmosphere is not constant but decreases rapidly at wavelengths below 5000\AA. The 
change of the transmission across the $U$ and $B$ bands is large enough to cause stars with different 
spectral energy distributions (SED) to experience slightly different extinction. Thus, the atmospheric extinction 
in $U$ and $B$ bands is not constant but it is a function of the object SED. It can be parametrized as:
\begin{eqnarray}
k_B &=& k'_B+k''_B\,(B-V) \\
k_U &=& k'_U+k''_U\,(U-B),
\label{eq:ext}
\end{eqnarray}
where $k'_{U,B}$ and $k''_{U,B}$ 
are the first-order and the second-order extinction. The second-order extinction can be neglected in $VRI$. Unfortunately, in the majority of the photometric works on SNe Ia the second-order extinction term has been neglected.
Here we discuss the uncertainties that are caused by this. 

The usual practice in SN photometry is to calibrate the magnitudes of the stars in the field of the
SN in several photometric nights. This is usually done by deriving the parameters in Eqs.~\ref{eq:tran}
by observing \cite{landolt92,2009AJ....137.4186L} standard fields that contain both blue and red stars
 in photometric conditions, and applying the equations to the field stars. Then in nonphotometric nights the SN magnitudes are calibrated using the field stars as tertiary standards.  Simple rearrangement of Eqs.~\ref{eq:tran},  taking $B$ band as an example gives:
\begin{equation}
B=b+(ct_B-k''_B\,X)\,(B-V)-k'_B\,X+zp_B,
\label{eq:tran1}
\end{equation}
If $k''_B$ is not fitted (or fixed to a reasonable value $\sim-0.03$), one would
not measure the true color term $ct_B$ but rather $ct_B-k''_B\,X$, meaning that the measured color-term will depend on the 
airmass of the observations. The second-order extinction  in $U$ and $B$ has been measured and it is about $-0.03$. 
Using the spectrophotometric atlas of \citet{specphot}, various $U$ and $B$ filters, and atmospheric transmission measured at different 
observatories we estimated $k''_B$ and $k''_U$ by means of synthetic photometry to be $\sim-0.03$ as well.
Therefore, two standard fields observed airmass $X=1$ and 2 will give color terms 
that differ by $\sim0.03$.

Another problem arises because the $B-V$ color index of SNe Ia changes from $\sim-0.1$ around maximum light to 1.3 about a month 
later. If a SN is observed at airmass 1.5 the correction for the second-order extinction at maximum and at +30 days will be 
$\sim0.0$ and 0.058 mag, respectively. When the second-order extinction term is not included in the calibration, 
this will not be corrected for. While around maximum the effect will be negligible, one to two months after maximum the $B$ magnitudes maybe 
as much as 0.06 mag in error. This will directly affect the $B-V$ color index, which is often used to estimate the host galaxy extinction. Therefore, the estimation of the peak magnitudes can be affected by as much as 0.18 mag for Milky Way type dust with $R_V=3.1$.

\section{S-corrections}

\subsection{Photometric systems responses}

To recover the effective system responses we follow the methods described in \cite{stan_03du}.
The relevant information was either taken from \cite{stan_03du} or from the corresponding 
observatories. Whenever possible measured filter transmission and CCD QEs were used; if unavailable, specifications 
from the manufacturer were used.
The total system responses were computed by multiplying the filter
transmissions by (a) the CCD quantum efficiency (QE), (b) the
reflectivity of at least two aluminum surfaces, (a) a telluric
absorption spectrum at airmass 1.3.  The lens and window transmissions
were not included because this information was unavailable. 
We note that the transmission of the Earth atmosphere was not included in the calculation of the 
effective $U$ and $B$ bands. The reason is that when computing the instrument color terms the second-order extinction 
has been taken into account. The effect of the atmospheric transmission on the $U$ and $B$ bands is to shift the bands effective 
wavelength redward with increasing the airmass of the observations. Taking into account the  second-order extinction meant that 
the computed color terms correspond to observations above the atmosphere.

Synthetic magnitudes were calculated from \citet{specphot}
spectrophotometry of Landolt standard stars,
\begin{equation}
m^\mathrm{syn}=-2.5\log\left(\int\,f_\lambda^\mathrm{phot}(\lambda)R^\mathrm{nat}(\lambda)d\lambda\right).
\label{eq:synthphot}
\end{equation}
The difference between the synthetic and the observed photometry was fitted as a function of the 
observed color indices to compute synthetic color-terms ($ct^\mathrm{syn}$), for example, for $B$ we have
\begin{equation}
B^\mathrm{std}-B^\mathrm{syn}=ct^\mathrm{syn}(B^\mathrm{std}-V^\mathrm{std})+zp^{R^\mathrm{nat}}.
\label{eq:synthcol}
\end{equation}
For the $VRI$-bands, the $ct^\mathrm{syn}$'s were close to the observed
ones $ct^\mathrm{obs}$. In some cases small differences exceeding the uncertainty were
accounted for by shifting the filter transmissions until $ct^\mathrm{syn}$
matched $ct^\mathrm{osb}$. $zp^{R^\mathrm{nat}}$ is the ZP for synthetic photometry with the band 
$R^\mathrm{nat}$.

As in \citet{stan_03du} we found that the
synthetic $U$ and $B$ bands were always too blue. This is, to some
extent, to be expected because the neglected optical elements such as lenses or windows, anti-reflection and other coatings will tend to reduce the
system sensitivity shortward of $\sim$4000\,\AA.  To account for the net effect of these 
uncertainties we modified the $U$ and $B$ bands by multiplying them with a  Sigmoid function:
\begin{equation}
F(\lambda;\lambda_0,\Delta)=\frac{1}{1+\exp(-(\lambda-\lambda_0)/\Delta)}.
\label{eq:sig}
\end{equation}
The parameters  $\lambda_0$ and $\Delta$  were selected so that $ct^\mathrm{syn}$ matched
$ct^\mathrm{obs}$ for both $U$ and $B$. 

Following \citet{stan_03du} we also rederived the standard Johnson-Cousins system responses with the
revised filter functions of \citet{2012PASP..124..140B}. This procedure also gave the ZPs
$zp^{R^\mathrm{std}}$ for synthetic photometry with the standard system responses 
$R^\mathrm{std}$.

\subsection{Computing the S-corrections}

The procedure that we used to calculate the $S$-corrections was the same as that in \citet{stan_03du}, but 
slightly modified to account for the fact that the $U$ and $B$ bands did not include the atmospheric 
transmission.
The calibration correction from natural magnitudes at airmass $X$, $m^\mathrm{nat}_X$, to standard magnitudes $m^\mathrm{std}$ is computed by 
means of synthetic photometry
\begin{equation}
m^\mathrm{std}  =  m^\mathrm{nat} + S,
\label{eq:corr}
\end{equation}
where $S$ is the $S$-correction:
\begin{eqnarray}
S  &= -2.5\log\left(\int\,f_\lambda^\mathrm{phot}(\lambda)R^\mathrm{std}(\lambda)d\lambda\right)  + zp^{R^\mathrm{std}} \nonumber\\
 & +2.5\log\left(\int\,f_\lambda^\mathrm{phot}(\lambda)R^\mathrm{nat}(\lambda)p(\lambda)^X\,d\lambda\right) - zp_X^{R^\mathrm{nat}}.
\label{eq:scorra}
\end{eqnarray}
Here  $f_\lambda^\mathrm{phot}(\lambda)$ is the photon spectral energy distributions (SED) of the SN,
$R^\mathrm{nat}(\lambda)$ and  $R^\mathrm{std}(\lambda)$ are the responses of the natural and the standard systems, 
respectively, and  $p(\lambda)$ is the transmission of the atmosphere at airmass $X=1$ above the site of observation.
Here $zp_X^{R^\mathrm{nat}}$ is the ZP obtained by fitting Eq.~\ref{eq:synthcol} to the synthetic magnitudes $m_X^\mathrm{syn}$ computed with 
the natural passband multiplied by the Earth's atmosphere transmission at airmass $X$, for example
\begin{equation}
m_X^\mathrm{syn}=-2.5\log\left(\int\,f_\lambda^\mathrm{phot}(\lambda)R^\mathrm{nat}(\lambda)p(\lambda)^X\,d\lambda\right).
\label{eq:synthphot1}
\end{equation}
Comparing Eq.~\ref{eq:tran1} and Eq.~\ref{eq:corr} we see that  the term $ct_B(B-V)-k''_B\,X(B-V)$ is simply an approximation
of the $S$-correction:
\begin{equation}
S\simeq ct_B(B-V)-k''_B\,X(B-V).
\end{equation}
By the way the ZPs in Eq.~\ref{eq:scorra} are derived, $S\simeq0.0$ for A0\,V
stars with all color indexes zero, for example, Vega. This ensures that for normal stars the 
synthetic $S$-correction will work the same way as the linear color-term corrections.

The best way to compute the $S$-corrections for a SN is to use spectrophotometry of the same object. If that is not possible the mean 
spectrum of SNe Ia should be used. In our work we used \cite{2007ApJ...663.1187H} spectral template, corrected for the Milky Way 
reddening and the redshift of each SN. 

\section{$s_\mathrm{opt}-\Delta M_{15}$ relation}

\label{ap:sdm15}

\begin{figure}[!t]
\includegraphics*[width=8.8cm]{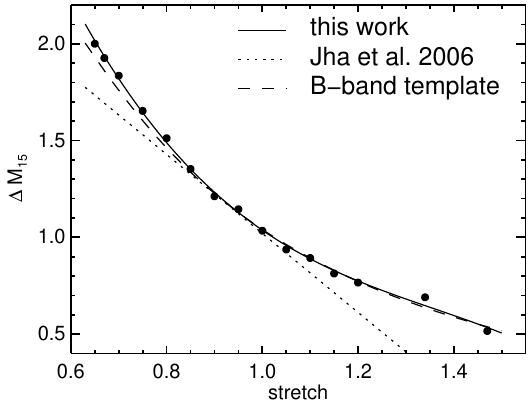}
\caption{$s_\mathrm{opt}-\Delta M_{15}$ relation from this work. }
\label{f:s_dm15}
\end{figure}

To derive the new $s_\mathrm{opt}-\Delta M_{15}$ relation for normal SNe~Ia a subsample of $\sim200$ SNe with well-sampled LCs and observations within two days from $t_{B_\mathrm{max}}$ was used. The observed $B$-band LC were corrected to rest-frame $B$ and normalized to zero peak magnitude. The LCs were divided into 0.05 stretch bins and let $s_\mathrm{c}$ be the stretch in the middle of a given bin. The time axes of the LCs that fall into given bin were converted to phases, corrected to the stretch in the middle of the bin by 
\begin{equation}
\phi=s_\mathrm{c}\,\frac{t-t_{B_\mathrm{max}}}{s_\mathrm{opt}\,(1+z)},
\label{eq:ph:ap}
\end{equation}
 and stacked together. Smoothing spline was then used to estimate the time maximum and the $\Delta M_{15}$ parameter. The resulting $s_\mathrm{opt}-\Delta M_{15}$  pairs are shown in Fig.~\ref{f:s_dm15} together with the the third-order polynomial fit (the solid line).  It is seen that our relation is close to the one derived by simply stretching the $B$-band LC template, indicating that the stretched template describes the LC of normal SNe~Ia rather well. On the other hand the linear relation derived by \cite{2006AJ....131..527J}, which has been used by many authors, is applicable only in the limited stretch range 0.85--1.05. The parameters of the third-order polynomials to convert from $s_\mathrm{opt}$ to $\Delta M_{15}$ and vice versa are given in Table~\ref{t:s-dm15}.

\begin{table}[!t]
\centering
\caption{Parameters of the third-order polynomials to convert from stretch to $\Delta M_{15}$ and vice versa.} 
\label{t:s-dm15}
\begin{tabular}{l|c|c}
\hline
\hline
 Parameter &    $s\rightarrow \Delta M_{15}$    &  $\Delta M_{15}\rightarrow s$ \\
\hline
$c_0$      & 7.3078781    & 2.4279136 \\
$c_1$      & $-$13.043647 & $-$2.3871091\\
$c_2$      & 8.9733595    & 1.2189403 \\
$c_3$      & $-$2.2005882 & $-$0.23502774\\
\hline
RMS        & 0.019          & 0.018         \\
\hline
\end{tabular}
\end{table}

\section{NIR $K$-corrections}

\label{sec:kcor}

To compute the NIR $K$-corrections we used spectrophotometry of nearby SNe Ia that also had NIR photometry  
available (Table~\ref{t:kcor}). For the majority of the available spectra the $J$,$H$ and $K$ parts were obtained separately with different, nonoverlapping
instrument settings. While the relative flux calibration of each individual part is probably accurate enough, the cross-band flux calibration may not be. Besides, the flux across the strong telluric absorption bands at $\sim1.38\,\mu$m and $\sim1.88\,\mu$m was not available. Therefore, before computing the $K$-corrections the spectra had to be "prepared". The following procedure was applied.

The noisy parts at the edges of the telluric bands were removed and the flux across the bands was linearly interpolated. Next, 
$JHK$ synthetic magnitudes were computed with the appropriate effective system responses\footnote{The effective system responses were computed by multiplying the filter transmission, the detector quantum efficiency (QE), two aluminum reflections and high-resolution atmosphere transmission model for airmass 1.25 computed with ATRAN software \citep{atran} for mid-altitude observatory with representative precipitable water vapor. When available the filter transmissions and the detector QEs were obtained form the instrument web pages, if not, generic filter transmissions and QEs appropriate for the used instrument were used. For accurate integration of the narrow features in the atmosphere transmission, the responses were sampled at 1~\AA.} 
and were compared with the observed ones.  When necessary the $J$, $H,$ and $K$ parts were scaled, the interpolation across the telluric bands was repeated and new synthetic photometry was computed\footnote{This is needed because in some cases the filters extend over the interpolated parts of the spectrum. As the $J$, $H,$ and $K$ parts were scaled, the interpolation changes, which can also affect the computed synthetic magnitudes to some extent.}. The procedure was iterated until  the synthetic and the observed colors matched. In the cases when the whole NIR spectral range was observed simultaneously we always found a good match between the synthetic and the observed colors, and only interpolation across the telluric bands was necessary. The spectra were corrected for dust reddening in the Milky Way and the SN host galaxies using $E(B-V)$ values from the 
Galactic dust map \citep{ebv} and from the corresponding photometric papers, and corrected to rest-frame wavelengths. The spectra at similar phases were plotted together and examined. Several spectra with poor S/N, inadequate interpolation across the telluric bands or other problems most likely related to the flux calibration were identified and discarded. Details of the final set of spectra are given in Table~\ref{t:kcor}. We note that a few spectra do not cover all $JHK$ bands. In addition, a few spectra had issues in one band, but were of sufficient quality in the others.

\begin{figure}[!t]
\includegraphics*[width=8.8cm]{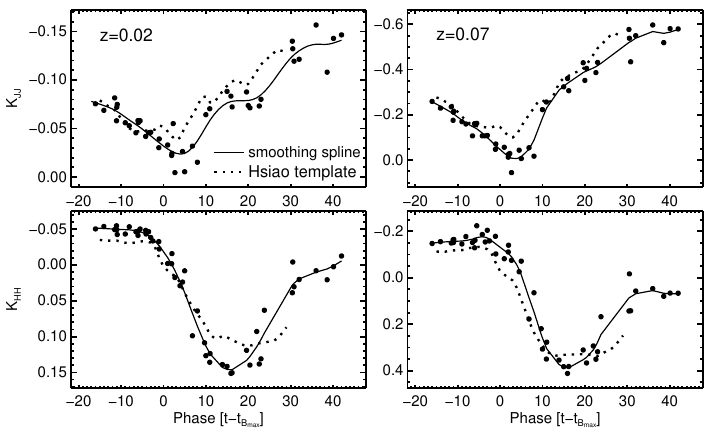}
\caption{Example of computing the NIR $J$ and $H$ $K$-corrections at $z=0.02$ (left panels) and $z=0.07$ (right panels). }
\label{f:kcorr}
\end{figure}

To compute $K$-corrections at any SN phase smoothing splines were used to interpolate the $K$-corrections computed from the individual spectra. The  smoothing splines have great flexibility to describe the data to any pre-defined rms around the fit. We have computed $J-J$ and $H-H$ $K$-corrections for several redshifts between $z=0.002$ and $z=0.2$ and visually examined the interpolation to tabulate the 
optimal smoothing parameter as a function of the redshift. This is a somehow subjective procedure and we tried to select the smoothing parameters in such a way that the data was well-fitted, but with special care to avoid over-fitting.  Examples of the outcome of this procedure are shown in Fig.~\ref{f:kcorr} for two redshifts of $z=0.02$ and $z=0.07$. The scatter around the smooth curves increases with the redshift and also depends on partucular filters use (see below). The scatter at redshift $z=0.02$ is $\sim0.01$ mag in all bands, at $z=0.08$ it increases to $\sim0.033$ and 0.04 mag, for $J$ and $H$, respectively, and at $z=0.14$ the $J$ and $H$ scatter is $\sim0.05$ and 0.08 mag. These values and the observed increase with the redshift are similar to the results of \cite{2014PASP..126..324B}. We note however, that we did not color-correct the spectra to follow single color curve as in 
 \cite{2007ApJ...663.1187H} and \cite{2014PASP..126..324B}. As a comparison in Fig.~\ref{f:kcorr} are also shown the  $K$-corrections computed from  \cite{2007ApJ...663.1187H}  template. One can see that there are considerable differences for some SN phases, including before +10 days, which is of interest for our study. \cite{2014PASP..126..324B} argue that the dominating part of the scatter is due to intrinsic differences in the energy distribution of different SNe. Thus, the scatter of the individual $K$-corrections around the smoothing spline function between phases $-$10 and +10 days was tabulated as a function of redshift and added in quadrature to the uncertainties of the peak magnitudes.

\begin{figure}[!t]
\includegraphics*[width=8.8cm]{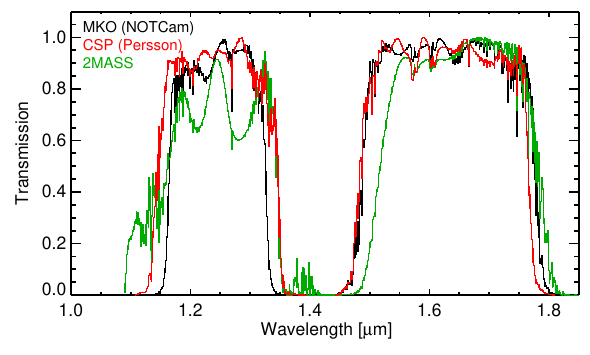}
\caption{$J$ and $H$ filters responses used to calculate the NIR $K$-corrections. }
\label{f:filt_nir}
\end{figure}

There is no standard photometric system in the NIR, and often slightly different filter sets are employed by different observatories. In our analysis we decided to apply $K$-corrections from the different observed filter systems to a single $JHKs$ photometric system, which we chose to be the one defined by the  Mauna Kea Observatories Near-Infrared Filter Set (MKO) \citep{irfilt}. As a practical realization of the MKO system we selected the filter and detector  combination used by the NOTCam instrument at the NOT. The observed filter systems that were used in this work are 
\begin{itemize}
 
\item[(i)] the MKO filters for the 16 new SNe from this work and the SNe from \citet{2012MNRAS.425.1007B} (BN12 set). 
 
\item[(ii)] 2MASS \citep{2003AJ....126.1090C} for the CfA data set \citep{2008ApJ...689..377W}

\item[(iii)] the filters used by The Carnegie Supernova Project (CSP),  which are similar to \cite{persson98}. The CSP filters were used for the CSP data set \citep{2010AJ....139..519C,2011AJ....142..156S}, the data set from K. Krisciunas and collaborators \citep{kri_04,kri_ir_temp,kri_01el,kri_01,kri_00,2007AJ....133...58K} (the Kri set) and the following SNe from various sources (the Var set)  SN~1998bu \citep{her98bu,1998IAUC.6907....2M,1999ApJS..125...73J}, SN~1999ee \citep{mario99ee}, SN~2000E  \citep{2003ApJ...595..779V},  SN~2002dj \citep{2008MNRAS.388..971P}, SN~2003cg \citep{nancy06}, SN~2003du \citep{stan_03du}, SN~2003hv \citep{2009A&A...505..265L}, SN~2004eo \citep{2007MNRAS.377.1531P}, SN~2005cf \citep{2009ApJ...697..380W,2007MNRAS.376.1301P}, and SN~2008Q (Stanishev et al., in preparation).
 
\item[(iv)] the filters of WHIRC instrument were used for SN~2011fe \citep{2012ApJ...754...19M}.
 
\end{itemize} 
and the employed MKO, 2MASS and CSP filter responses are shown in Fig.~\ref{f:filt_nir}. $K$-corrections from MKO to MKO, CSP and 2MASS for several redshifts and phases between $-16$ and $+40$ days are also provided in electronic form.

Strictly speaking the above procedure will work accurately only if the observations are obtained with filters closely resembling those of the photometric system to which the photometry is tied. 
This is the case only for the CSP, CfA and BN12 data sets, and to a large extent for the Kri data set. The observations of the other SNe were obtained with filters that differ from those of the photometric systems to which the photometry was calibrated.  Most of our new photometry is tied to 2MASS system. The rest of the SN photometry is tied to \cite{persson98} or \citet{irstan} systems, which are similar to each other. Because of this mismatch, it is expected that the photometry will contain small systematic uncertainties. To estimate the magnitude of this uncertainty we computed synthetic photometry with the filters of the four photometric systems considered in this work at two epochs from the Hsiao spectral template -- at NIR maxima and a week after $t_{B_\mathrm{max}}$. We found that the errors should be smaller than 0.02 mag. The only exception was $H$ band a  week after $t_{B_\mathrm{max}}$. In this case the error could be as big as 0.06 mag for most SNe and only for WHIRC observations of SN~2011fe it could reach 0.1 mag.

\begin{table}[!t]

\caption{NIR spectra of SNe Ia used for computing $K$-corrections. The phases of the spectra are
relative to the time of the $B$ band maximum light. The sources of the NIR photometry are given in the text.} 
\label{t:kcor}
\begin{tabular}{@{}lccr@{}}
\hline
\hline\noalign{\smallskip}
SN      &  Phase & $s_\mathrm{opt}$ & Ref.  \\
\hline\noalign{\smallskip}
SN 1998bu & $-$3,15,22,36                                   & 0.973     & 1,2 \\
SN 1999ee & $-$9,0,2,5,8,15,20,23,32,42                     & 1.084     & 2   \\
SN 2000ca &  2.7,9.7,17.7,22.6,30.8,38.6                    & 1.065     & 3   \\
SN 2002bo & 11                                              & 0.928     & 4   \\
SN 2002dj & $-$5.9                                          & 0.948     & 5   \\
SN 2003cg & $-$6.5,$-$4.1,$-$3.5,4,14,19.6                  & 0.937     & 6   \\
$\cdots$ &  24,30.5,40                                      &           & \\
SN 2003du & $-$11,$-$11.5,$-$5.5,$-$2.5,2.3,3.4             & 1.018     & 7   \\
$\cdots$ &  4.5,12.4,16.2,20.4,30.4                         &           & \\
SN 2004S  & 15                                              & 0.992     & 8   \\
SN 2005cf & $-$11,$-$9,$-$6,$-$1,1,10,32,42                 & 0.979     & 9   \\
SN 2011fe & $-$16,$-$14,$-$11,$-$8,$-$4,$-$1,2,7,11,16      & 0.905     & 10 \\
\hline 
\end{tabular}
\tablebib{(1)~\citet{1999ApJS..125...73J}; (2)~\citet{mario99ee}; (3)~Stanishev et al., in prep.; (4)~\citet{02bo}; (5)~\citet{2008MNRAS.388..971P}; (6)~\citet{nancy06};  (7)~\citet{stan_03du}; (8)~\citet{2012MNRAS.427..994G};  (9)~\citet{2007MNRAS.376.1301P};  (10)~\citet{2013ApJ...766...72H}.
}
\end{table}

\section{$(B-V)$ color at $B$ band maximum}
\label{sec:colbv0}

\citet{phil99} were the first to suggest that the intrinsic $B-V$ color of SNe~Ia at maximum $(B-V)_{B_{\mathrm{max},0}}$ is a linear function of $\Delta M_{15}$ with a slope of $\sim0.11$. The parameters of the $(B-V)_{B_\mathrm{max},0}- \Delta M_{15}$ relation were determined by linear fit of the bluest objects \citep[see also][]{2010AJ....139..120F}. The disadvantage of this approach is that it requires a prior knowledge of which SNe are likely to be unreddened and that only a small fraction of all available data is used. Here we followed a different path to study the distribution of observed $(B-V)_{B_\mathrm{max}}$ colors  with the much larger sample available today. \cite{nobili08} have shown that the intrinsic SN~Ia $(B-V)_{B_\mathrm{max}}$ color has a dispersion of about 0.05 mag. The observed colors may be further reddened by interstellar dust along the line of sight, evidence for which is clearly seen as an extended red tail of the color distribution in the nearby SN~Ia sample. Let us assume as in \cite{2007ApJ...659..122J}  that (i) the intrinsic $(B-V)_{B_\mathrm{max}}$ colors follow normal distribution with mean $c_0$ and standard deviation $\sigma_\mathrm{c}$, and (ii) that the reddening $E(B-V)$ follows exponential distribution with scale $\tau_\mathrm{e}$ for $E(B-V)\geq0$ mag and is zero otherwise. Under these assumptions the probability density function (PDF) of the observed $(B-V)_{B_\mathrm{max}}$ color, $\rho((B-V)_{B_\mathrm{max}};c_0,\sigma_\mathrm{c},\tau_\mathrm{e})$, is the convolution of the two functions \citep{2007ApJ...659..122J,2011ApJ...740...72M}:\begin{equation}
\rho(c;c_0,\sigma_\mathrm{c},\tau_\mathrm{e})\propto\int_{0}^{+\infty}\exp\left(-\frac{(c-c_0-\tau)^2}{2\sigma_\mathrm{c}^2}\right)\exp\left(-\frac{\tau}{\tau_\mathrm{e}}\right)d\tau,
\label{eq:colprob}
\end{equation}
where $c$ is the observed $(B-V)_{B_\mathrm{max}}$ color and the integration is over the color excess from extinction, $\tau$.

\begin{figure}
\includegraphics*[width=8.8cm]{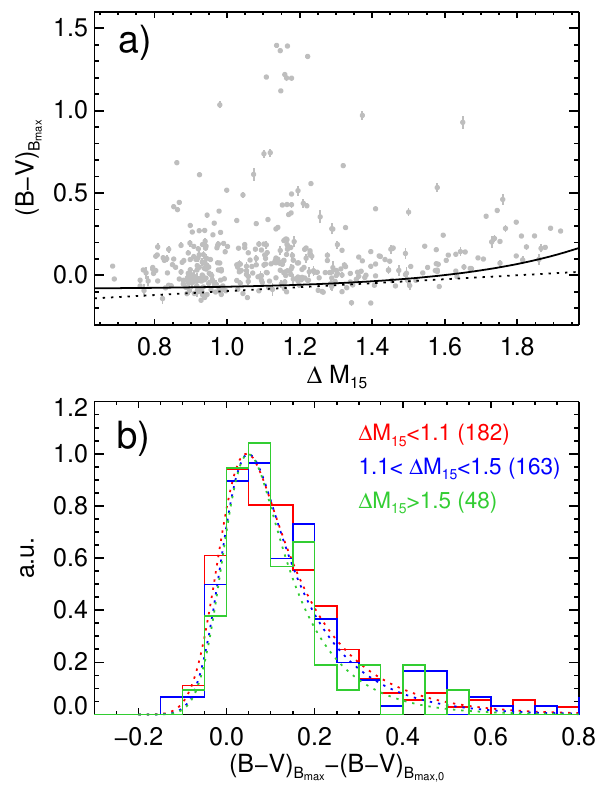}
\caption{(a) $(B-V)_{B_\mathrm{max}}$ color of the SNe used in this work as function of $\Delta M_{15}$. The solid line shows the relation defined by Eq.~\ref{eq:bv0}. The dashed line has a slope of 0.12 and is shown for a comparison; (b) Histograms of $(B-V)_{B_\mathrm{max}}-(B-V)_{B_\mathrm{max},0}$ for SNe with $\Delta M_{15}<1.1$ (182 objects), $1.1<\Delta M_{15}<1.3$ (163 objects) and $\Delta M_{15}>1.5$ (48 objects) with the corresponding fits with Eq.~\ref{eq:colprob}.}
\label{f:bv_max}
\end{figure}

Figure~\ref{f:bv_max}a shows the $(B-V)_{B_\mathrm{max}}$ colors measured with uncertainty less than 0.04 mag from 393 LCs. The dotted line has a slope of 0.12. It describes the blue edge of the distribution of the SNe with $\Delta M_{15}\leq1.4$ well, but beyond that the colors are clearly increasingly redder than the line. It is clear that a nonlinear function is need to describe the average intrinsic $(B-V)_{B_\mathrm{max}}$ color as a function of $\Delta M_{15}$. We tried different functions and found that a simple exponential function worked best, while a second-order polynomial was not flexible enough to describe the data. 

The parameters of the exponential function that describes $(B-V)_{B_\mathrm{max},0}$ were determined by fitting Eq.~\ref{eq:colprob} to the histogram of $E(B-V)_\mathrm{obs}=(B-V)_{B_\mathrm{max}}-(B-V)_{B_\mathrm{max},0}$ computed in 0.05 mag wide bins. Unfortunately, there is no unique way to define the quantity to be minimized. We tried several approaches, which are appropriate under the reasonable assumption that the intrinsic scatter $\sigma_\mathrm{c}$ is independent on $\Delta M_{15}$. First the histogram of the whole data set was fitted and the parameters that give the lowest $\sigma_\mathrm{c}$ was selected. The reasoning behind this is that with incorrect parameters the histogram will be additionally blurred and hence a larger $\sigma_\mathrm{c}$ will be derived; the correct parameters should thus give the lowest $\sigma_\mathrm{c}$. However, this approach was unsuccessful and the blue edge of the distribution of the SNe with $\Delta M_{15}\geq1.4$ was not described well. The likely reason for this is that the number of SNe with $\Delta M_{15}>1.4$ is small and the final solution is mostly driven by the SNe with $\Delta M_{15}<1.4$. The approach that we found to work best to address this problem was to split the sample into three bins -- SNe with $\Delta M_{15}<1.1$ (182 points), $1.1\leq\Delta M_{15}<1.5$ (163 points) and $\Delta M_{15}\geq1.5$ (48 points) -- and fit the histogram of $E(B-V)_\mathrm{obs}$ in each bin independently. In the fitting the $c_0$ parameter in Eq.~\ref{eq:colprob} was treated as an unknown constant. It is easy to see that with the correct $(B-V)_{B_\mathrm{max},0}- \Delta M_{15}$ relation, $\sigma_\mathrm{c}$ should be the same in all three bins and the $c_0$'s should be close to zero. Thus, the quantity to minimize was the scatter of $\sigma_\mathrm{c}$ around the mean in the three bins and the scatter of $c_0$ around zero added in quadrature. The $\tau_\mathrm{e}$ parameters were not required to be the same in the three bins because the distribution of the reddening may be different. For example, the SNe with larger $\Delta M_{15}$ are fainter and this may lead to bias against discovering reddened SNe. 

With the above procedure, the following function was derived
\begin{equation}
 (B-V)_{B_\mathrm{max},0}=0.036\,\exp\left[\frac{\Delta M_{15}-1.3}{0.3455}\right]-0.084 (\pm0.01).
 \label{eq:bv0}
\end{equation}
This relation is shown with the solid line in Fig.~\ref{f:bv_max}a and the histograms in the three bins and the best fits in Fig.~\ref{f:bv_max}b. The mean value of $\sigma_\mathrm{c}$ in the three bins is 0.045 mag with estimated uncertainty of 0.01 mag. The uncertainty of the free term in Eq.~\ref{eq:bv0}, $\pm0.01$ mag, was estimated from the fits of the histograms, which give uncertainty of $\sim0.01$ mag of $c_0$. The constant 1.3 was derived empirically. Several values between 1.1 and 1.4 were tried and the one providing the solution with smallest $\sigma_\mathrm{c}$ was selected. Regarding $\tau_\mathrm{e}$, the following values were derived in the three bins, $\tau_\mathrm{e}=0.14\pm0.02$ mag, $\tau_\mathrm{e}=0.13\pm0.02$ mag and $\tau_\mathrm{e}=0.11\pm0.04$ mag.  We note that the results are somehow dependent on the binning on $\Delta M_{15}$ and the color, and color error cut (0.04 mag in our case). When a larger sample of SNe, especially with $\Delta M_{15}>1.4$, becomes available it will be possible to estimate the parameters more accurately. 

The values of $\sigma_\mathrm{c}$ and $\tau_\mathrm{e}$ that we find are close to those of \cite{2007ApJ...659..122J}, \cite{2010ApJS..190..418G} and \cite{2010AJ....139..120F}, which were estimated from the late-time SN~Ia colors at $+35$ days using the Lira-relation \citep{lira,phil99}. The Lira-relation is considered as one of the best methods to estimate the host galaxy reddening of SNe Ia. In another analysis of late-time $B-V$ color curves of SNe~Ia \cite{2014ApJ...789...32B} introduced a new stretch parameter of the $B-V$ curves, $s_{BV}$. The authors estimated the intrinsic scatter of the $B-V$ color to be $0.06-0.09$ mag. The intrinsic scatter of $(B-V)_{B_\mathrm{max}}$ that we estimate is smaller than the scatter of $B-V$ at late times derived by the above authors. This indicates that at least for normal SNe Ia the $(B-V)_{B_\mathrm{max}}$ color should provide as good estimate of the reddening as the Lira-relation. This may be important for high-redshift studies where the observations do not extend to late phases and the extinction must be estimated from the colors at maximum.

As a final note, we would like to point out that the width of the observed color distribution is not determined only by the intrinsic color scatter $\sigma_\mathrm{c}$ and the distribution of the reddening (here parametrized by  $\tau_\mathrm{e}$). Several other other factors, such as the observational errors, the fact that $S$-corrections were not applied to all data and possible deviation of the fitting LC templates from the real SN LCs will all increase the width of the observed distribution and lead to overestimate of $\sigma_\mathrm{c}$ and $\tau_\mathrm{e}$. Some of these sources of errors may be different for the different samples and therefore $\sigma_\mathrm{c}$ should be regarded as a mean value for the whole nearby SN~Ia sample. The mean uncertainty  of $(B-V)_{B_\mathrm{max}}$ in the sample used in this analysis is $\sim0.015$ mag and taking it into account the true intrinsic scatter $\sigma_\mathrm{c}$ is likely $\sim0.04$~mag.  

\end{appendix}

\end{document}